\def\figbzk{
\begin{figure*}
 \epsscale{1}
 \plotone{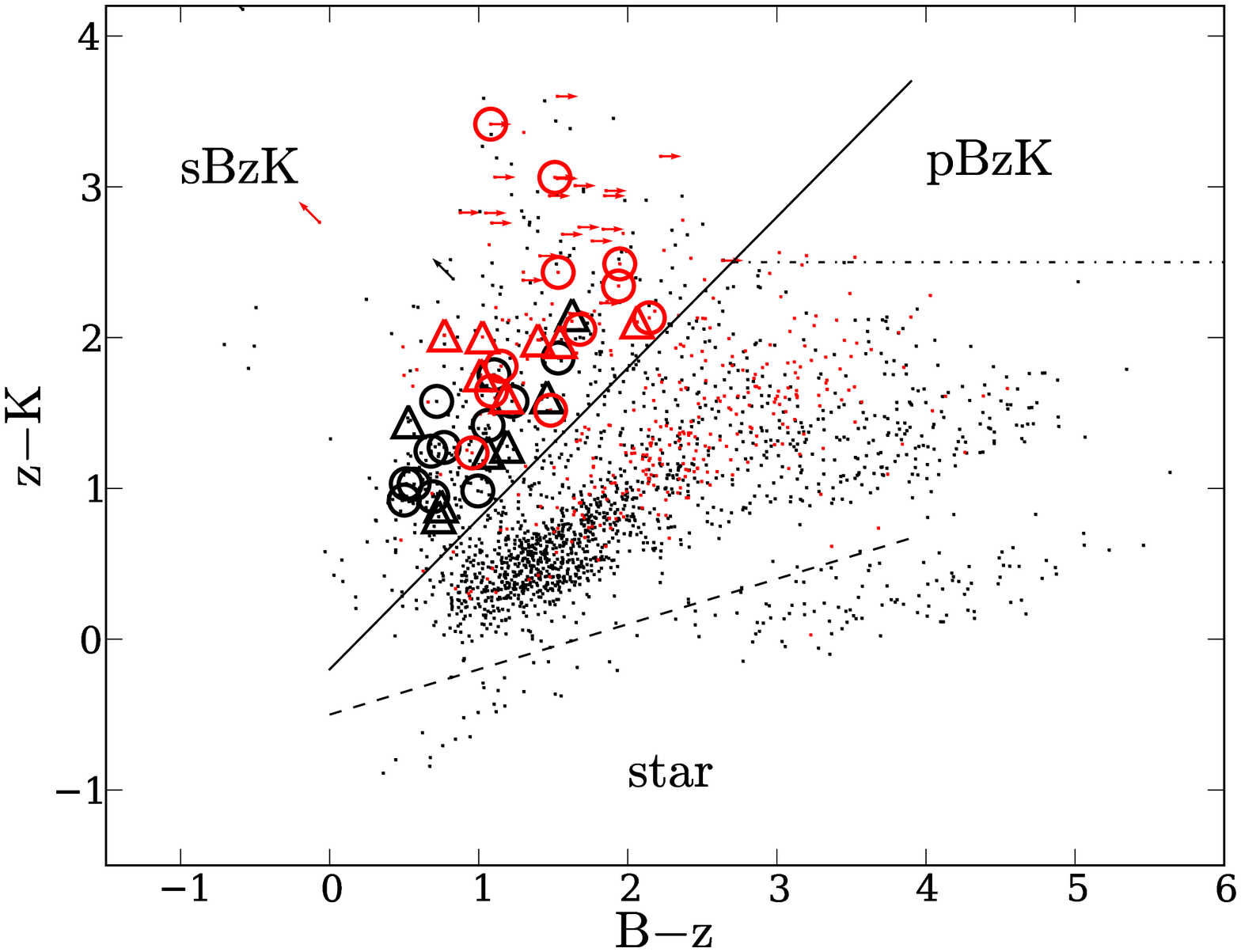} 
 \caption{BzK color-color diagram of the $K_s < 23$ objects from the
   $K_s$-band selected catalog of the MOIRCS Deep Survey (MODS;
   \citealp{Kajisawa2009}). Red symbols show the MIPS sources listed
   in the public catalog (Chary et al. in preparation). The sample for
   the present work is denoted by circles and triangles for the
   objects with emission lines detected and not detected,
   respectively. Arrows show upper/lower limits for sBzK
   galaxies. \label{fig:bzk}}
\end{figure*}
}
\def\fighistbzk{
\begin{figure*}
 \epsscale{1}
 \plottwo{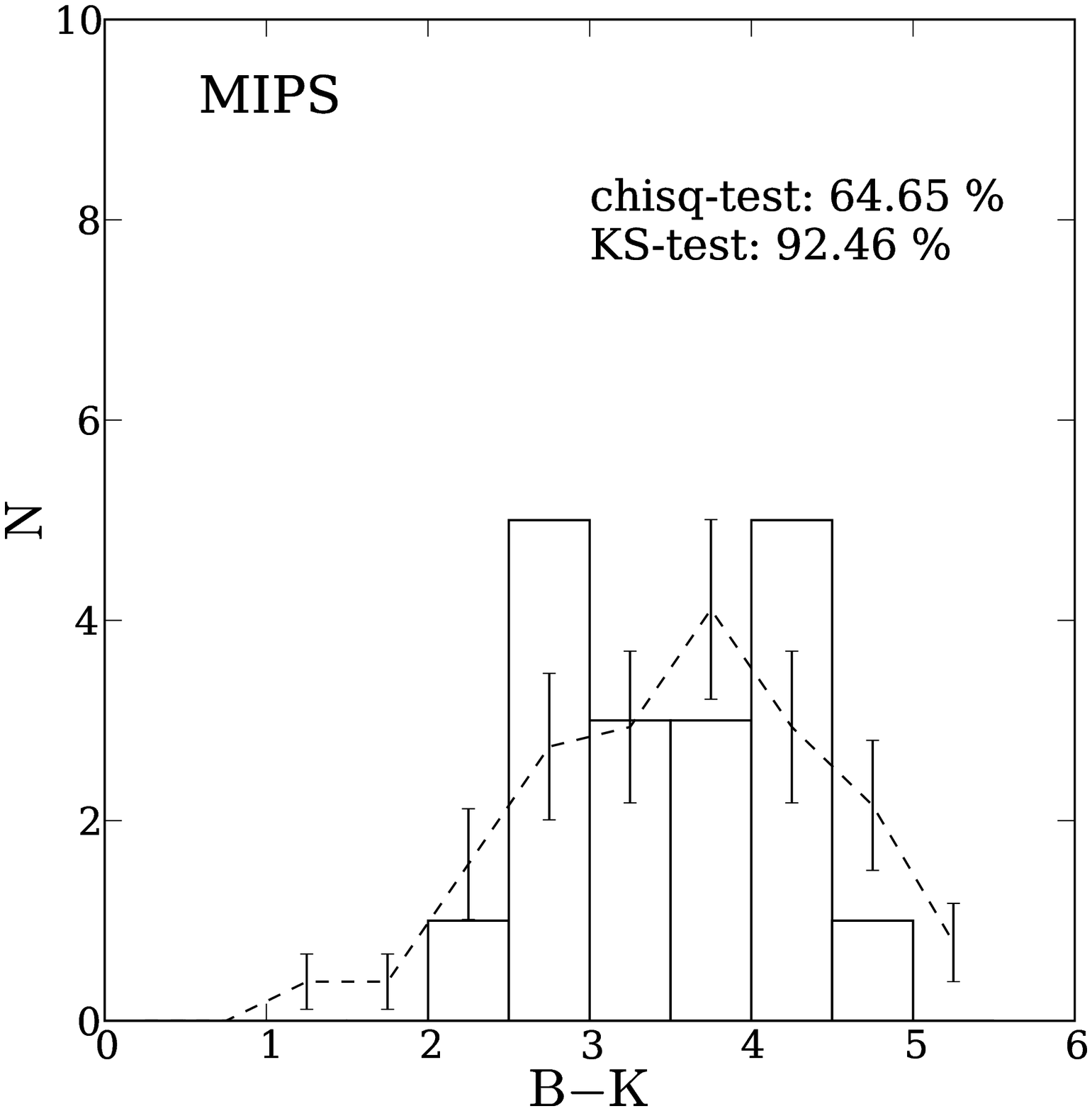}{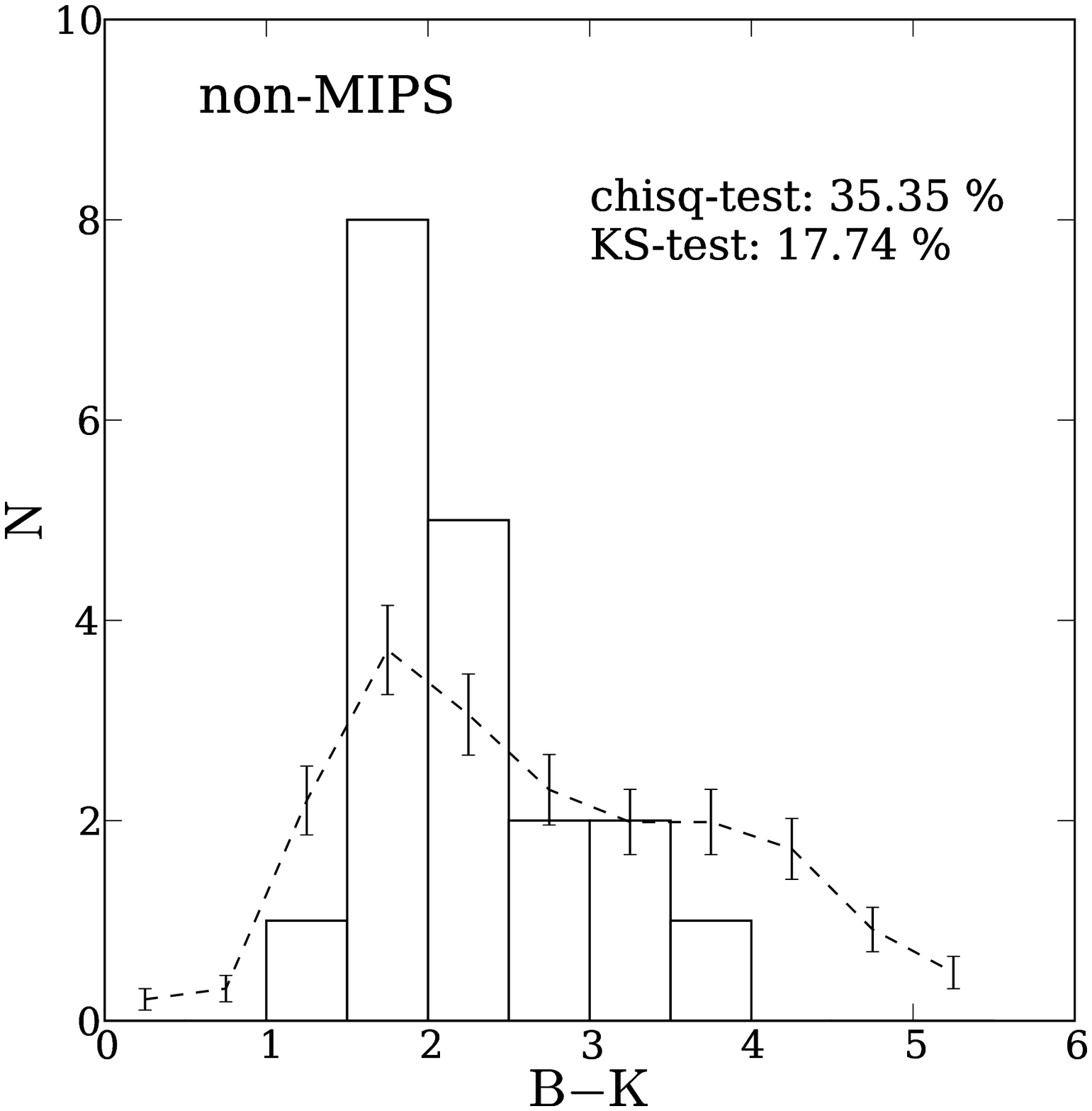}
 \caption{ $B-K$ color distribution of sBzK-MIPS ({\it left}) and
   sBzK-non-MIPS ({\it right}) galaxies (see text). The bar plots show
   our observed sample. The dashed line with error bars shows the
   scaled color distribution of all of the $K_s<23$ sBzK-MIPS and
   sBzK-non-MIPS galaxies, respectively. The error bars show the
   Poisson errors at each bins. \label{fig:hist_bzk}}
\end{figure*}
}
\def\fighaspec{
\begin{figure*}
 \epsscale{1}
 \plotone{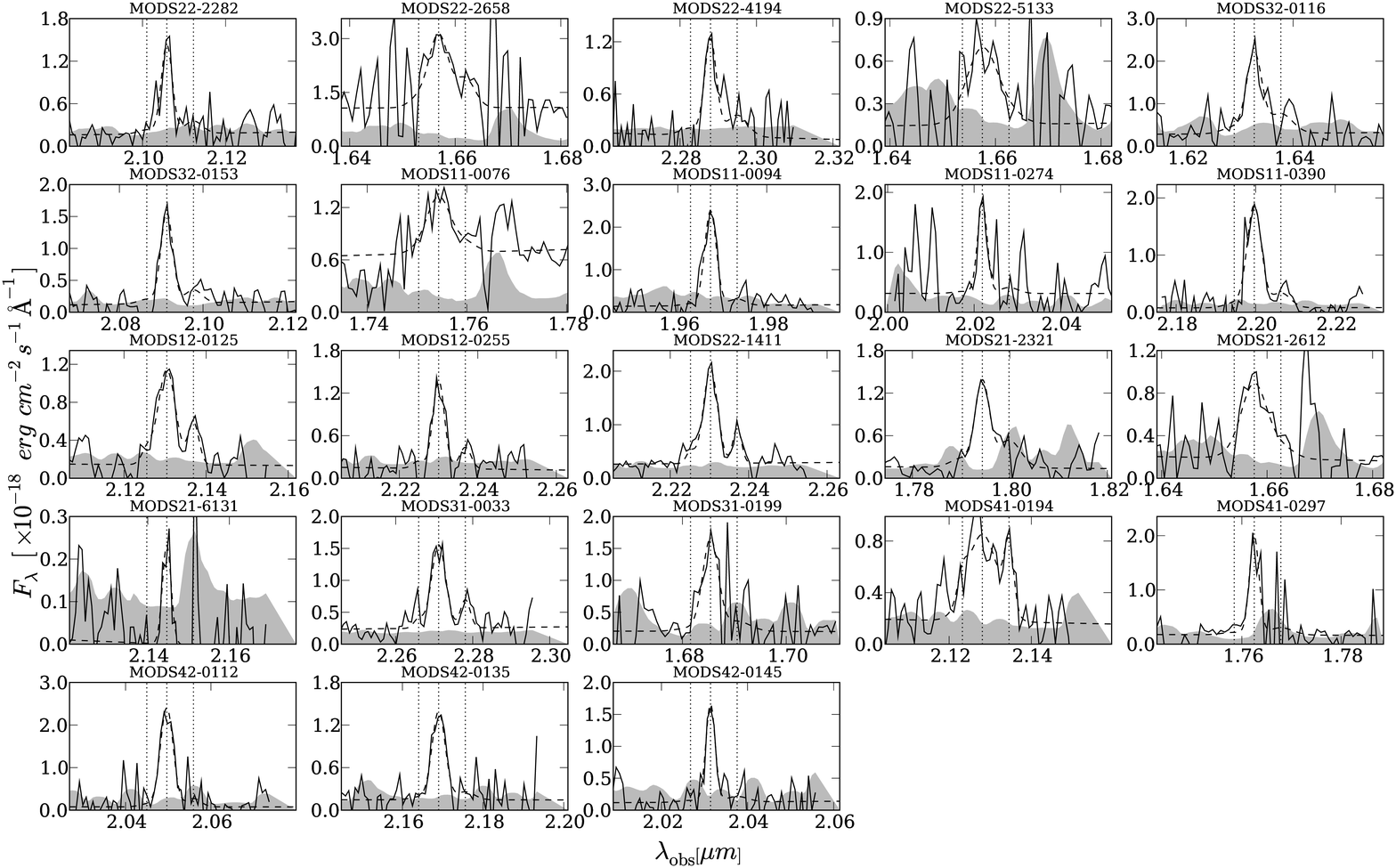}
 \caption{ \ntwo-H$\alpha$ spectra of the emission-line galaxies.
   Shaded region shows the size of the error as a function of
   wavelength. The dashed line shows the spectrum of the best-fit
   model. The vertical dotted lines in each panel show, from left to
   right, the locations of \ntwo$\lambda$6548, H$\alpha$,
   \ntwo$\lambda$6583.
 \label{fig:haspec}}
\end{figure*}
}
\def\figothreespec{
\begin{figure*}
 \epsscale{1}
 \plotone{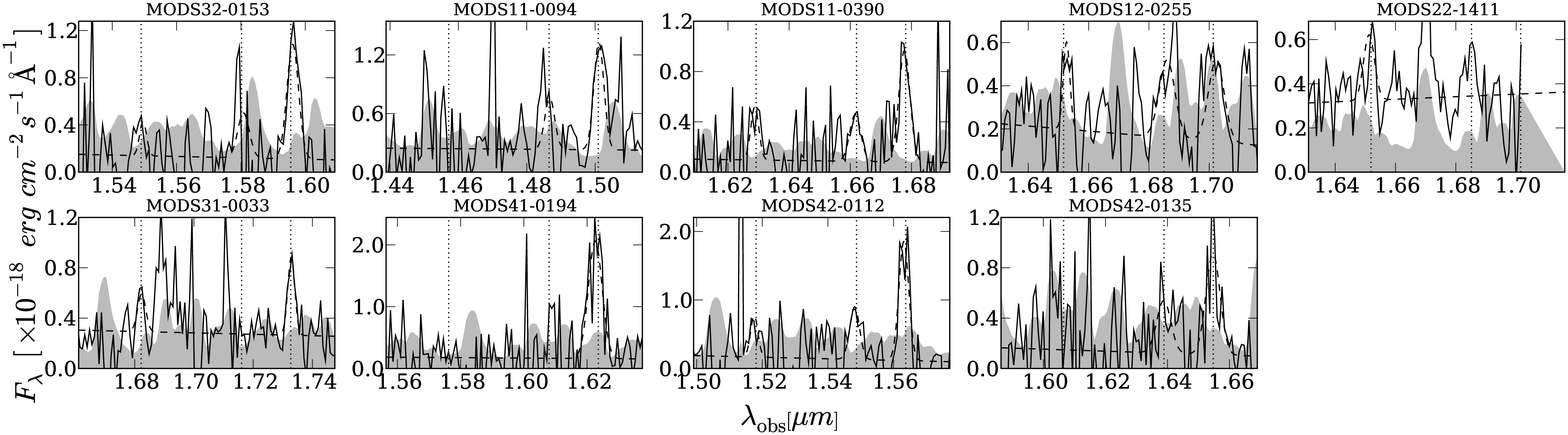}
 \caption{ Same as Figure \ref{fig:haspec}, but for \othree-H$\beta$
   spectra of the emission line galaxies with S/N$>$3.  The vertical
   dotted lines in each panel show, from left to right, the locations
   of H$\beta$, \othree$\lambda$4959, \othree$\lambda$5007.
 \label{fig:o3spec}}
\end{figure*}
}
\def\figothreestack{
\begin{figure*}
 \epsscale{1}
 \plottwo{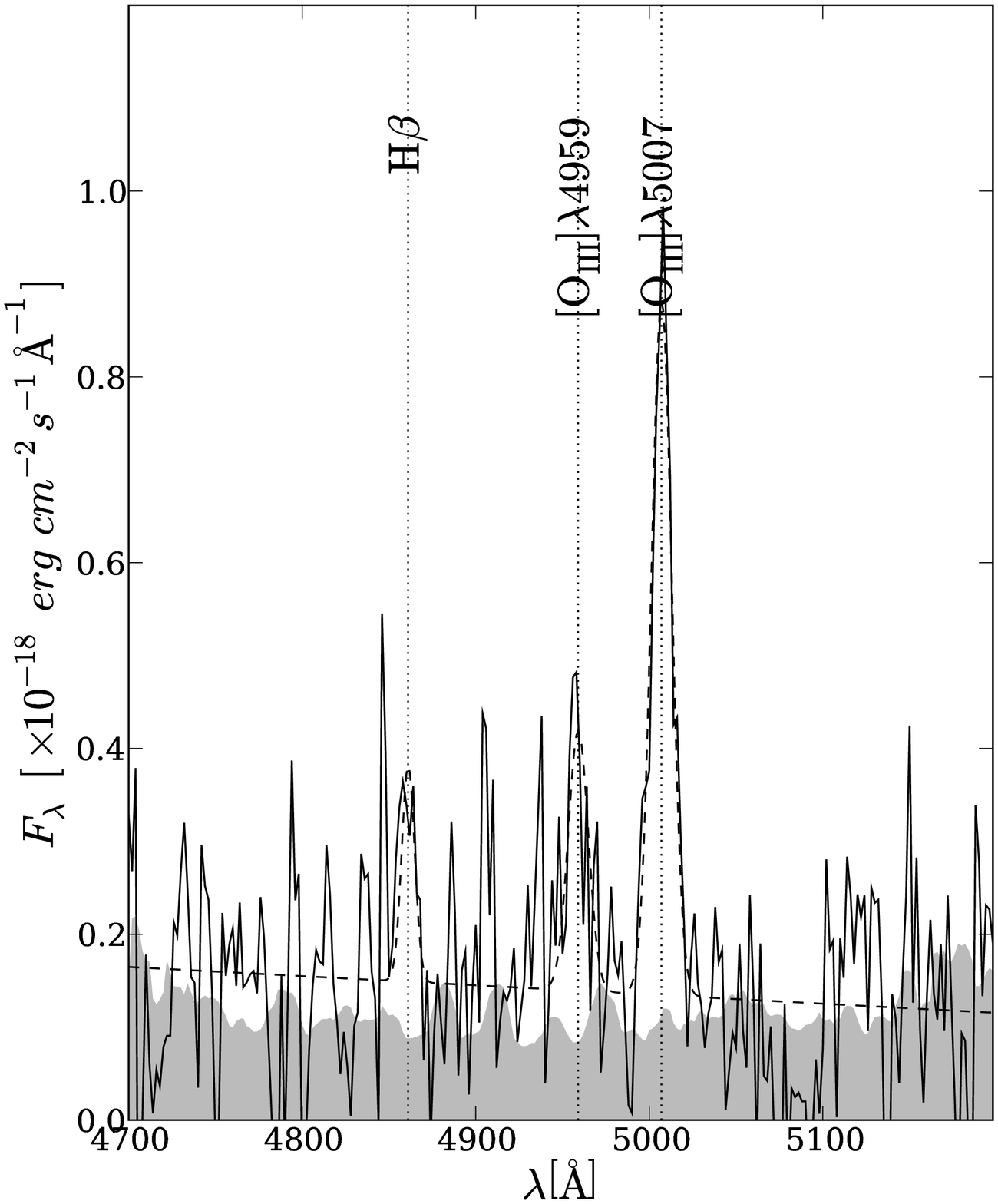}{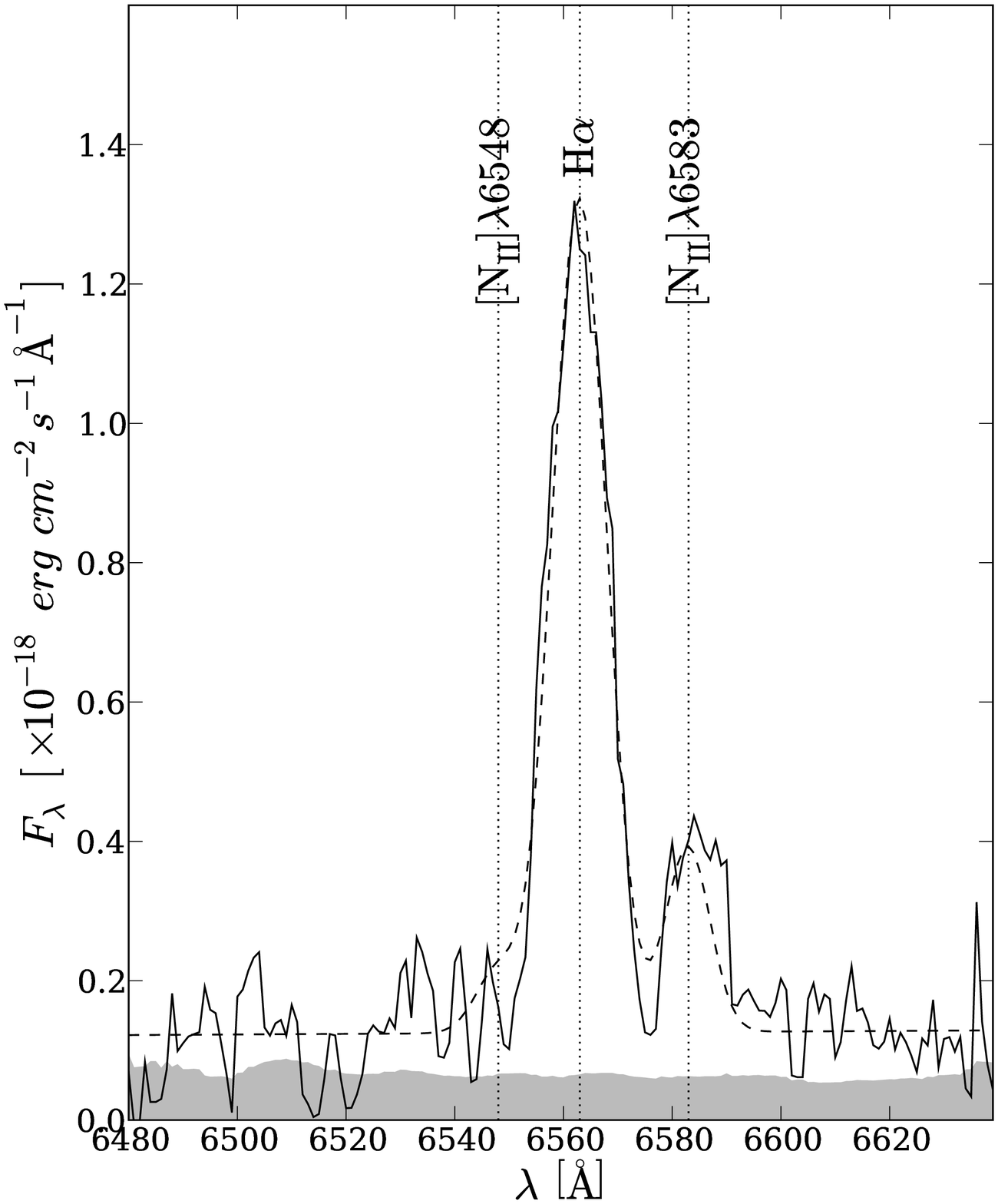}
 \caption{ Same as Figure \ref{fig:haspec}, but for the stacked
   \othree-H$\beta$ ({\it left}) and the stacked \ntwo-H$\alpha$ ({\it
     right}) spectra. The objects which have \othree\, and/or H$\beta$
   emission are stacked for both spectra.
 \label{fig:o3stack}}
\end{figure*}
}
\def\figsed{
\begin{figure*}
 \epsscale{1}
 \plottwo{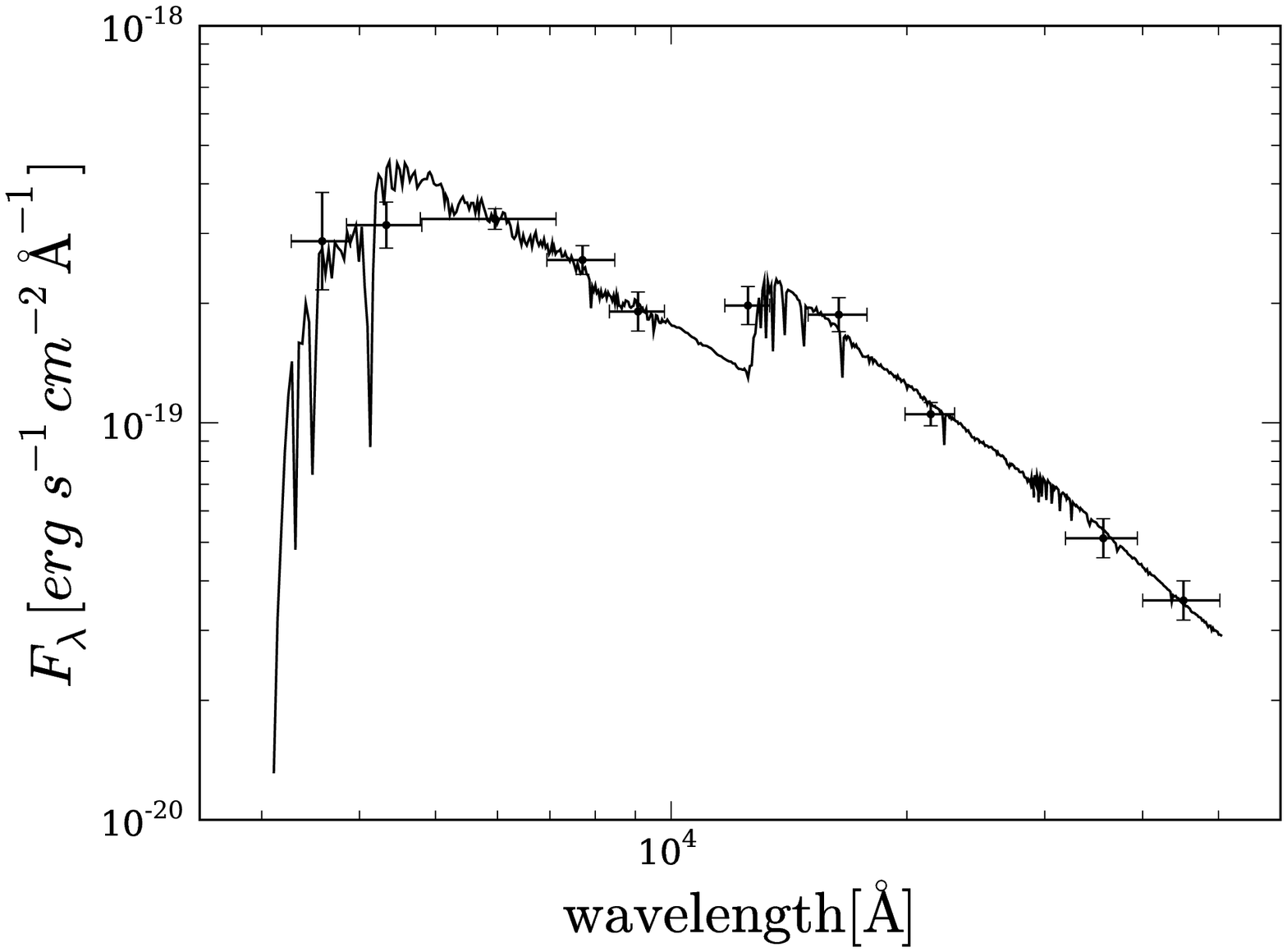}{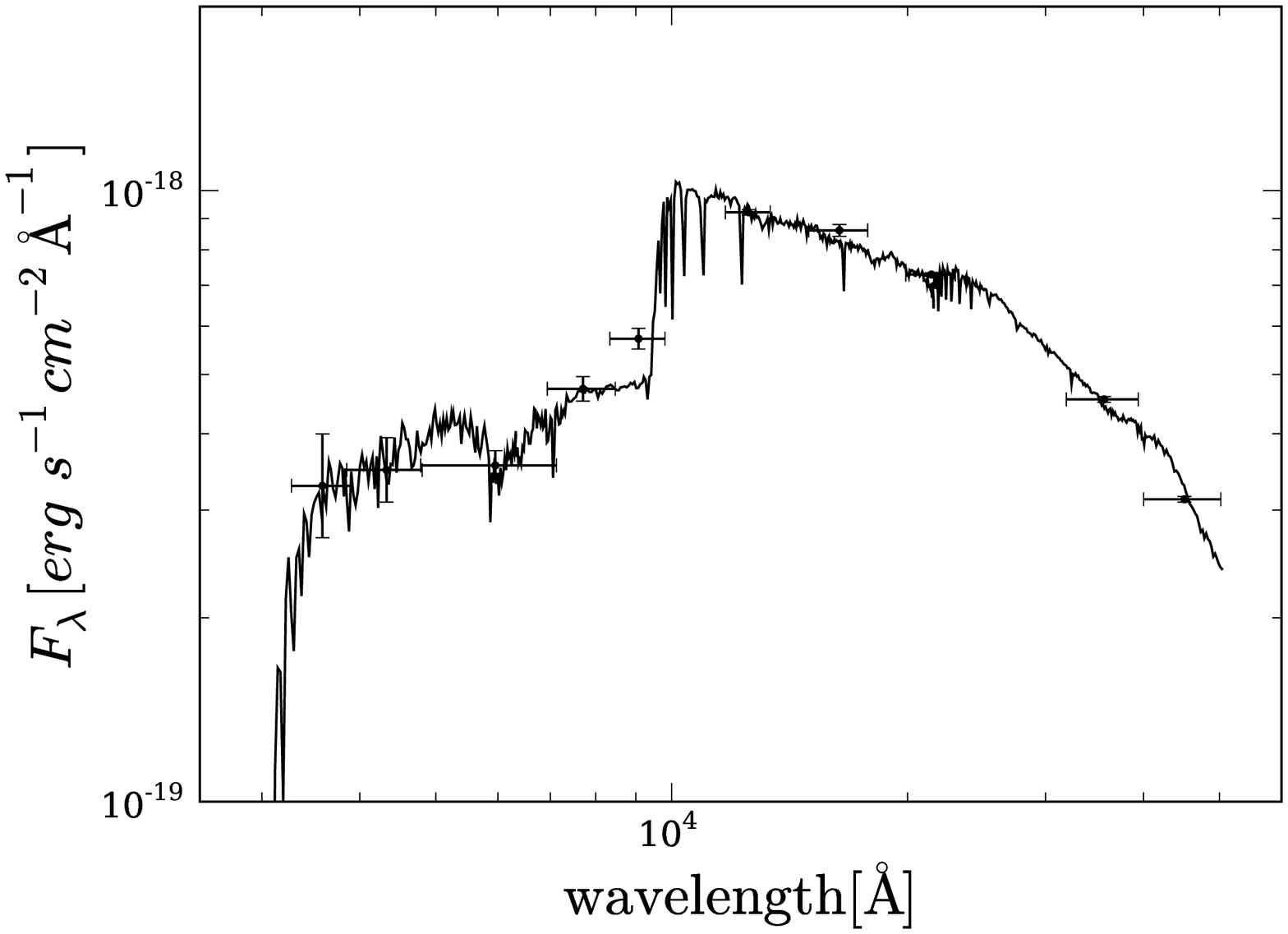}
 \plottwo{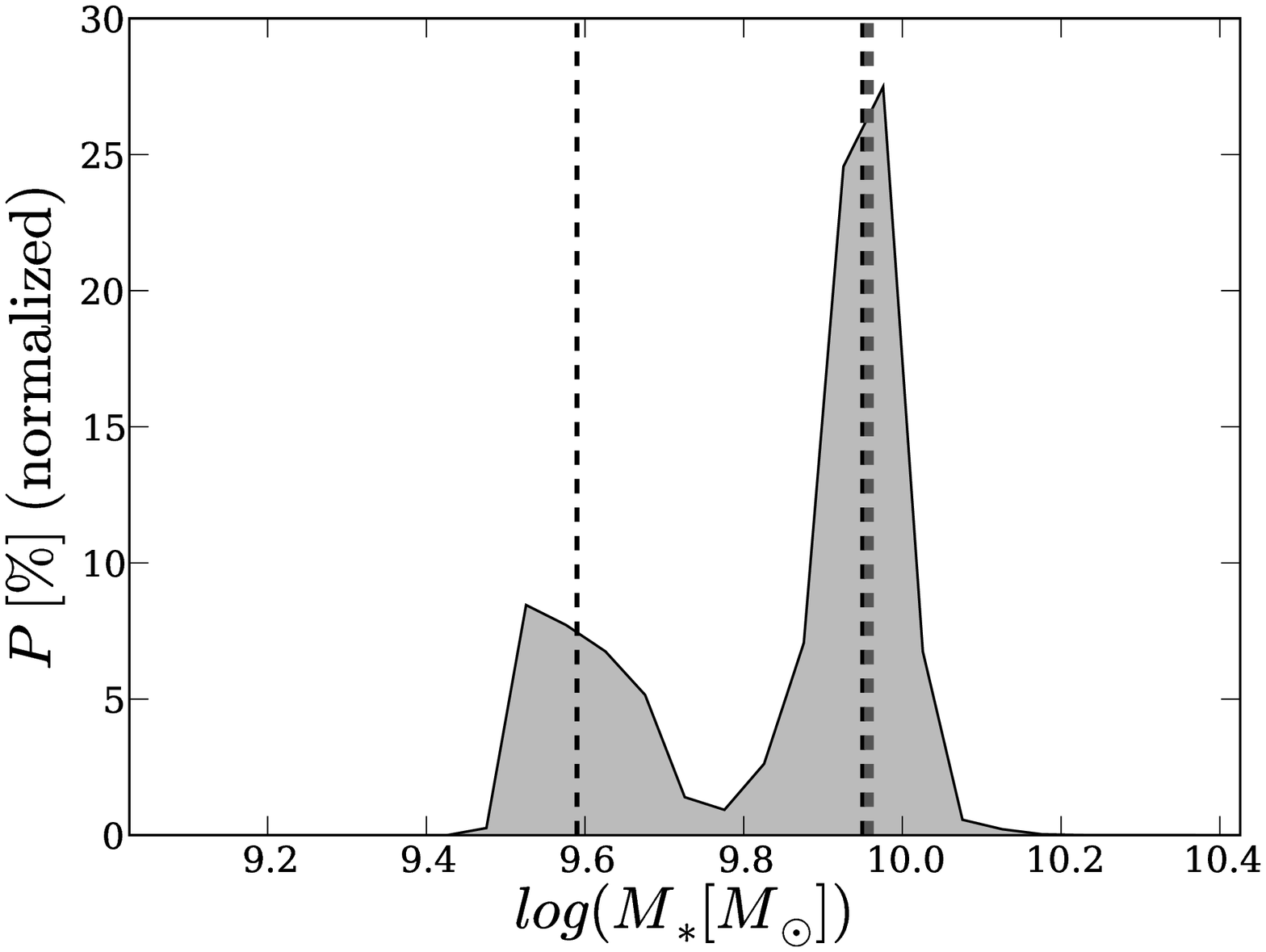}{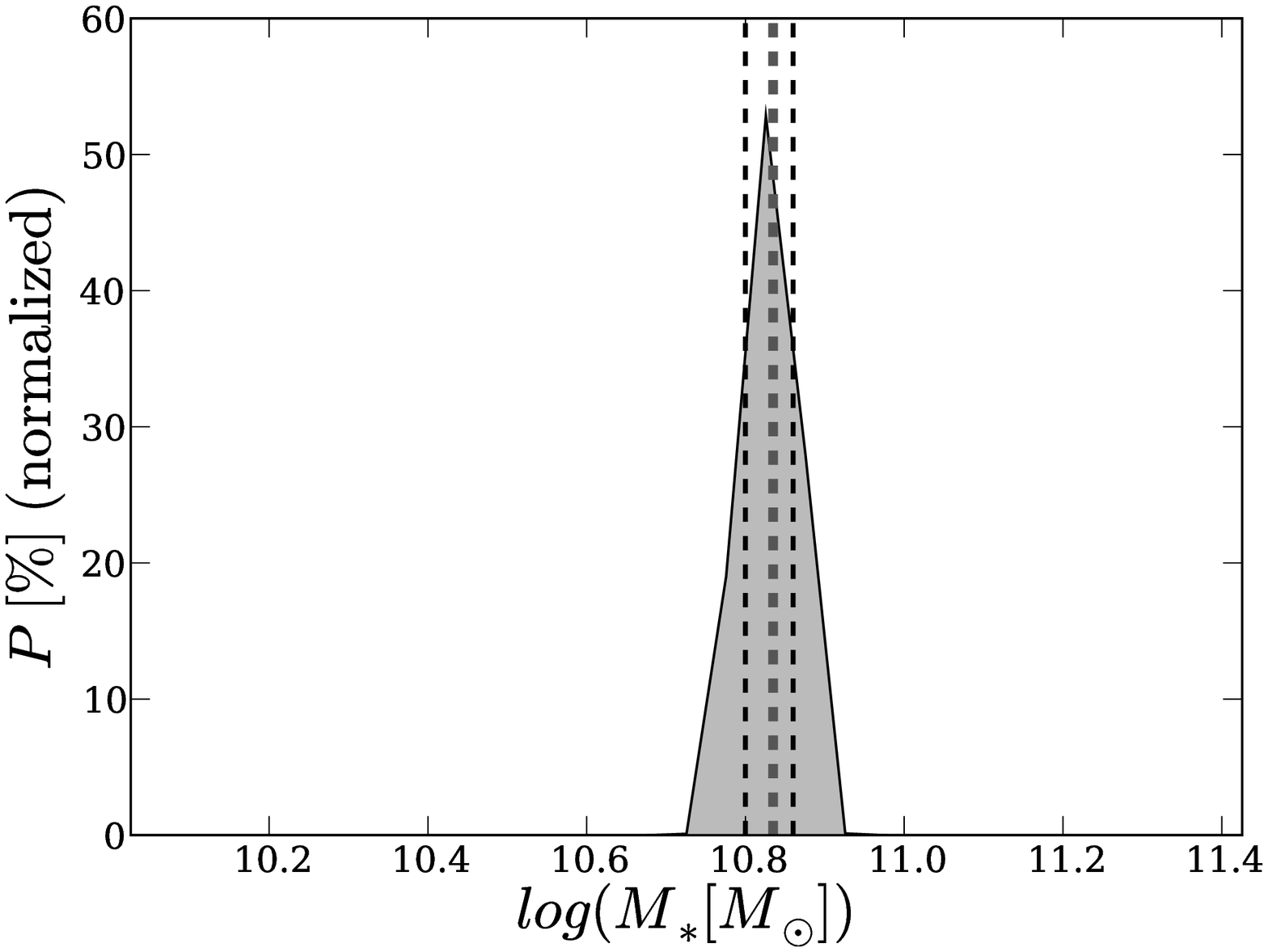}
 \plottwo{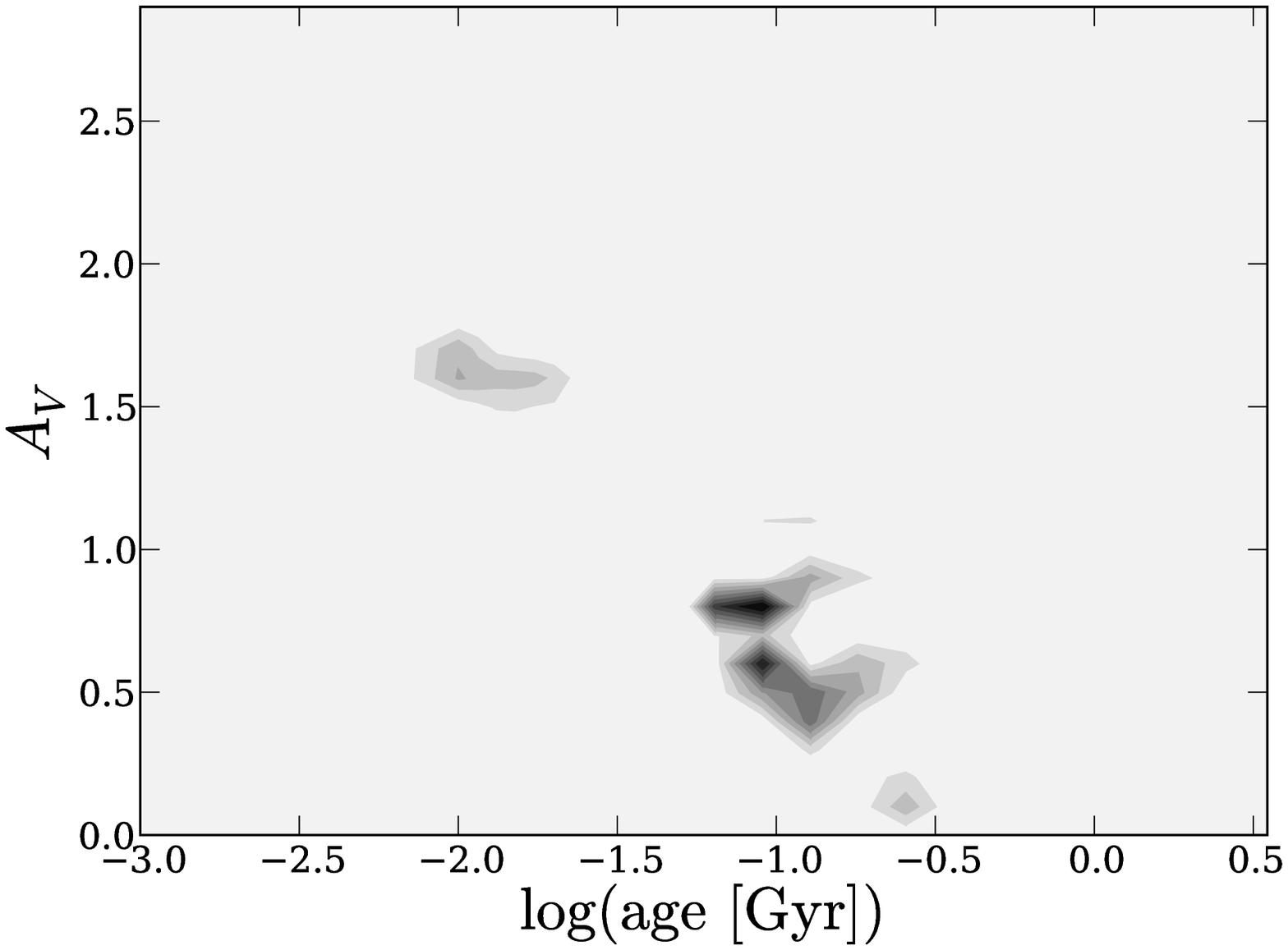}{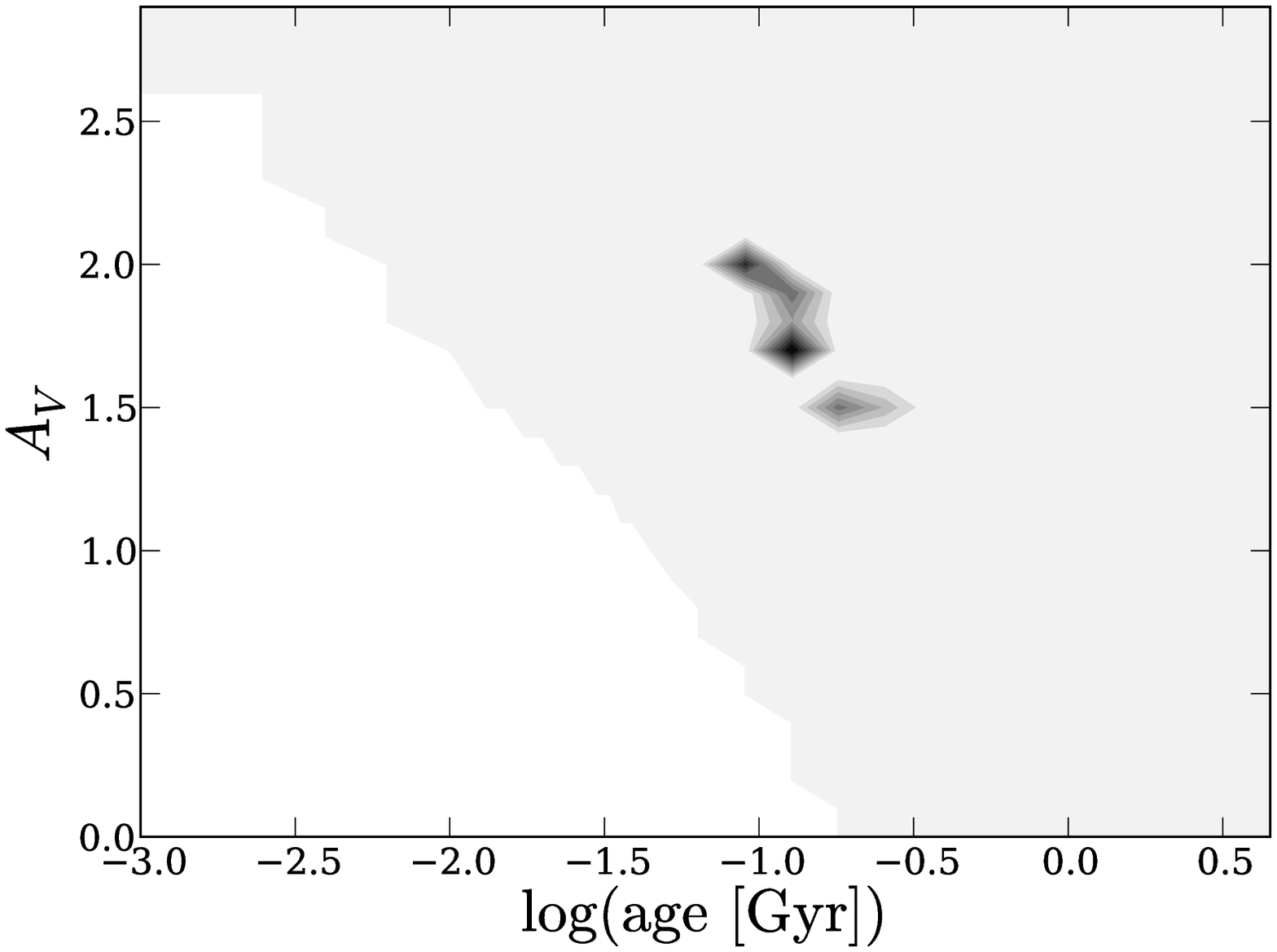}
 \caption{Examples of best fit SEDs ({\it top}) and probability distribution against
 stellar mass ({\it middle}) and age versus the amount of extinction ({\it
 bottom}). {\it Left} and {\it Right} panels show MODS12-1025 (non-MIPS) and
 MODS22-2658 (MIPS), respectively.
 In the middle diagrams, the dashed lines show the range of 68\%
 confidence intervals (1$\sigma$ error), and the thick dashed line shows
 the best-fit stellar mass.
 \label{fig:sed}} 
\end{figure*}
}
\def\fighalum{
\begin{figure*}
 \epsscale{1}
 \plottwo{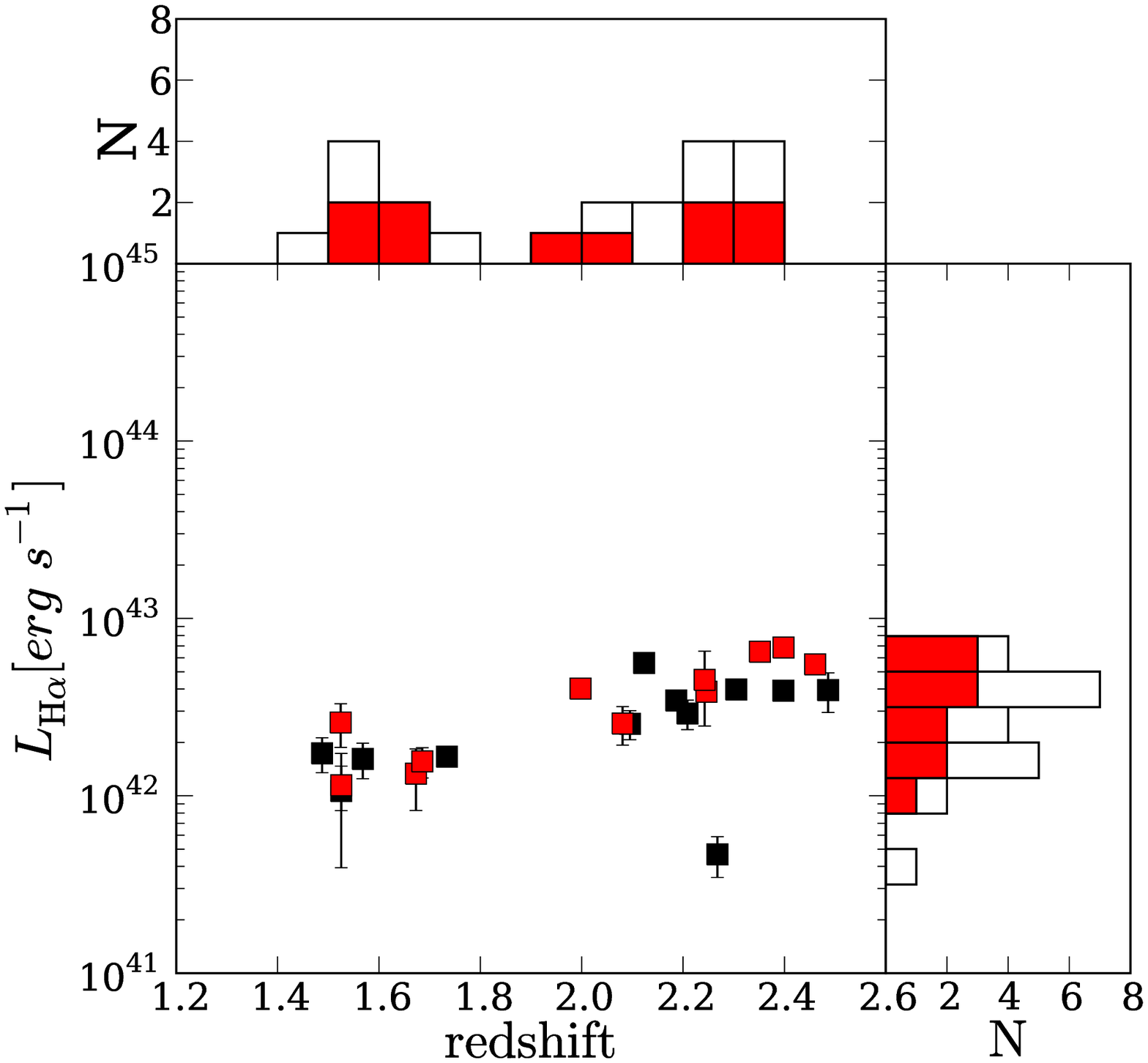}{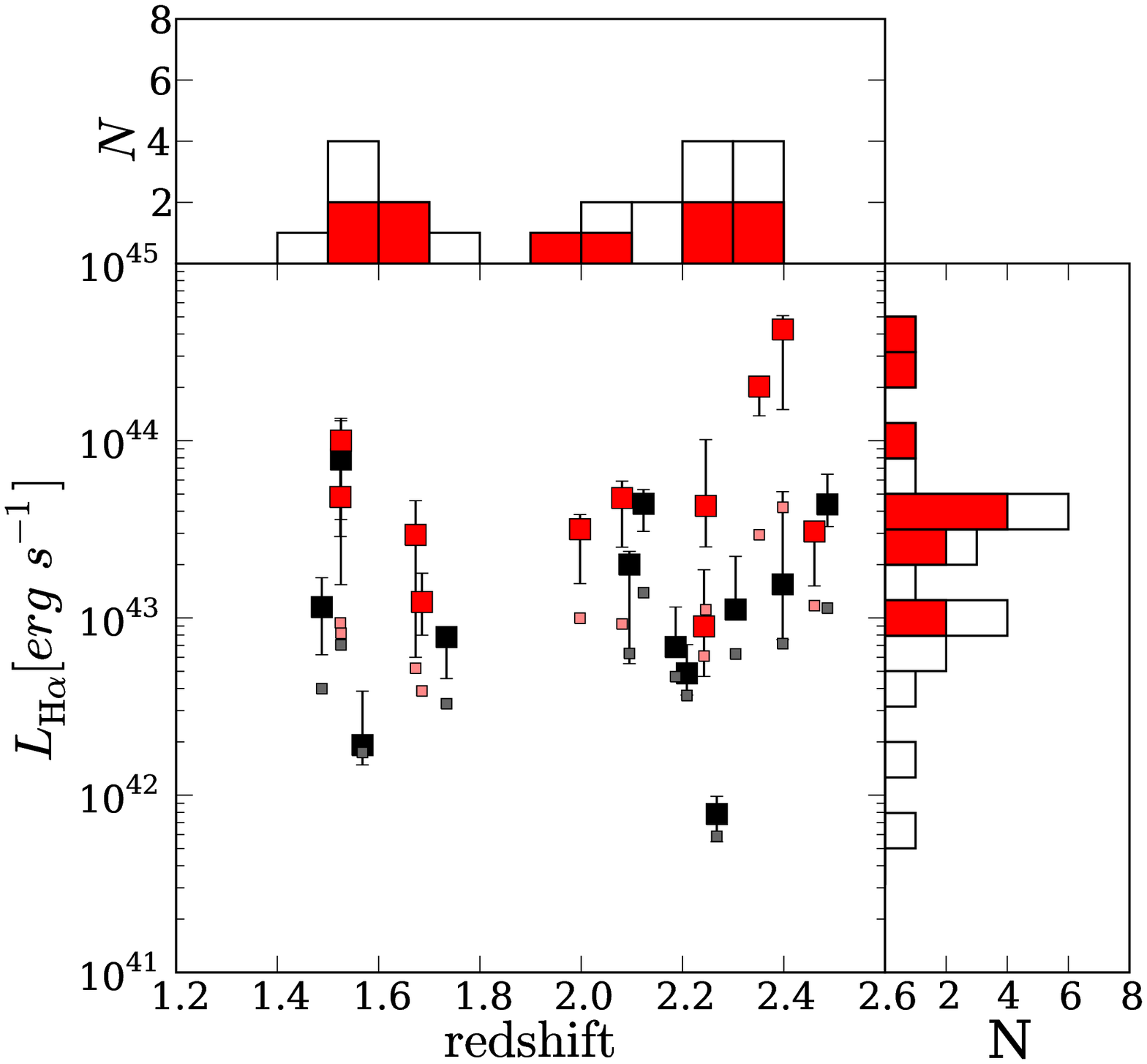}
 \caption{Luminosity of H$\alpha$ emission lines and redshift
   distribution of the H$\alpha$-detected sBzK galaxies before ({\em
     left}) and after ({\em right}) extinction correction. Red and
   black symbols show sBzK-MIPS and sBzK-non-MIPS galaxies,
   respectively. In the right figure, small symbols show the results
   of the equal-extinction case correction to the stellar continuum.
\label{fig:halum}}
\end{figure*}
}
\def\figbpt{
\begin{figure}
 \epsscale{1}
 \plotone{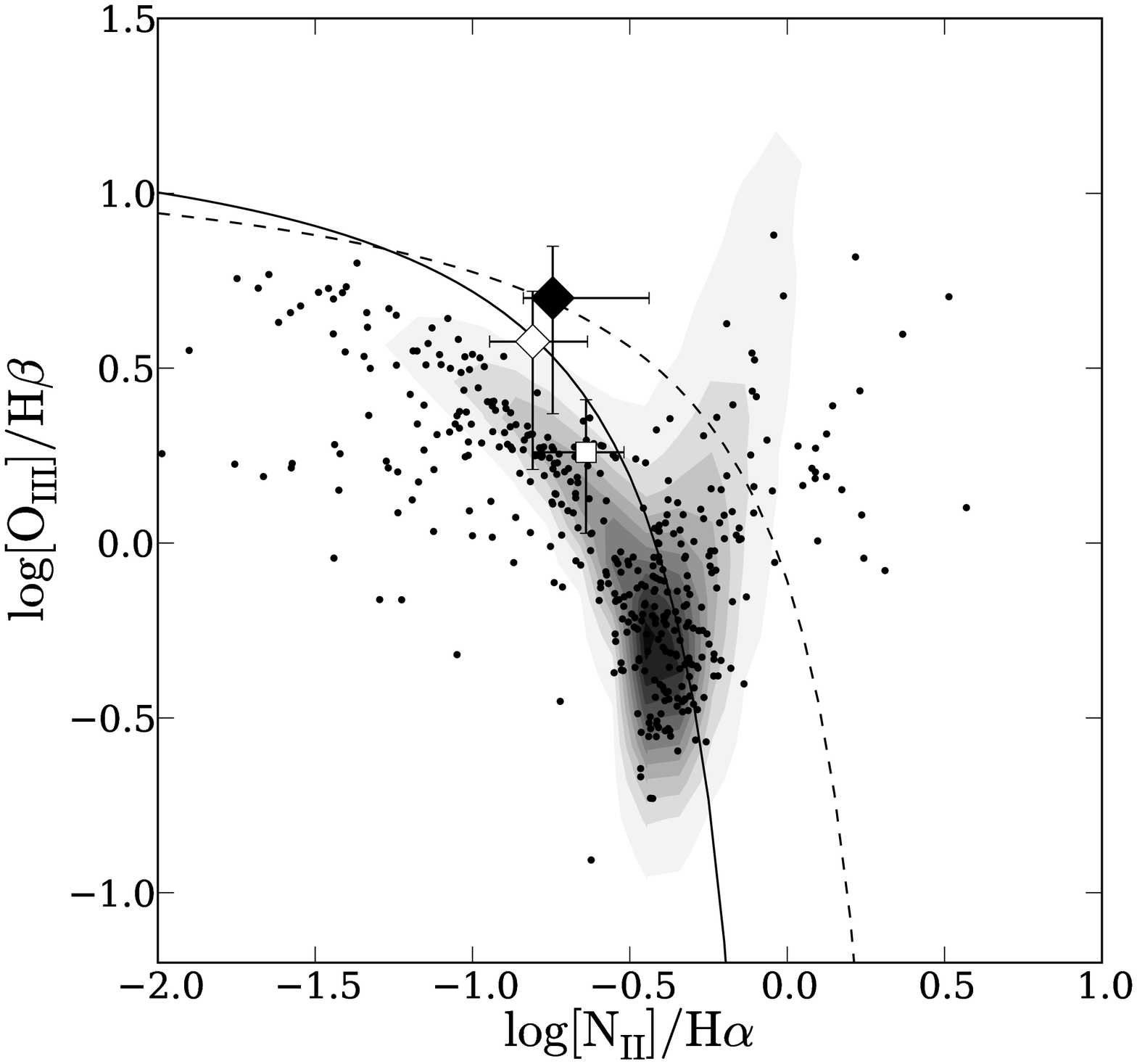}
 \caption{\ntwo-H$\alpha$ and \othree-H$\beta$ line ratios. The
 black diamond shows the line ratios of the stacked spectra, while the
 white diamond is the stacked spectra with MODS41-0194 (a possible AGN
 candidate; see text) excluded. The open square is MODS31-0033, in
 which the four emission lines are detected.  The contours in gray
 scale show the distribution of $\sim347000$ objects from SDSS DR6,
 while the black dots show local starburst galaxies analyzed
 by \citet{Moustakas2006a}. The solid line is the empirically
 determined classification line derived for the SDSS objects
 by \citet{Kauffmann2003}. Objects below and to the left of this line
 are normal starburst galaxies. The dashed line is the classification
 boundary theoretically determined by \citet{Kewley2001a}.
 \label{fig:bpt}}
\end{figure}
}
\def\figextinction{
\begin{figure}
 \epsscale{1}
 \plotone{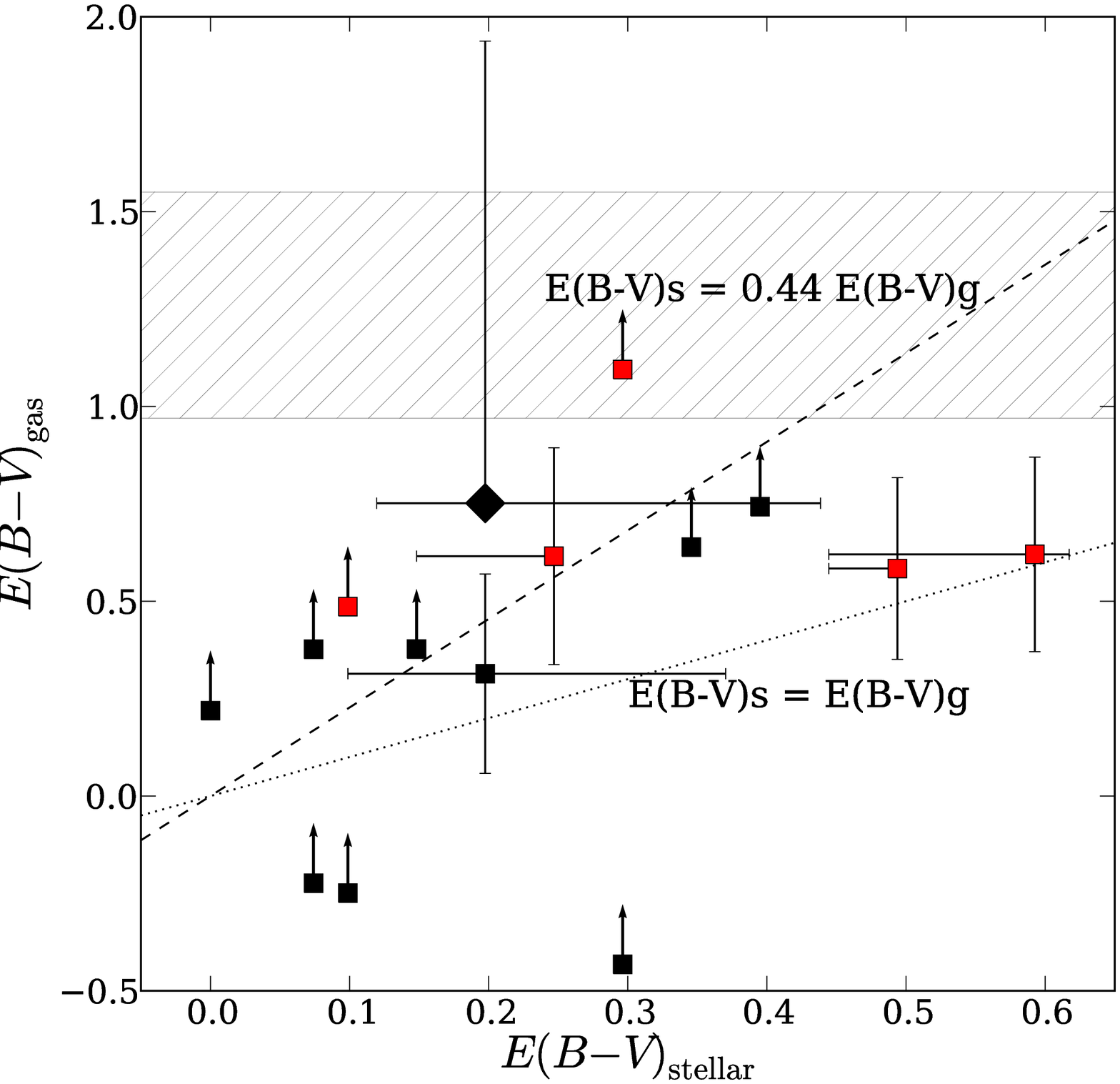}
 \caption{Comparison of dust extinction derived from an SED fit to
   that derived from the Balmer decrement. The black diamond shows the
   stacked spectrum. Red and black squares show sBzK-MIPS and
   sBzK-non-MIPS galaxies, respectively. Squares with error bars are
   the galaxies with an H$\beta$ emission line detected (MODS11-0390,
   MODS12-0255, MODS22-1411, MODS31-0033), while those with upper
   arrows are the galaxies with an H$\beta$ upper limit
   (3$\sigma$). The hatched region shows the 3$\sigma$ detection limit
   of H$\beta$ estimated from H$\alpha$ flux. The height of the region
   is calculated from the 68\% distribution of the H$\alpha$
   fluxes. The dotted and dashed lines illustrate the equal-extinction
   case and the relation given by \citet{Calzetti2001} (Equation
   \ref{eq:ebv}), respectively.
 \label{fig:extinction}}
\end{figure}
}
\def\figsedms{
\begin{figure*}
 \epsscale{1}
 \plottwo{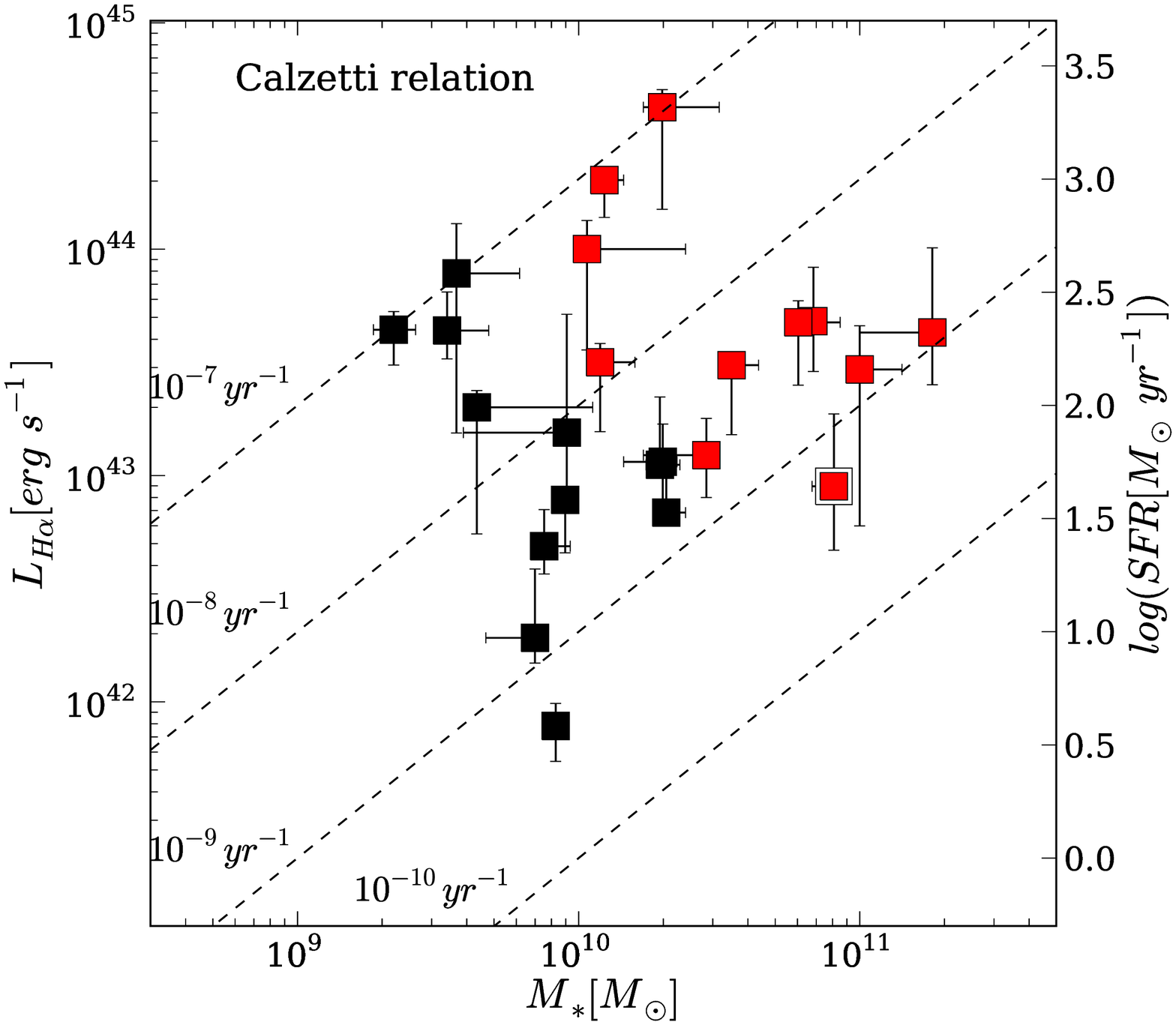}{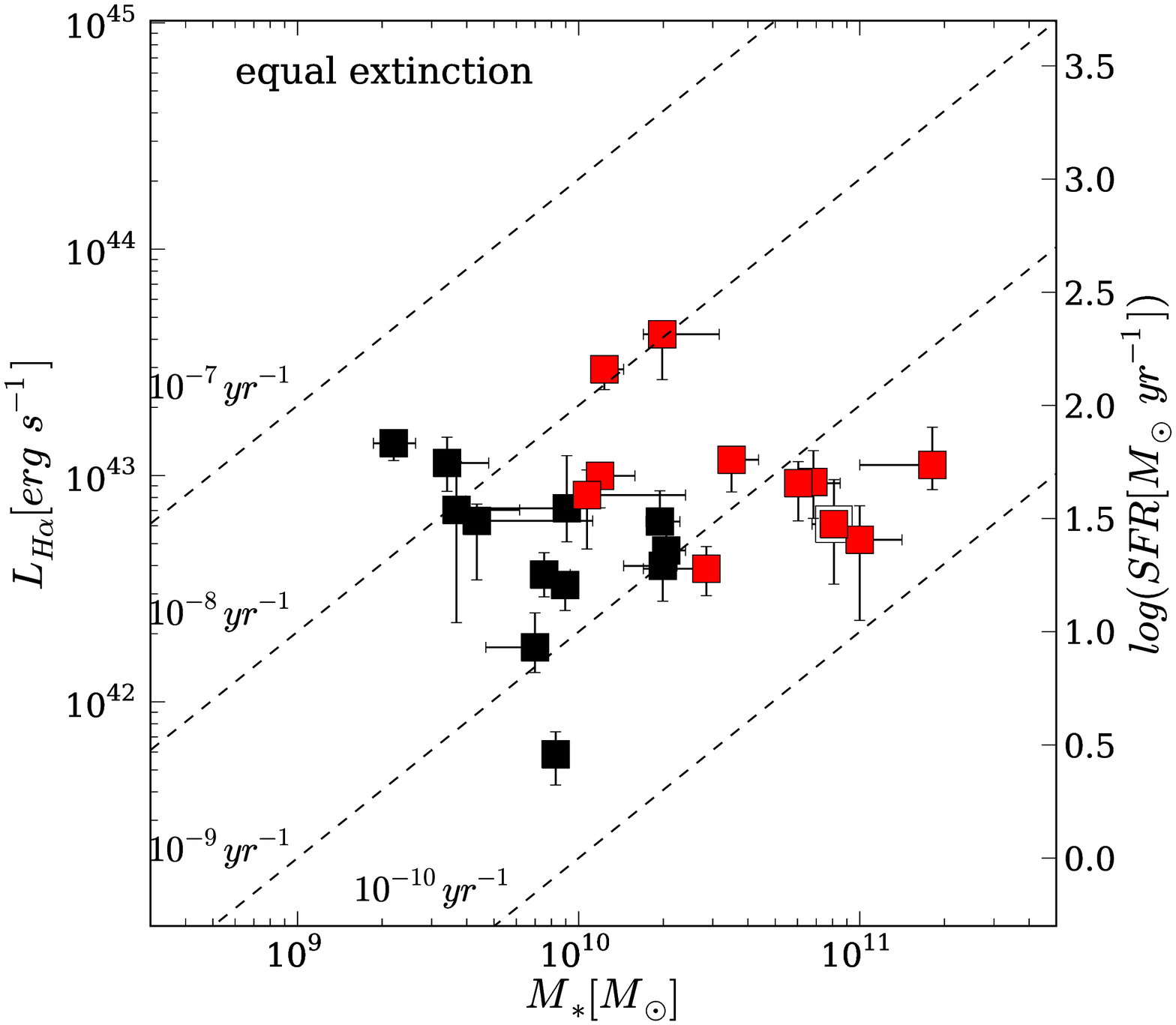}
 \caption{ H$\alpha$ luminosity with extinction corrected using the
   Calzetti relation ({\it left}) and the equal-extinction case ({\it
     right}) as a function of stellar mass derived from SED
   fitting. Red and black squares show sBzK-MIPS and sBzK-non-MIPS
   galaxies, respectively. MODS41-0194, which is the possible AGN
   candidate discussed in \S \ref{subsec:bpt}, is denoted by an open
   square.  ${\rm SFR}_{H\alpha}$ (the right axis) is calculated by
   Equation. \ref{eq:hasfr}.  The dashed lines indicate SSFRs.
 \label{fig:sed_ms}}
\end{figure*}
}
\def\figagests{
\begin{figure*}
 \epsscale{1}
 \plottwo{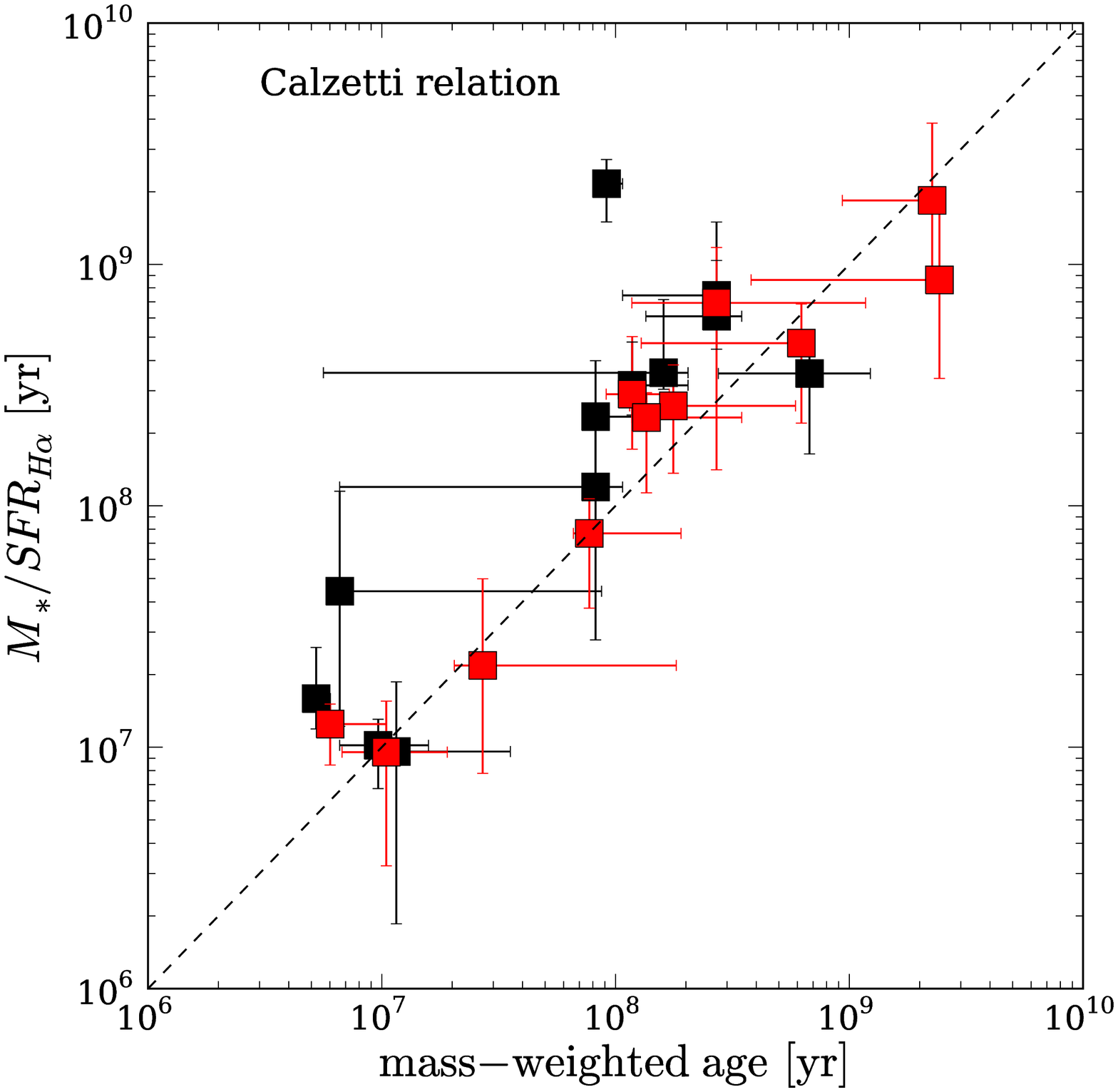}{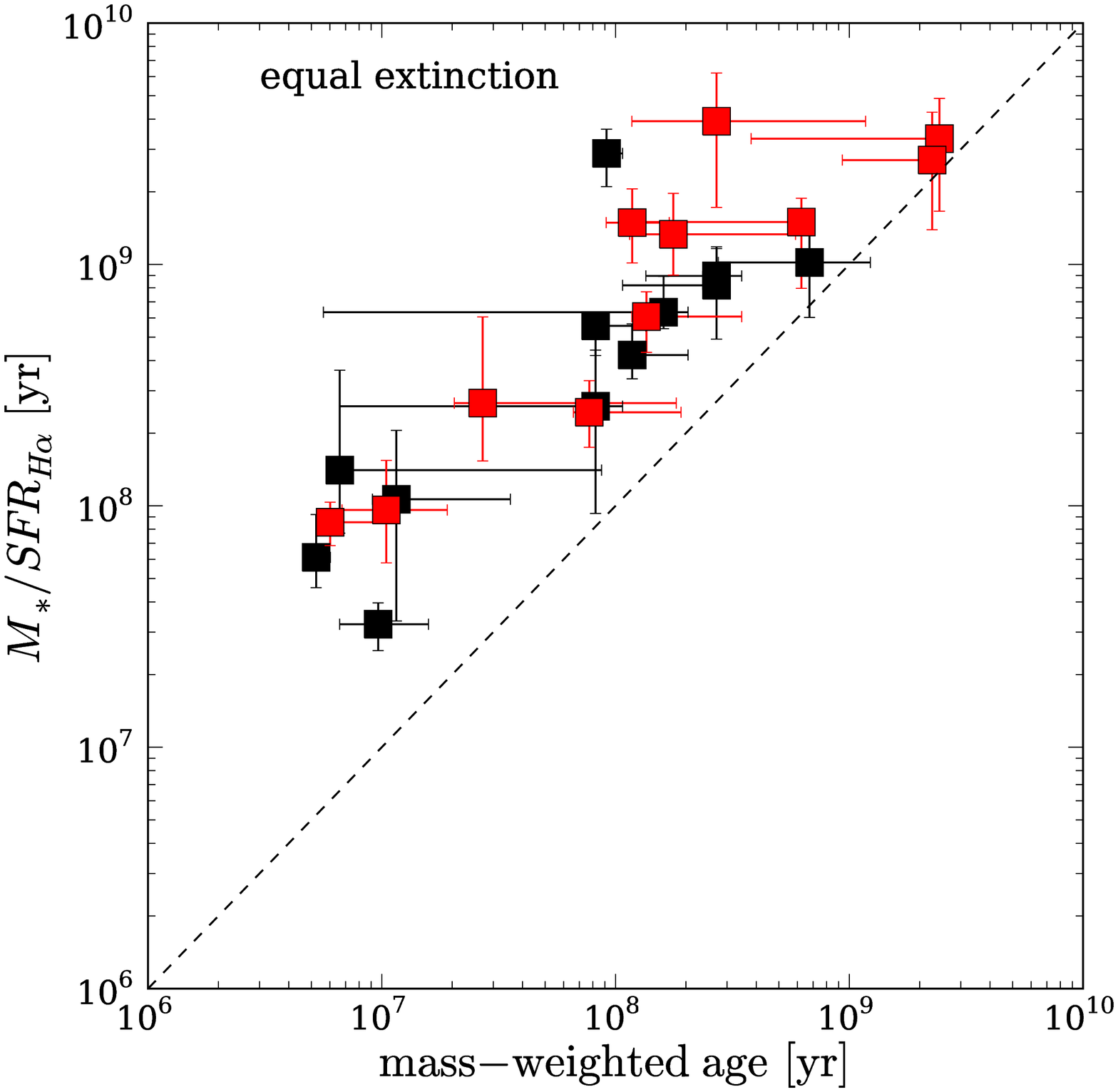}
 \caption{ SSFR age (${\rm SSFR}^{-1}$) with extinction corrected
   using the Calzetti relation ({\it left}) and the equal-extinction
   case ({\it right}) as a function of mass weighted age. The dashed
   line shows the equal age. Symbols are same as Figure
   \ref{fig:sed_ms}.
 \label{fig:age_sts}}
\end{figure*}
}
\def\figsfrcompuv{
\begin{figure*}
 \epsscale{1}
 \plottwo{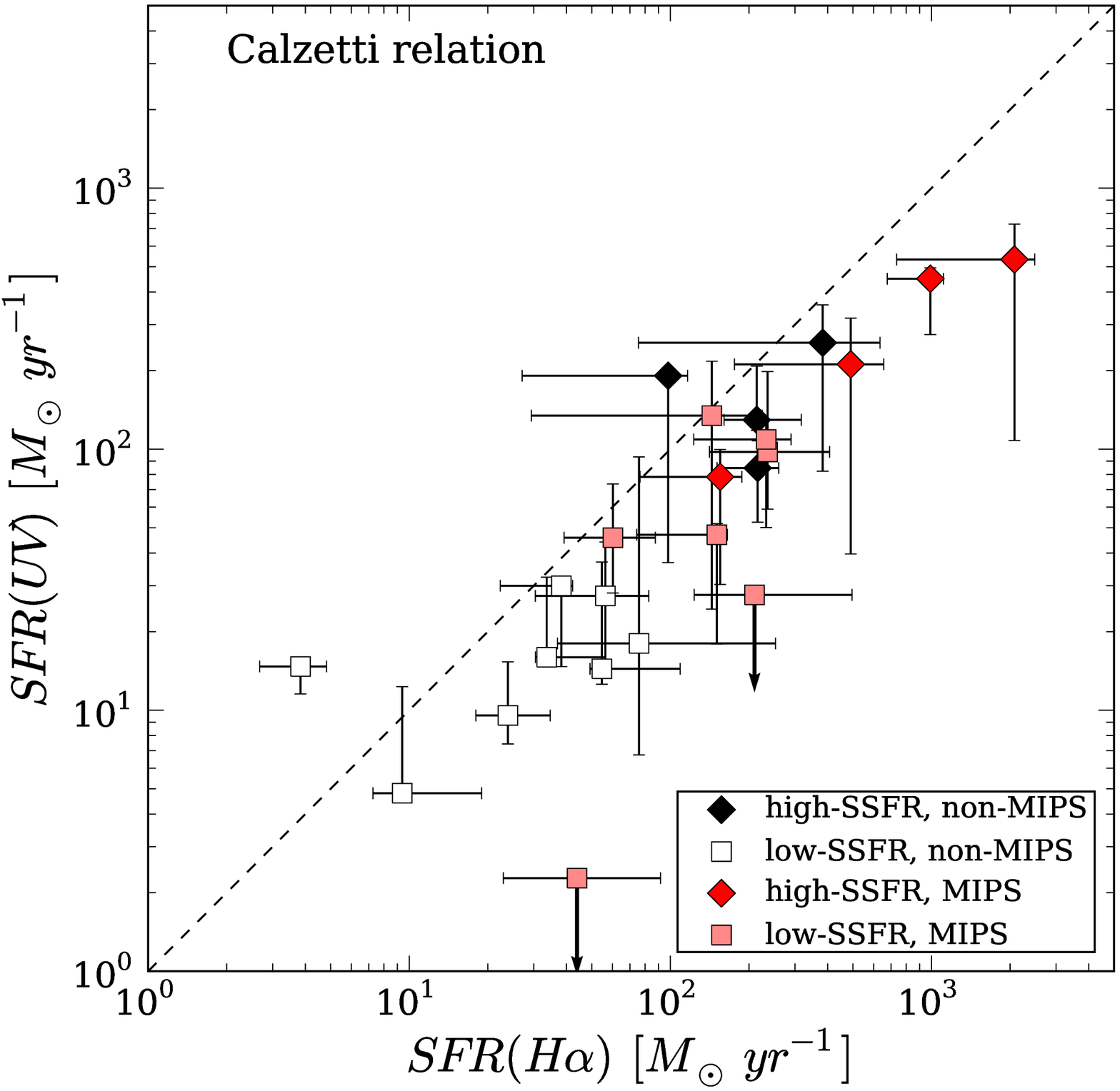}{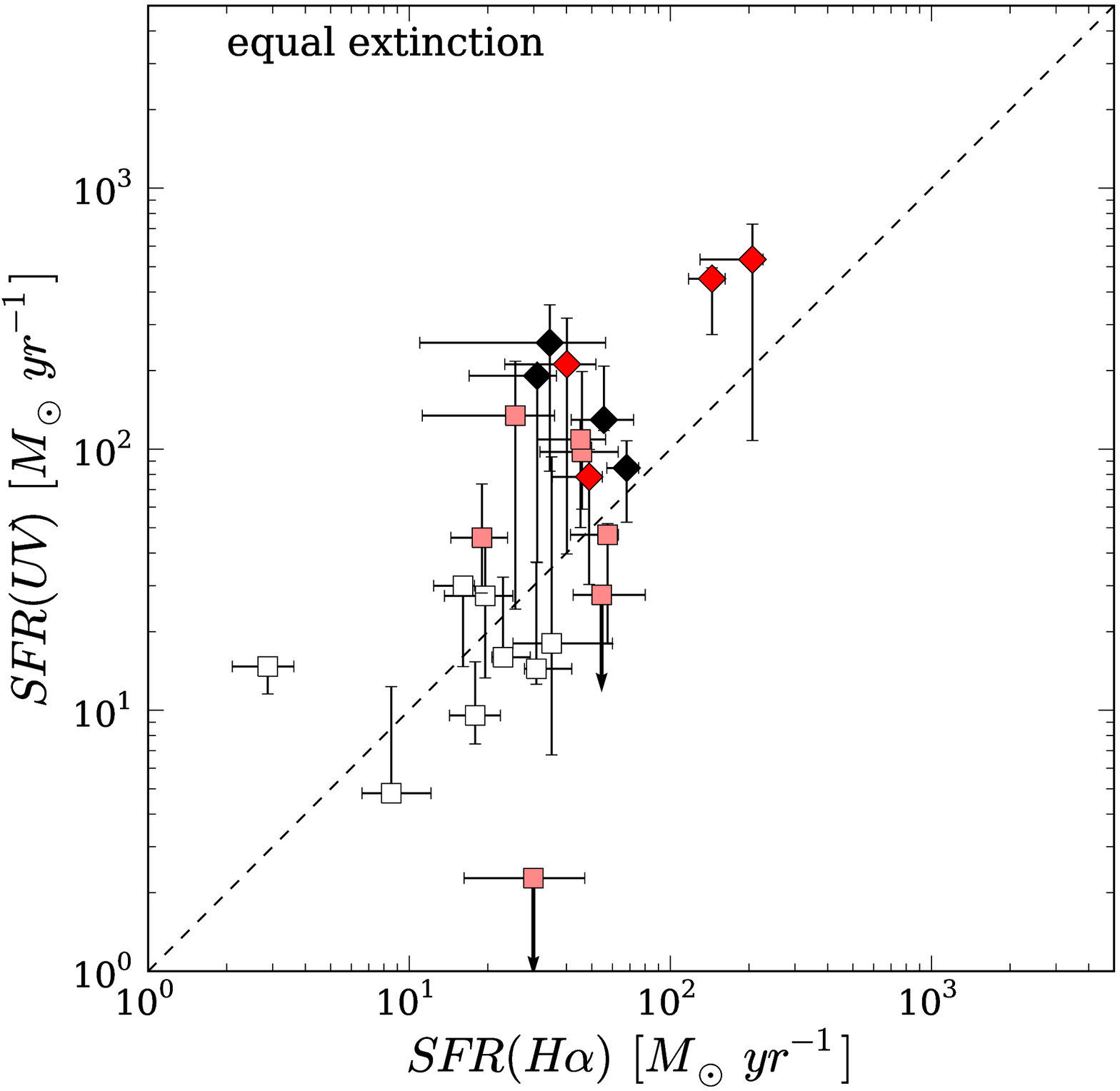}
 \caption{ ({\it left}) Comparison of SFRs inferred from H$\alpha$
   luminosity (SFR(H$\alpha$)) and UV luminosity (SFR(UV)). Diamonds
   and squares are high-SSFR (${\rm SSFR}>10^{-8}\,{\rm yr}^{-1}$) and
   low-SSFR (${\rm SSFR}<10^{-8}\,{\rm yr}^{-1}$ galaxies,
   respectively.  Red symbols are MIPS-selected samples. The SFR(UV)
   is calculated by Equation \ref{eq:uvsfr}, with UV luminosity, which
   is measured with $\phi$1\farcs5 in $B$-band with an aperture
   correction obtained by $K_s$-band total magnitude. The dust
   extinction is corrected with that derived from SED fit. ({\it
     right}) Same as left, but \citet{Calzetti2001} relation is not
   used for extinction correction for H$\alpha$ luminosity.
 \label{fig:sfrcompuv}}
\end{figure*}
}
\def\figsfrcomp{
\begin{figure*}
 \epsscale{1}
 \plottwo{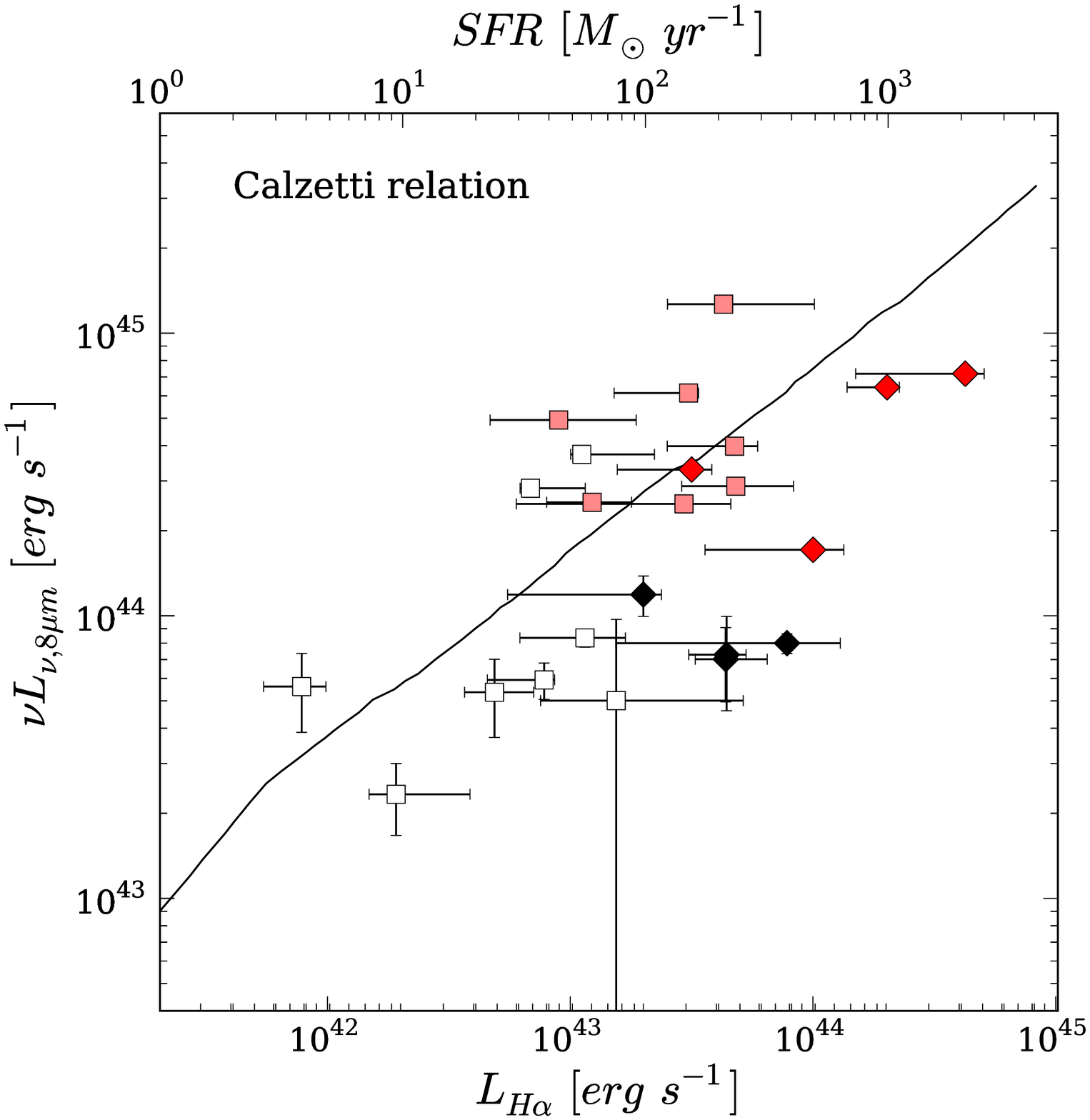}{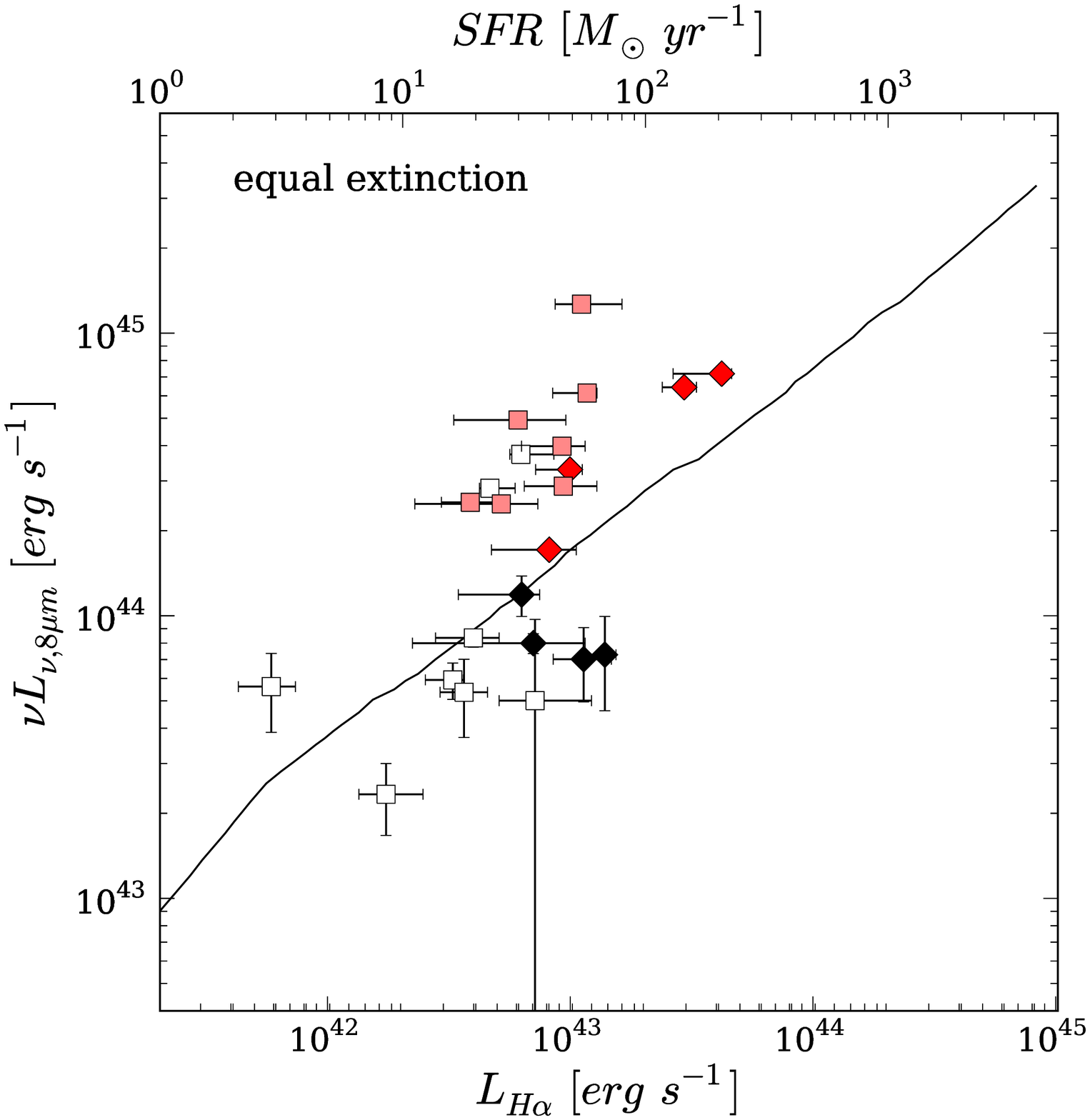}
 \caption{ ({\it left}) Comparison of H$\alpha$ emissions with
   rest-8\,$\mu$m luminosity. Symbols are same as in Figure
   \ref{fig:sfrcompuv}. The 24\,$\mu$m fluxes are measured from a {\it
     Spitzer} MIPS image (Dickinson et al. in preparation) with
   IRAF/DAOPHOT package \citep{Stetson1992} (see text). The solid line
   shows the expected relation between $\nu L_{\nu,8\mu m}$ and
   H$\alpha$ luminosity (see text). The error bars of $\nu L_{\nu,8\mu
     m}$ are estimated from photometric error of 24\,$\mu$m flux and
   the residual of fitting.  ({\it right}) Same as left, but
   \citet{Calzetti2001} relation is not used for extinction correction
   for H$\alpha$ luminosity.
 \label{fig:sfrcomp}}
\end{figure*}
}
\def\figsedmscomp{
\begin{figure*}
 \epsscale{1}
 \plottwo{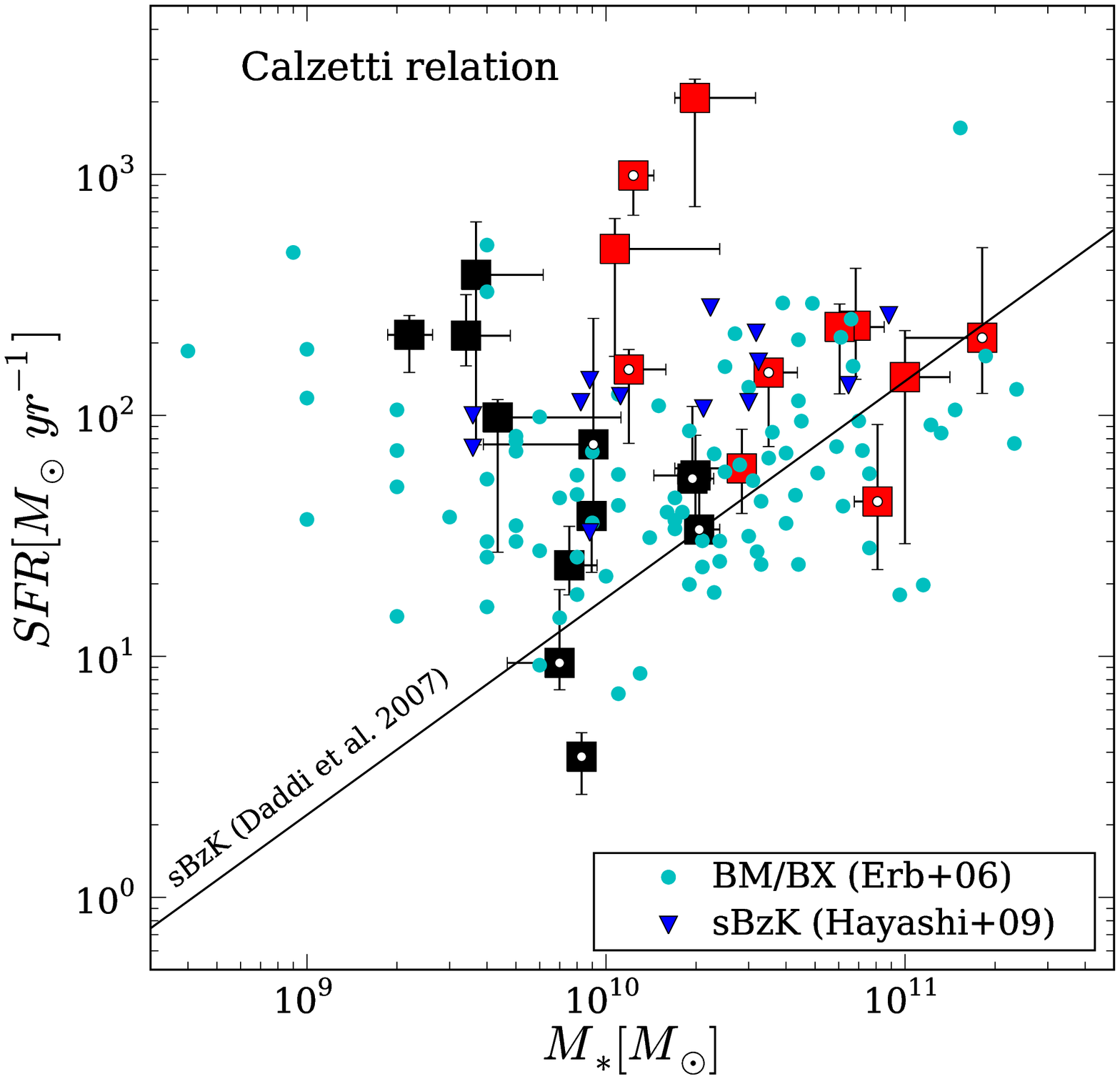}{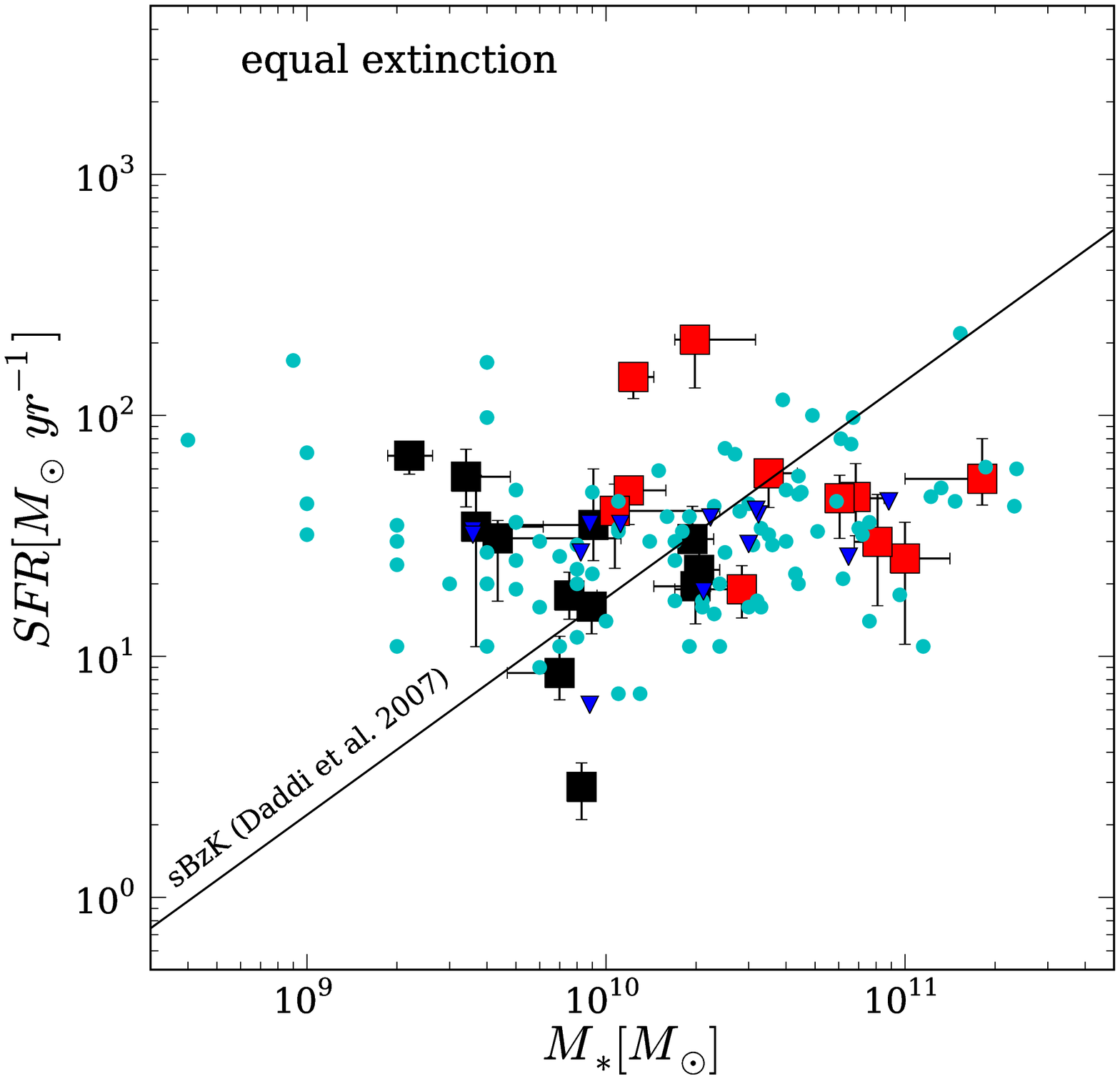}
 \caption{ Same as Figures \ref{fig:sed_ms}, but plotted with the
   results from literature. The black solid line shows the correlation
   between SFR and stellar mass of sBzK galaxies \citep{Daddi2007a}.
   The results of \citet{Erb2006c} and \citet{Hayashi2009} are shown
   by cyan circles and blue triangles, respectively. White dots denote
   MIPS-excess galaxies.
   \label{fig:sed_ms_comp} }
\end{figure*}
}
\def\figsedmsbz{
\begin{figure*}
 \epsscale{1}
 \plottwo{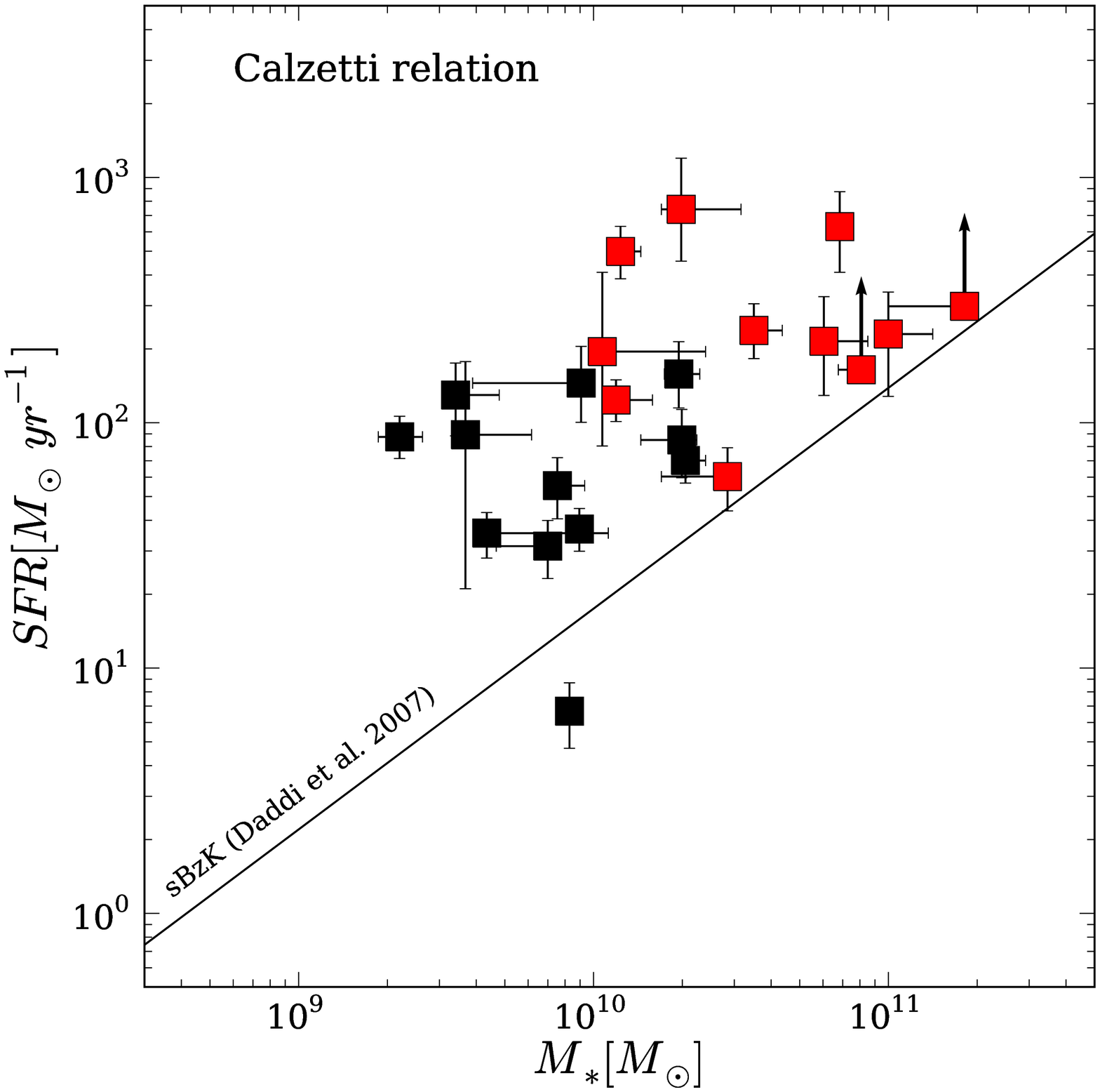}{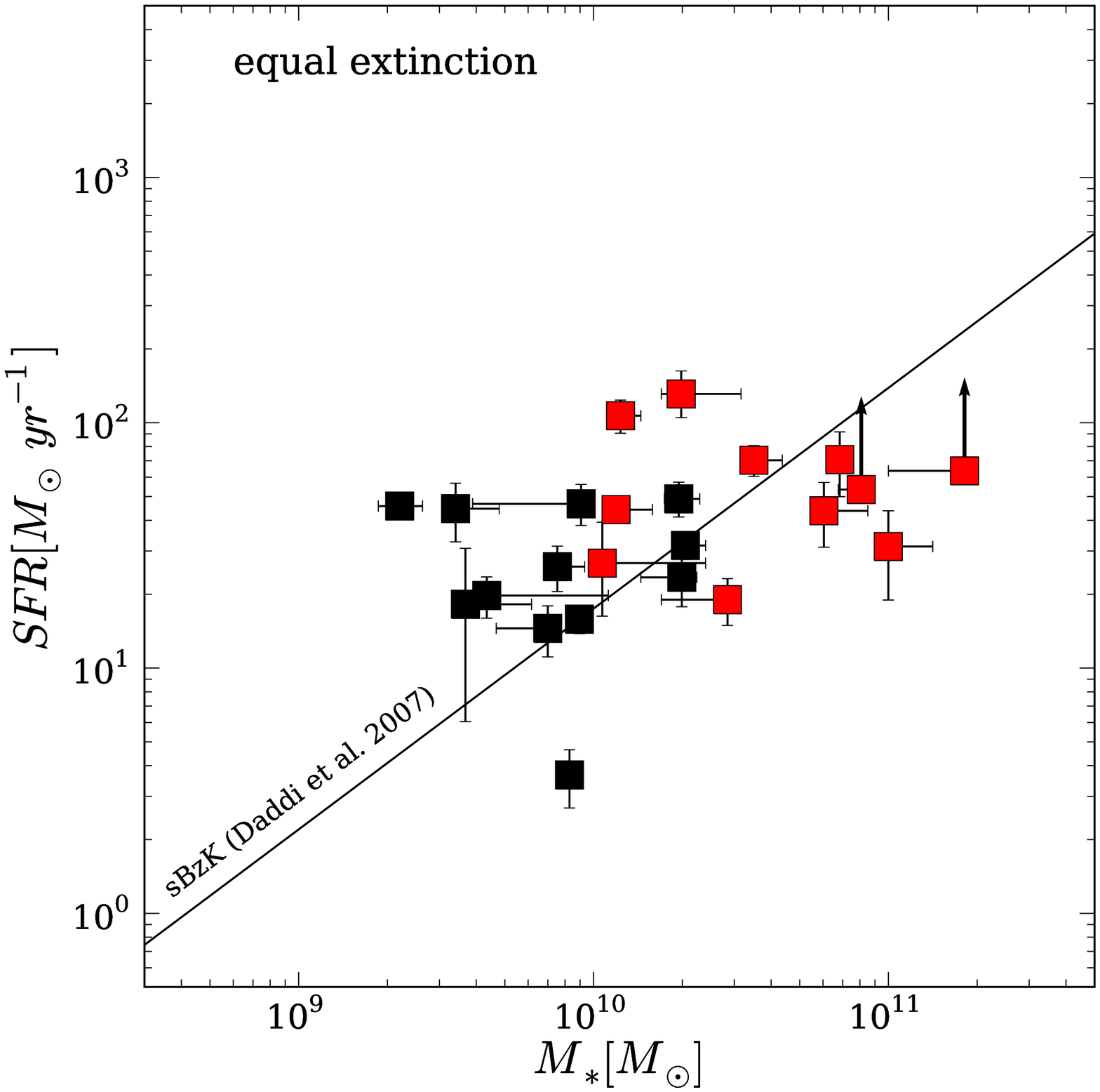}
 \caption{ Same as Figures \ref{fig:sed_ms}, but the amout of
   extinction is estimated from $B-z$ color, as \citet{Daddi2007a}.
   \label{fig:sed_ms_bz} }
\end{figure*}
}
\def\figebvbz{
\begin{figure}
 \epsscale{1}
 \plotone{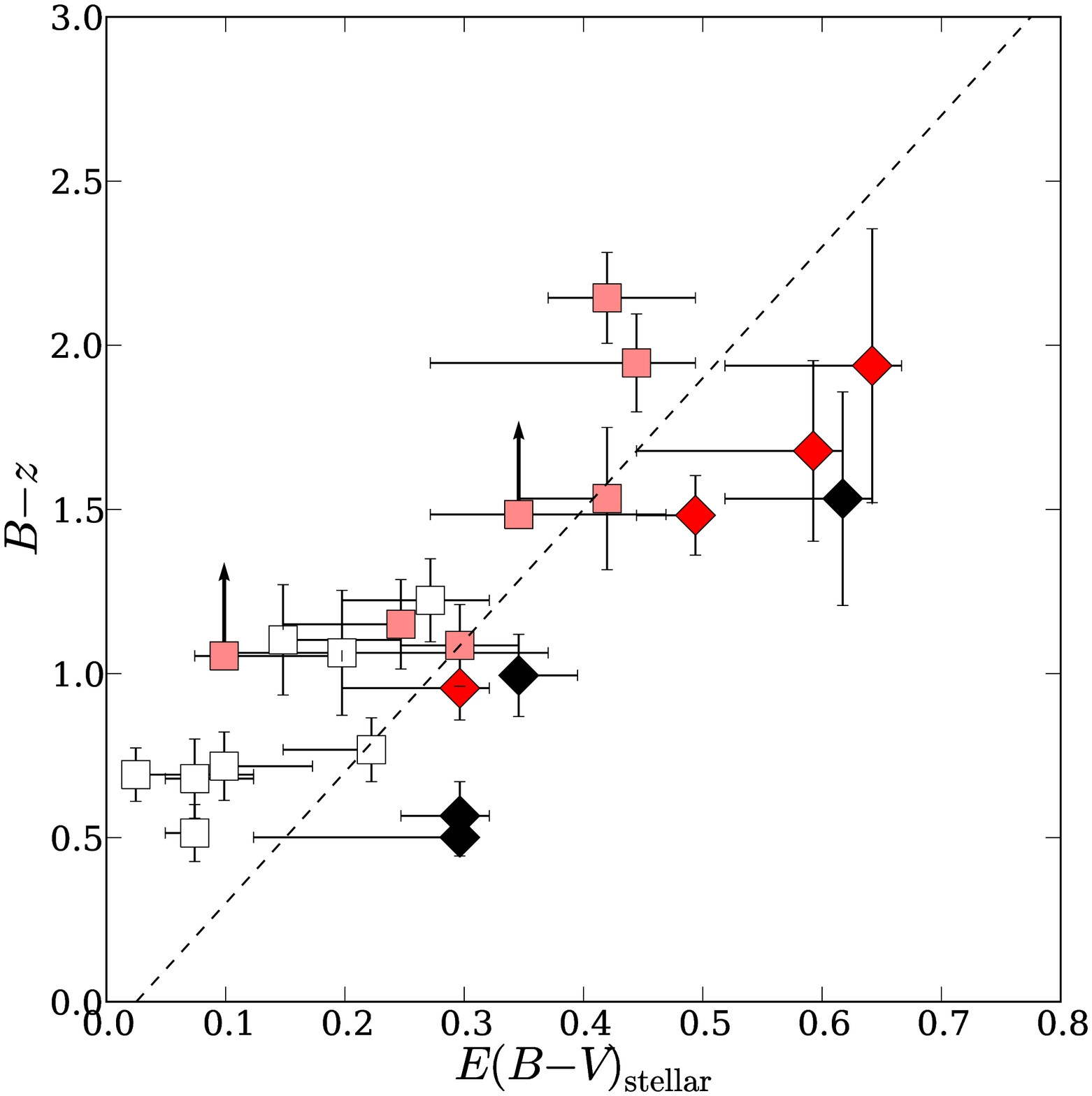}
 \caption{ Comparison of $E(B-V)$ with $B-z$ color for the present
   sample. Symbols are same as in Figure \ref{fig:sfrcompuv}. The
   Dashed line shows the linear correlation between $B-z$ color and
   $E(B-V)$ by \citet{Daddi2004b}.
\label{fig:ebv_bz}}
\end{figure}
}
\def\figmsebv{
\begin{figure}
 \epsscale{1}
 \plotone{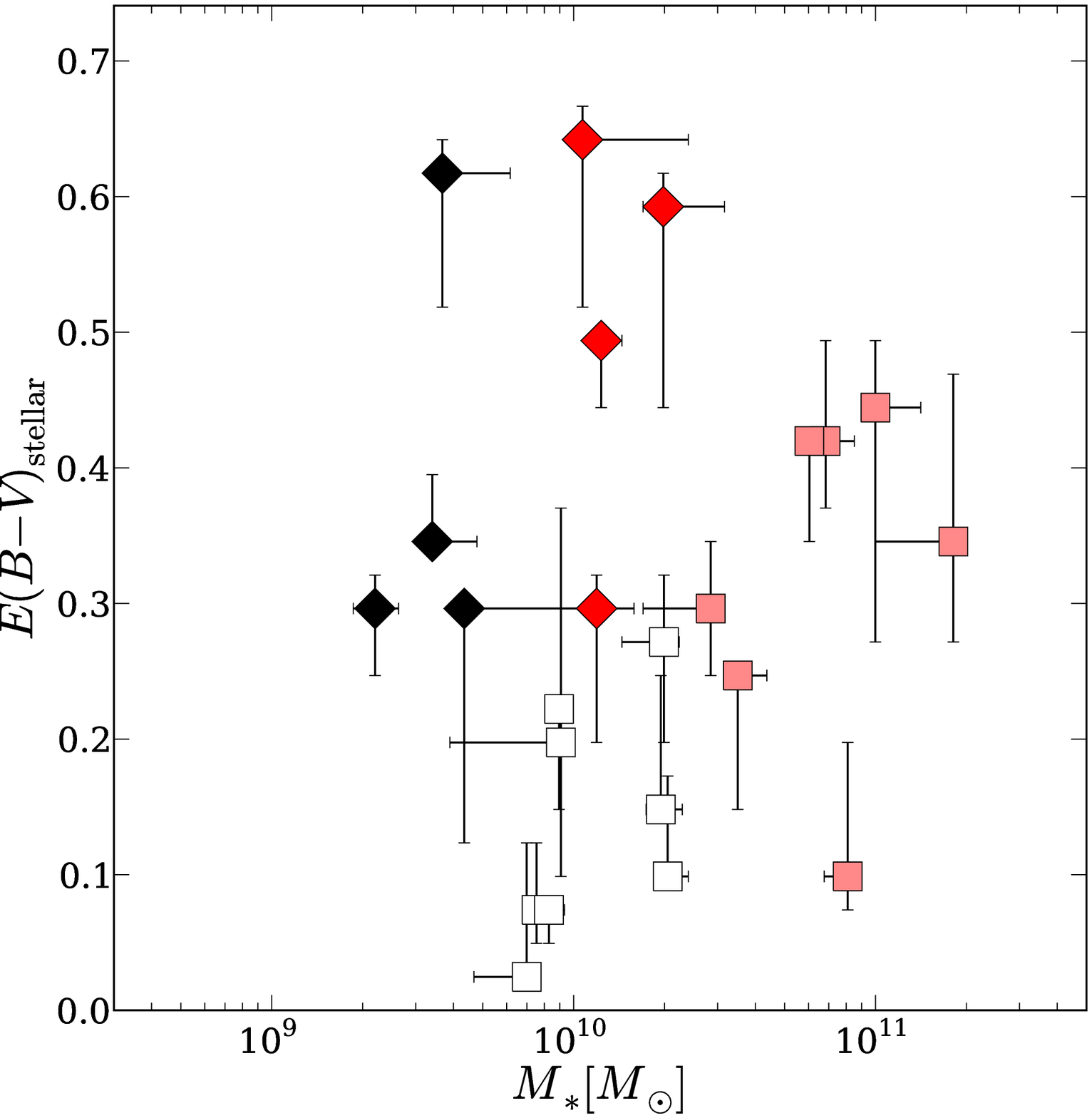}
 \caption{ Comparison of stellar mass with $E(B-V)$ for the present
   sample. Symbols are same as in Figure \ref{fig:sfrcompuv}.
\label{fig:ms_ebv}}
\end{figure}
}
\def\figmzr{
\begin{figure}
 \epsscale{1}
 \plotone{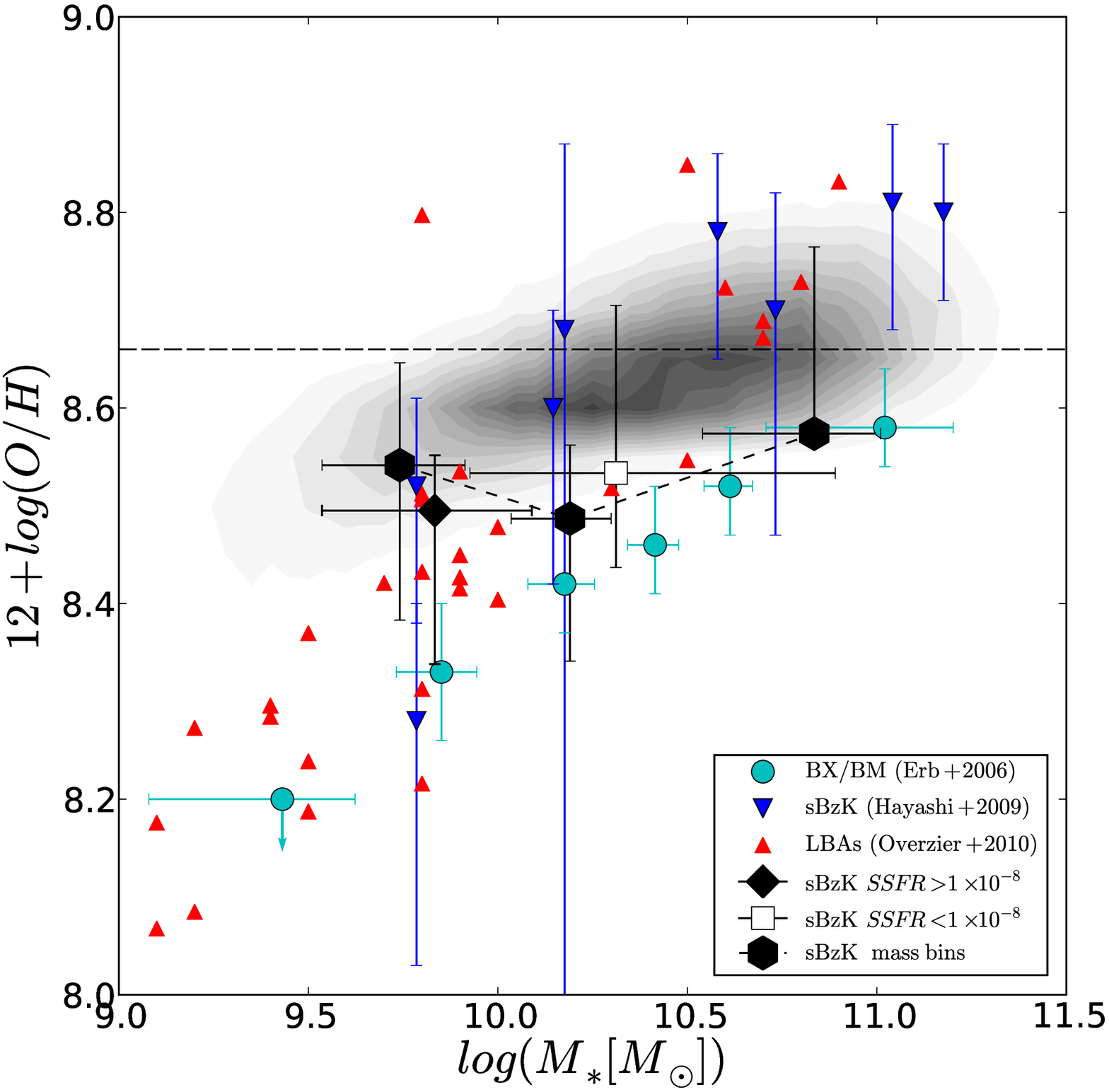}
 \caption{ Oxygen abundance derived from the N2 index for the stacked
   [N\,{\sc ii}]-H$\alpha$ spectra, which are grouped by (1) stellar
   mass at $9\times10^9\,M_\odot$ and $2.5\times10^{10}\,M_\odot$
   (black hexagons) and (2) SSFR at $10^{-8}\,{\rm yr}^{-1}$ (black
   diamond and open square), as a function of the median stellar mass
   of each group. Horizontal error bars indicate the 64\% quantiles of
   stellar masses in the respective bins, while vertical bars indicate
   the error of the N2 index measurements derived from bootstrap
   sampling. The results of the stacked spectra of BM/BX galaxies by
   \citet{Erb2006a} are plotted with cyan circles. The sBzK galaxies
   by \citet{Hayashi2009} and the local Lyman Break Analog (LBA)
   galaxies by \citet{Overzier2010} are plotted with blue downward
   triangles and red upward triangles, respectively. Gray contours
   show the plots of $\sim395000$ objects from SDSS DR4 with the same
   calibration of oxygen abundance. The horizontal dashed line shows
   solar metallicity \citep[$12+\log(\rm O/H)=8.66;$][]{Asplund2004},
   where the N2 index calibration saturates.
 \label{fig:mzr}}
\end{figure}
}

\def\tabobscat{
\begin{deluxetable*}{cccccccccccc}
\tabletypesize{\scriptsize}
\tablewidth{0pt}
\tablecaption{Properties of Galaxies Observed\label{tab:obscat}}
\tablehead{
 \colhead{ID} &  \colhead{RA} &  \colhead{Dec} &  \colhead{$K_s$\tablenotemark{a}} & 
 \colhead{$B-z$\tablenotemark{b}} &  \colhead{$z-K$\tablenotemark{b}} &
 \colhead{$S_{24}$\tablenotemark{c}} & \colhead{MASK ID}&
 \colhead{$z_{\rm literature}$\tablenotemark{d}} &
 \colhead{$z_{\rm phot}$\tablenotemark{e}} &
 \colhead{$z_{\rm H\alpha}$\tablenotemark{e}} &
 \colhead{emission line\tablenotemark{f}} \\
   &  \colhead{(J2000.0)} &  \colhead{(J2000.0)} &  \colhead{(mag)} & 
 \colhead{(mag)} &  \colhead{(mag)} &
 \colhead{($\mu$Jy)} & & & & &
}
\startdata
MODS11-0076 & 12 36 18.50 & 62 09 03.5 & 20.5 & 1.9 & 2.5 & 130.1 & CDFN2 & - & $1.48_{-0.02}^{+0.02}$ & 1.6731 & H$\alpha$\\
MODS11-0094 & 12 36 24.29 & 62 08 27.9 & 22.2 & 1.0 & 1.2 & 106.8 & CDFN2 & - & $2.11_{-0.18}^{+0.08}$ & 1.9976 & H$\alpha$, [OIII]\\
MODS11-0274 & 12 36 25.29 & 62 10 35.6 & 21.4 & 1.5 & 2.4 & 106.4 & CDFN2 & - & $2.10_{-0.20}^{+0.10}$ & 2.0808 & H$\alpha$\\
MODS11-0299 & 12 36 35.64 & 62 09 41.4 & 22.9 & 1.2 & 1.3 & 38.6 & CDFN2 & - & $1.31_{-0.06}^{+0.06}$ & - & -\\
MODS11-0390 & 12 36 39.37 & 62 10 06.6 & 21.8 & 1.5 & 1.5 & 127.2 & CDFN2 & - & $2.22_{-0.06}^{+0.06}$ & 2.3516 & H$\alpha$, H$\beta$, [OIII]\\
MODS12-0125 & 12 36 11.50 & 62 10 33.7 & 21.7 & $>1.5$ & 3.1 & 282.1 & CDFN2 & - & $2.12_{-0.12}^{+0.06}$ & 2.2461 & H$\alpha$, [NII]\\
MODS12-0255 & 12 36 18.38 & 62 11 39.1 & 22.8 & 1.1 & 1.4 & 9.7 & CDFN2 & 2.390 & $2.31_{-0.06}^{+0.09}$ & 2.3979 & H$\alpha$, H$\beta$, [OIII]\\
MODS21-2005 & 12 37 01.04 & 62 10 16.8 & 22.5 & 1.0 & 2.0 & 210.7 & CDFN3 & - & $1.98_{-0.11}^{+0.17}$ & - & -\\
MODS21-2022 & 12 36 51.80 & 62 11 14.8 & 22.6 & 0.7 & 0.8 & 0.0 & CDFN3 & - & $1.47_{-0.06}^{+0.04}$ & - & -\\
MODS21-2047 & 12 36 54.95 & 62 10 56.4 & 22.0 & 1.4 & 2.0 & 204.2 & CDFN3 & 0.999 & $2.17_{-0.06}^{+0.05}$ & - & -\\
MODS21-2321 & 12 36 44.65 & 62 12 27.5 & 22.3 & 0.8 & 1.3 & 26.6 & CDFN3 & 1.879 & $1.56_{-0.05}^{+0.06}$ & 1.7341 & H$\alpha$\\
MODS21-2612 & 12 37 01.07 & 62 10 54.3 & 22.1 & 1.9 & 2.3 & 99.0 & CDFN3 & - & $1.55_{-0.14}^{+0.07}$ & 1.5256 & H$\alpha$\\
MODS21-4647 & 12 37 11.85 & 62 11 15.6 & 22.4 & 0.7 & 0.9 & 7.2 & CDFN3 & - & $1.41_{-0.04}^{+0.06}$ & - & -\\
MODS21-5277 & 12 37 02.91 & 62 14 04.6 & 21.9 & 1.1 & 1.2 & 97.6 & CDFN3 & 1.244 & $1.24_{-0.05}^{+0.05}$ & - & -\\
MODS21-6131 & 12 37 08.77 & 62 12 57.9 & 22.6 & 0.5 & 1.0 & 12.6 & CDFN3 & 2.269 & $2.21_{-0.52}^{+0.15}$ & 2.2678 & H$\alpha$\\
MODS22-1411 & 12 36 26.97 & 62 13 17.3 & 22.0 & 1.7 & 2.1 & 148.3 & CDFN2 & - & $2.19_{-0.04}^{+0.08}$ & 2.3983 & H$\alpha$, [NII], H$\beta$\\
MODS22-2282 & 12 36 21.74 & 62 14 52.9 & 22.6 & 0.7 & 1.2 & 12.7 & CDFN1 & 2.200 & $2.13_{-0.14}^{+0.08}$ & 2.2087 & H$\alpha$\\
MODS22-2658 & 12 36 30.08 & 62 14 27.9 & 20.6 & 2.1 & 2.1 & 181.1 & CDFN1 & 1.525 & $1.37_{-0.02}^{+0.06}$ & 1.5246 & H$\alpha$\\
MODS22-4194 & 12 36 33.67 & 62 15 33.0 & 22.8 & 1.0 & 1.0 & 13.4 & CDFN1 & 2.488 & $2.39_{-0.04}^{+0.04}$ & 2.4862 & H$\alpha$\\
MODS22-4806 & 12 36 57.22 & 62 14 29.9 & 22.2 & 1.0 & 1.8 & 114.2 & CDFN1 & 1.873 & $1.73_{-0.10}^{+0.20}$ & - & -\\
MODS22-5133 & 12 36 45.84 & 62 14 47.0 & 22.5 & 1.5 & 1.9 & 48.7 & CDFN1 & - & $1.61_{-0.06}^{+0.07}$ & 1.5255 & H$\alpha$\\
MODS22-5375 & 12 36 50.11 & 62 14 01.1 & 22.8 & 0.5 & 1.4 & 69.3 & CDFN1 & 2.231 & $2.23_{-0.29}^{+0.04}$ & - & -\\
MODS31-0033 & 12 37 20.05 & 62 12 22.8 & 21.9 & 1.2 & 1.8 & 113.3 & CDFN3 & 2.458 & $2.36_{-0.03}^{+0.08}$ & 2.4605 & H$\alpha$, [NII], H$\beta$, [OIII]\\
MODS31-0129 & 12 37 26.43 & 62 13 30.3 & 21.1 & 2.1 & 2.1 & 89.2 & CDFN3 & - & $1.45_{-0.03}^{+0.02}$ & - & -\\
MODS31-0199 & 12 37 19.39 & 62 16 21.1 & 22.5 & 0.7 & 0.9 & 13.8 & CDFN4 & 1.567 & $1.46_{-0.04}^{+0.04}$ & 1.5681 & H$\alpha$\\
MODS32-0035 & 12 36 37.88 & 62 16 45.3 & 21.7 & 1.5 & 2.0 & 117.8 & CDFN1 & - & $1.45_{-0.06}^{+0.05}$ & - & -\\
MODS32-0116 & 12 36 50.26 & 62 16 55.7 & 21.9 & 1.2 & 1.6 & 54.0 & CDFN1 & 1.315 & $1.44_{-0.10}^{+0.04}$ & 1.4878 & H$\alpha$\\
MODS32-0153 & 12 36 53.65 & 62 17 24.3 & 21.9 & 0.7 & 1.6 & 71.6 & CDFN1 & 2.186 & $2.02_{-0.10}^{+0.15}$ & 2.1865 & H$\alpha$, [OIII]\\
MODS32-0162 & 12 36 51.33 & 62 17 50.9 & 22.4 & 0.8 & 2.0 & 158.5 & CDFN1 & - & $1.83_{-0.07}^{+0.12}$ & - & -\\
MODS41-0194 & 12 37 34.42 & 62 18 12.1 & 21.9 & $>1.1$ & 3.4 & 118.9 & CDFN4 & 2.241 & $2.34_{-0.06}^{+0.08}$ & 2.2425 & H$\alpha$, [OIII]\\
MODS41-0297 & 12 37 24.15 & 62 16 11.7 & 21.9 & 1.1 & 1.6 & 123.6 & CDFN4 & 1.689 & $1.46_{-0.04}^{+0.03}$ & 1.6855 & H$\alpha$\\
MODS42-0112 & 12 37 16.34 & 62 19 20.5 & 22.8 & 0.6 & 1.0 & 19.8 & CDFN4 & - & $2.18_{-0.45}^{+0.05}$ & 2.1231 & H$\alpha$, [OIII]\\
MODS42-0135 & 12 37 20.04 & 62 19 23.2 & 22.1 & 1.1 & 1.8 & 78.1 & CDFN4 & 2.295 & $2.36_{-0.03}^{+0.05}$ & 2.3049 & H$\alpha$, [OIII]\\
MODS42-0145 & 12 37 24.11 & 62 19 04.9 & 22.1 & 0.5 & 0.9 & 32.0 & CDFN4 & 2.094 & $2.05_{-0.10}^{+0.08}$ & 2.0952 & H$\alpha$\\
MODS42-0162 & 12 37 25.80 & 62 19 14.2 & 20.8 & 1.6 & 2.2 & 61.4 & CDFN4 & - & $1.45_{-0.11}^{+0.02}$ & - & -\\
MODS42-0194 & 12 37 33.50 & 62 20 01.2 & 22.0 & 1.5 & 1.6 & 51.6 & CDFN4 & - & $1.36_{-0.05}^{+0.07}$ & - & -\\
MODS42-0228 & 12 37 22.28 & 62 20 39.1 & 21.9 & 1.2 & 1.6 & 244.3 & CDFN4 & - & $1.99_{-0.17}^{+0.15}$ & - & -
\enddata
\tablenotetext{a}{Total magnitude, MAG\_AUTO by SEXTRACTOR}
\tablenotetext{b}{Color in $\phi$1\farcs5 aperture centered at
 $K_s$ coordinate}
\tablenotetext{c}{24\,$\mu$m flux measured from the Spitzer 24\,$\mu$m data
(Dickinson et al. in preparation; see \S\ref{subsec:ext})}
\tablenotetext{d}{Spectroscopic redshift in literature
  \citep{Cohen2001,Cohen2000,Dawson2001,Wirth2004,Cowie2004,Treu2005,Chapman2005,Reddy2006b,Barger2008}
}
\tablenotetext{e}{Photometric redshift \citep{Kajisawa2009}}
\tablenotetext{f}{Our spectroscopic redshift derived from H$\alpha$ emission line}
\tablenotetext{g}{Emission lines detected with S/N$>3$. \ntwo\, and
 \othree\, is \ntwo$\lambda$6583 and \othree$\lambda$5007, respectively.}
\end{deluxetable*}
}
\def\tabobslog{
\begin{deluxetable*}{ccccccccc}
\tablewidth{0pt}
\tablecaption{Summary of the observation\label{tab:obslog}}
\tablehead{
 \colhead{mask ID} & \colhead{R.A.}     & \colhead{Dec.} &
 \colhead{P.A.\tablenotemark{a}} & \colhead{UT date}  & \colhead{grism} &
 \colhead{target}  & \colhead{exposure} & \colhead{seeing\tablenotemark{b}}\\ 
                   & \colhead{(J2000.0)}& \colhead{(J2000.0)}&
 \colhead{(degree)}&                    &                 &
 \colhead{(number)}& \colhead{(min)}    & \colhead{(arcsec)}
}
\startdata
 CDFN1 & 12 36 38.7 & 62 15 58 &   0 &
 2007 Mar 25 & HK500 & 10    & 316      & 0.52\\
 CDFN2 & 12 36 22.3 & 62 10 38 & 249 &
 2007 Mar 24 & HK500 & 8     & 160      & 0.64\\
 CDFN3 & 12 37 03.2 & 62 12 20 & 338 &
 2007 Mar 26 & HK500 & 10    & 310      & 0.68\\
 CDFN4 & 12 37 21.3 & 62 18 28 & 228 &
 2007 Mar 27 & HK500 & 9     & 274      & 0.69
\enddata
\tablenotetext{a}{Direction of the slits, from north to east}
\tablenotetext{b}{Average FWHM sizes of PSF in $K_s$ band during the observations}
\end{deluxetable*}
}
\def\tabsfrcat{
\begin{deluxetable*}{ccccccccc}
\tabletypesize{\footnotesize}
\tablewidth{0pt}
\tablecaption{Star Formation Rates\label{tab:sfrcat}}
\tablehead{
 \colhead{ID}  &
 \colhead{$z_{\rm H\alpha}$\tablenotemark{a}} &
 \colhead{$F_{\rm H\alpha}$\tablenotemark{b}} &
 \colhead{$L_{\rm H\alpha}$\tablenotemark{c}} &
 \colhead{Corrected $L_{\rm H\alpha}$\tablenotemark{d}} &
 \colhead{${\rm SFR}_{\rm H\alpha}$\tablenotemark{e}} & 
 \colhead{$F_{\rm 24\,\mu m}$\tablenotemark{f}} &
 \colhead{$\nu L_{\rm \nu, 8\,\mu m}$\tablenotemark{g}} &
 \colhead{${\rm SFR}_{\rm IR}$\tablenotemark{h}}
}
\startdata
MODS11-0076 & $1.6731\pm0.0009$ & $7.1\pm2.7$ & $1.3\pm0.5$ &
$29.4_{-23.4}^{+16.4}$ & $144.1_{-114.8}^{+80.6}$ & $130.1\pm5.5$ &
$2.5\pm0.1$ & $114.6\pm4.9$\\ MODS11-0094 & $1.9976\pm0.0002$ &
$13.9\pm1.3$ & $4.0\pm0.4$ & $31.6_{-16.0}^{+6.7}$ &
$155.1_{-78.5}^{+32.7}$ & $106.8\pm3.9$ & $3.3\pm0.1$ &
$165.6\pm6.1$\\ MODS11-0274 & $2.0808\pm0.0004$ & $8.0\pm2.0$ &
$2.6\pm0.6$ & $47.5_{-22.4}^{+11.7}$ & $232.7_{-109.8}^{+57.2}$ &
$106.4\pm3.9$ & $4.0\pm0.1$ & $219.8\pm8.1$\\ MODS11-0390 &
$2.3516\pm0.0005$ & $15.1\pm1.9$ & $6.5\pm0.8$ &
$202.0_{-63.8}^{+24.9}$ & $989.7_{-312.7}^{+122.0}$ & $127.2\pm3.9$ &
$6.4\pm0.2$ & $599.4\pm18.5$\\ MODS12-0125 & $2.2461\pm0.0003$ &
$10.0\pm0.9$ & $3.9\pm0.3$ & $42.8_{-17.7}^{+58.5}$ &
$209.9_{-86.6}^{+286.5}$ & $282.1\pm3.9$ & $12.7\pm0.2$ &
$1581.8\pm22.0$\\ MODS12-0255 & $2.3979\pm0.0003$ & $8.7\pm1.1$ &
$3.9\pm0.5$ & $15.5_{-7.9}^{+36.1}$ & $75.8_{-38.8}^{+176.9}$ &
$9.7\pm9.1$ & $0.5\pm0.5$ & $10.2\pm9.6$\\ MODS21-2321 &
$1.7341\pm0.0002$ & $8.1\pm0.9$ & $1.7\pm0.2$ & $7.8_{-3.2}^{+0.8}$ &
$38.2_{-15.9}^{+4.0}$ & $26.6\pm3.9$ & $0.6\pm0.1$ &
$14.6\pm2.2$\\ MODS21-2612 & $1.5256\pm0.0011$ & $7.7\pm2.2$ &
$1.1\pm0.3$ & $100.1_{-64.2}^{+33.8}$ & $490.6_{-314.7}^{+165.8}$ &
$99.0\pm3.9$ & $1.7\pm0.1$ & $77.7\pm3.1$\\ MODS21-6131 &
$2.2678\pm0.0005$ & $1.2\pm0.3$ & $0.5\pm0.1$ & $0.8_{-0.2}^{+0.2}$ &
$3.8_{-1.2}^{+1.0}$ & $12.6\pm3.9$ & $0.6\pm0.2$ &
$11.5\pm3.6$\\ MODS22-1411 & $2.3983\pm0.0002$ & $15.2\pm0.9$ &
$6.8\pm0.4$ & $424.0_{-273.9}^{+83.3}$ & $2077.4_{-1342.3}^{+408.2}$ &
$148.3\pm3.9$ & $7.2\pm0.2$ & $822.1\pm21.7$\\ MODS22-2282 &
$2.2087\pm0.0004$ & $7.9\pm1.5$ & $2.9\pm0.6$ & $4.9_{-1.2}^{+2.2}$ &
$23.9_{-5.9}^{+10.8}$ & $12.7\pm3.9$ & $0.5\pm0.2$ &
$10.2\pm3.2$\\ MODS22-2658 & $1.5246\pm0.0035$ & $17.4\pm4.8$ &
$2.6\pm0.7$ & $48.1_{-19.3}^{+35.1}$ & $235.6_{-94.5}^{+171.8}$ &
$181.1\pm3.9$ & $2.9\pm0.1$ & $165.6\pm3.6$\\ MODS22-4194 &
$2.4862\pm0.0007$ & $8.0\pm2.0$ & $3.9\pm1.0$ & $43.7_{-11.0}^{+21.0}$
& $214.2_{-53.7}^{+103.0}$ & $13.4\pm3.9$ & $0.7\pm0.2$ &
$20.1\pm5.9$\\ MODS22-5133 & $1.5255\pm0.0016$ & $7.2\pm4.5$ &
$1.1\pm0.7$ & $78.2_{-62.8}^{+51.4}$ & $383.1_{-307.6}^{+251.9}$ &
$48.7\pm3.9$ & $0.8\pm0.1$ & $33.1\pm2.7$\\ MODS31-0033 &
$2.4605\pm0.0003$ & $11.5\pm1.1$ & $5.5\pm0.5$ & $30.7_{-15.6}^{+3.0}$
& $150.6_{-76.3}^{+14.8}$ & $113.3\pm3.9$ & $6.1\pm0.2$ &
$717.0\pm24.8$\\ MODS31-0199 & $1.5681\pm0.0020$ & $10.1\pm2.3$ &
$1.6\pm0.4$ & $1.9_{-0.4}^{+1.9}$ & $9.4_{-2.1}^{+9.5}$ & $13.8\pm3.9$
& $0.2\pm0.1$ & $6.6\pm1.9$\\ MODS32-0116 & $1.4878\pm0.0005$ &
$12.4\pm2.8$ & $1.7\pm0.4$ & $11.5_{-5.3}^{+5.4}$ &
$56.3_{-26.0}^{+26.3}$ & $54.0\pm3.9$ & $0.8\pm0.1$ &
$35.9\pm2.6$\\ MODS32-0153 & $2.1865\pm0.0003$ & $9.6\pm0.9$ &
$3.4\pm0.3$ & $6.9_{-0.6}^{+4.7}$ & $33.6_{-3.2}^{+22.9}$ &
$71.6\pm3.9$ & $2.8\pm0.2$ & $129.9\pm7.1$\\ MODS41-0194 &
$2.2425\pm0.0007$ & $11.8\pm5.3$ & $4.5\pm2.0$ & $9.0_{-4.3}^{+9.7}$ &
$43.9_{-21.0}^{+47.7}$ & $118.9\pm3.9$ & $4.9\pm0.2$ &
$378.8\pm12.5$\\ MODS41-0297 & $1.6855\pm0.0003$ & $8.2\pm1.6$ &
$1.6\pm0.3$ & $12.3_{-4.3}^{+5.6}$ & $60.3_{-21.1}^{+27.4}$ &
$123.6\pm3.9$ & $2.5\pm0.1$ & $114.6\pm3.6$\\ MODS42-0112 &
$2.1231\pm0.0002$ & $16.7\pm1.3$ & $5.6\pm0.4$ & $44.1_{-13.3}^{+9.0}$
& $215.9_{-65.1}^{+43.9}$ & $19.8\pm7.3$ & $0.7\pm0.3$ &
$17.4\pm6.4$\\ MODS42-0135 & $2.3049\pm0.0003$ & $9.7\pm1.0$ &
$4.0\pm0.4$ & $11.2_{-1.1}^{+11.1}$ & $54.6_{-5.4}^{+54.3}$ &
$78.1\pm3.9$ & $3.7\pm0.2$ & $236.5\pm11.9$\\ MODS42-0145 &
$2.0952\pm0.0002$ & $7.8\pm1.5$ & $2.5\pm0.5$ & $20.0_{-14.5}^{+3.7}$
& $98.1_{-71.0}^{+18.3}$ & $32.0\pm5.2$ & $1.2\pm0.2$ & $33.1\pm5.4$
\enddata \tablenotetext{a}{Spectroscopic redshift measured from the
  H$\alpha$ emission line} \tablenotetext{b}{Observed flux of the
  H$\alpha$ emission line after correcting slit loss, in units of
  $10^{-17}$ ergs ${\rm s}^{-1} {\rm cm}^{-2}$}
\tablenotetext{c}{Observed luminosity of the H$\alpha$ emission line,
  in units of $10^{42}$ ergs ${\rm s}^{-1}$}
\tablenotetext{d}{Extinction corrected luminosity of the H$\alpha$
  emission line, in units of $10^{42}$ ergs ${\rm s}^{-1}$}
\tablenotetext{e}{SFR derived from extinction corrected luminosity of
  H$\alpha$ emission using Equation \ref{eq:hasfr}, in units of
  $M_\odot {\rm yr}^{-1}$}
\tablenotetext{f}{Observed flux density of MIPS 24 $\mu$m, in units of
  $\mu$Jy. The fluxes of the objects undetected in the public catalog
  are measured from the public image.}
\tablenotetext{g}{Rest-8$\mu$m luminosity inferred from 24 $\mu$m flux
  with $k$-correction by \citet{Chary2001} template, in units of
  $10^{44}$ ergs ${\rm s}^{-1}$.}
\tablenotetext{h}{SFR derived from MIPS 24 $\mu$m flux, in units of
  ${\rm M}_\odot {\rm yr}^{-1}$.}

\end{deluxetable*}
}
\def\tabsedcat{
\begin{deluxetable*}{ccccccc}
\tabletypesize{\scriptsize}
\tablewidth{0pt}
\tablecaption{Stellar Population Parameters\label{tab:sedcat}}
\tablehead{
 \colhead{ID}  & \colhead{Z}         & \colhead{$\tau$}  &
 \colhead{age} & \colhead{mass weighted age} & \colhead{$A_V$}   &
 \colhead{$M_\star$} \\
               & \colhead{$Z_\odot$} & \colhead{Gyr}     &
 \colhead{Myr} & \colhead{Myr}       &                   &
 \colhead{$10^9 M_\odot$}
}
\startdata
MODS11-0076 & 0.2 & 0.10 & $360.2_{-179.7}^{+1339.8}$ & $270.4_{-152.9}^{+904.5}$ & $1.8_{-0.7}^{+0.2}$ & $99.9_{-2.2}^{+41.4}$\\
MODS11-0094 & 0.4 & 0.10 & $127.8_{-63.7}^{+232.4}$ & $77.2_{-11.1}^{+113.3}$ & $1.2_{-0.4}^{+0.1}$ & $11.9_{-0.7}^{+3.9}$\\
MODS11-0274 & 0.2 & 0.10 & $255.0_{-127.2}^{+463.7}$ & $176.7_{-61.9}^{+412.1}$ & $1.7_{-0.3}^{+0.0}$ & $60.4_{-2.9}^{+24.7}$\\
MODS11-0390 & 0.2 & 5.00 & $11.5_{-0.0}^{+8.5}$ & $6.0_{-0.1}^{+4.5}$ & $2.0_{-0.2}^{+0.0}$ & $12.3_{-0.9}^{+2.1}$\\
MODS12-0125 & 0.2 & 2.00 & $3500.0_{-2991.2}^{+0.0}$ & $2430.2_{-2050.1}^{+0.0}$ & $1.4_{-0.3}^{+0.5}$ & $181.1_{-81.1}^{+0.0}$\\
MODS12-0255 & 0.4 & 0.02 & $90.5_{-77.3}^{+37.3}$ & $82.1_{-75.5}^{+25.0}$ & $0.8_{-0.4}^{+0.7}$ & $9.1_{-5.2}^{+0.0}$\\
MODS21-2321 & 0.4 & 0.02 & $90.5_{-0.0}^{+90.0}$ & $82.1_{-0.9}^{+52.7}$ & $0.9_{-0.3}^{+0.0}$ & $9.0_{-0.8}^{+0.6}$\\
MODS21-2612 & 0.4 & 0.01 & $36.0_{-6.0}^{+472.8}$ & $27.0_{-6.6}^{+155.0}$ & $2.6_{-0.5}^{+0.1}$ & $10.7_{-0.5}^{+13.3}$\\
MODS21-6131 & 0.2 & 0.01 & $90.5_{-0.0}^{+37.3}$ & $91.5_{-0.3}^{+15.6}$ & $0.3_{-0.1}^{+0.0}$ & $8.3_{-0.3}^{+0.2}$\\
MODS22-1411 & 0.2 & 5.00 & $20.0_{-6.8}^{+19.0}$ & $10.5_{-3.7}^{+8.6}$ & $2.4_{-0.6}^{+0.1}$ & $19.8_{-2.8}^{+11.8}$\\
MODS22-2282 & 0.2 & 0.01 & $127.8_{-0.0}^{+127.2}$ & $117.8_{-10.6}^{+86.4}$ & $0.3_{-0.1}^{+0.2}$ & $7.5_{-0.1}^{+1.8}$\\
MODS22-2658 & 1.0 & 0.01 & $127.8_{-37.3}^{+52.7}$ & $117.8_{-26.6}^{+52.0}$ & $1.7_{-0.2}^{+0.3}$ & $68.4_{-5.3}^{+4.1}$\\
MODS22-4194 & 1.0 & 5.00 & $10.0_{-0.0}^{+1.5}$ & $5.2_{-0.1}^{+0.8}$ & $1.4_{-0.0}^{+0.2}$ & $3.4_{-0.0}^{+1.4}$\\
MODS22-5133 & 1.0 & 0.50 & $22.9_{-7.8}^{+29.6}$ & $11.5_{-2.4}^{+23.9}$ & $2.5_{-0.4}^{+0.1}$ & $3.7_{-0.3}^{+2.5}$\\
MODS31-0033 & 0.4 & 0.05 & $180.5_{-0.0}^{+328.3}$ & $135.6_{-18.1}^{+211.1}$ & $1.0_{-0.4}^{+0.0}$ & $35.0_{-2.6}^{+8.7}$\\
MODS31-0199 & 1.0 & 0.10 & $360.2_{-232.4}^{+0.0}$ & $270.4_{-163.3}^{+0.0}$ & $0.1_{-0.0}^{+0.4}$ & $7.0_{-2.3}^{+0.0}$\\
MODS32-0116 & 0.2 & 1.00 & $1015.2_{-655.0}^{+1284.8}$ & $676.0_{-400.5}^{+554.3}$ & $1.1_{-0.3}^{+0.2}$ & $19.9_{-5.5}^{+2.5}$\\
MODS32-0153 & 0.2 & 0.10 & $360.2_{-179.7}^{+148.6}$ & $270.4_{-135.5}^{+76.3}$ & $0.4_{-0.0}^{+0.3}$ & $20.5_{-0.1}^{+3.5}$\\
MODS41-0194 & 0.2 & 0.50 & $2600.0_{-1584.8}^{+0.0}$ & $2261.9_{-1328.7}^{+0.0}$ & $0.4_{-0.1}^{+0.4}$ & $80.9_{-13.3}^{+6.2}$\\
MODS41-0297 & 0.4 & 2.00 & $1015.2_{-887.4}^{+418.8}$ & $623.7_{-494.9}^{+37.0}$ & $1.2_{-0.2}^{+0.2}$ & $28.5_{-11.5}^{+0.0}$\\
MODS42-0112 & 0.2 & 0.05 & $17.4_{-5.9}^{+12.6}$ & $9.7_{-3.0}^{+6.2}$ & $1.2_{-0.2}^{+0.1}$ & $2.2_{-0.3}^{+0.4}$\\
MODS42-0135 & 0.2 & 0.02 & $180.5_{-90.0}^{+74.5}$ & $160.5_{-154.9}^{+43.6}$ & $0.6_{-0.0}^{+0.4}$ & $19.5_{-2.1}^{+3.5}$\\
MODS42-0145 & 0.2 & 0.02 & $11.5_{-0.0}^{+116.3}$ & $6.6_{-0.0}^{+80.5}$ & $1.2_{-0.7}^{+0.0}$ & $4.3_{-0.1}^{+6.9}$
\enddata
\end{deluxetable*}
}

\documentclass[apj]{emulateapj}
\usepackage{apjfonts}

\newcommand{\ntwo}{[N\,{\sc ii}]}
\newcommand{\othree}{[O\,{\sc iii}]}

\shorttitle{MOIRCS Deep Survey VI}
\shortauthors{Yoshikawa et al.}

\begin{document}

\title{MOIRCS Deep Survey. VI. Near-Infrared Spectroscopy of
  $K$-Selected Star-Forming Galaxies at
  \lowercase{$z\sim2$}\altaffilmark{1}}
\altaffiltext{1}{This study is based on data collected at Subaru
  Telescope, which is operated by the National Astronomical
  Observatory of Japan.  }
\author{
Tomohiro Yoshikawa\altaffilmark{2}\altaffilmark{3}\altaffilmark{4},
Masayuki Akiyama\altaffilmark{3},
Masaru Kajisawa\altaffilmark{3},
David M. Alexander\altaffilmark{5},
Kouji Ohta\altaffilmark{6},
Ryuji Suzuki\altaffilmark{4},
Chihiro Tokoku\altaffilmark{3},
Yuka K. Uchimoto\altaffilmark{7}, 
Masahiro Konishi\altaffilmark{7},
Toru Yamada\altaffilmark{3},
Ichi Tanaka\altaffilmark{4},
Koji Omata\altaffilmark{4},
Tetsuo Nishimura\altaffilmark{4},
Anton M. Koekemoer\altaffilmark{8},
Niel Brandt\altaffilmark{9},
and Takashi Ichikawa\altaffilmark{3}
}
\altaffiltext{2}{Koyama Astronomical Observatory, Kyoto Sangyo University,
Motoyama, Kamigamo, Kita-ku, Kyoto 603-8555, Japan; tomohiro@cc.kyoto-su.ac.jp}
\altaffiltext{3}{Astronomical Institute, Tohoku University, Aramaki,
  Aoba-ku, Sendai 980-8578, Japan}
\altaffiltext{4}{Subaru Telescope, 650 North A'ohoku Place, Hilo, HI 96720}
\altaffiltext{5}{Department of Physics, Durham University, South Road,
Durham, DH1 3LE, UK}
\altaffiltext{6}{Department of Astronomy, Kyoto University, Kyoto 606-8502}
\altaffiltext{7}{Institute of Astronomy, the University of Tokyo, 2-21-1, Osawa, Mitaka,
Tokyo 181-8588, Japan}
\altaffiltext{8}{Space Telescope Science Institute, 3700 San Martin
Drive, Baltimore, MD 21218, USA}
\altaffiltext{9}{Department of Astronomy and Astrophysics, The Pennsylvania
State University, 525 Davey Lab, University Park, PA 16802 USA}

\begin{abstract}
We present the results of near-infrared multi-object spectroscopic
observations for 37 $BzK$-color-selected star-forming galaxies
conducted with MOIRCS on the Subaru Telescope. The sample is drawn
from the $K_s$-band selected catalog of the MOIRCS Deep Survey (MODS)
in the GOODS-N region. About half of our samples are selected from the
publicly available 24\,$\mu$m-source catalog of the Multiband Imaging
Photometer for {\it Spitzer} on board the {\it Spitzer Space
  Telescope}. H$\alpha$ emission lines are detected from 23 galaxies,
of which the median redshift is 2.12. We derived the star formation
rates (SFRs) from extinction-corrected H$\alpha$ luminosities. The
extinction correction is estimated from the SED fitting of multi-band
photometric data covering UV to near-infrared wavelengths. The Balmer
decrement of the stacked emission lines shows that the amount of
extinction for the ionized gas is larger than that for the stellar
continuum. From a comparison of the extinction corrected H$\alpha$
luminosity and other SFR indicators we found that the relation between
the dust properties of stellar continuum and ionized gas is different
depending on the intrinsic SFR (differential extinction). We compared
SFRs estimated from extinction corrected H$\alpha$ luminosities with
stellar masses estimated from SED fitting. The comparison shows no
correlation between SFR and stellar mass. Some galaxies with stellar
mass smaller than $\sim10^{10}\,{\rm M}_\odot$ show SFRs higher than
$\sim100\,{\rm M}_\odot\,{\rm yr}^{-1}$. The specific SFRs (SSFRs) of
these galaxies are remarkably high; galaxies which have SSFR higher
than $\sim10^{-8}\,{\rm yr}^{-1}$ are found in 8 of the present
sample. From the best-fit parameters of SED fitting for these high
SSFR galaxies, we find that the average age of the stellar population
is younger than 100\,Myr, which is consistent with the implied high
SSFR. The large SFR implies the possibility that the high SSFR
galaxies significantly contribute to the cosmic SFR density of the
universe at $z\sim2$. When we apply the larger extinction correction
for the ionized gas or the differential extinction correction, the
total SFR density estimated from the H$\alpha$ emission line galaxies
is $0.089-0.136\,{\rm M}_\odot{\rm yr}^{-1}{\rm Mpc}^{-3}$, which is
consistent with the total SFR densities in the literature. The
metallicity of the high-SSFR galaxies, which is estimated from the N2
index, is larger than that expected from the mass-metallicity relation
of UV-selected galaxies at $z\sim2$ by \citet{Erb2006a}.
\end{abstract}

\keywords{galaxies: evolution --- galaxies: high-redshift --- infrared:
galaxies}

\section{Introduction}
\label{sec:intro}

Recent studies revealed that a significant fraction of stars in the
present-day galaxies were formed between redshifts 1 and 3
\citep[e.g.,][]{Hopkins2004,PerezGonzalez2005,Hopkins2006,Wang2006,Dahlen2007,Caputi2007}.
For example, results of deep near-infrared observations show the
average stellar mass density at $0.5<z<1$ is $\sim 53-72\%$ of that in
the present universe, while at $2.5<z<3.5$ it is just $\sim4-9\%$ of
the present value \citep{Kajisawa2009}. Although such studies
demonstrate the total star formation rate (SFR) density in the
universe as a function of cosmic time, it is still uncertain which
populations of galaxies contribute to the active star formation
between redshift 1 and 3. The evolution of the stellar mass-SFR
relation constrains theoretical views of how galaxies accumulate the
stellar mass \citep{Dave2008}. Therefore, as a next step, in order to
understand the build up of stellar mass in galaxies at this redshift
range, it is crucial to examine the relationship of SFR to stellar
mass.

SFRs are often estimated from photometric indicators that are closely
related to the emission from massive stars
\citep{Kennicutt1998}. Massive stars have short lifetimes, so they
represent stars which were born recently. The UV continuum
(1500--2800\,\AA) directly reflects the number of massive stars, and
therefore can be used as an indicator of SFR for galaxies in this
redshift range. Additionally, far-infrared emission is another
indicator of SFR; the UV emission of young massive stars is absorbed
by interstellar dust and reradiated in the far-infrared at
$\sim$8--1000\,$\mu$m. Fluxes at 24\,$\mu$m wavelength taken with the
Multiband Imaging Photometer for {\it Spitzer} (MIPS) are commonly
used to estimate far-infrared emission of galaxies at high redshifts,
because observations in longer wavelengths are too shallow to detect
such galaxies.  \citet{Daddi2007a} showed that $K$-selected galaxies
at $z\sim2$ have a tight correlation ($\sim0.2$\,dex) between stellar
mass and SFR estimated from UV and IR luminosities. The SFRs of these
galaxies are roughly proportional to their stellar masses. However,
the relation is not still conclusive. Other studies show different
relationships between SFRs and stellar masses. For example, rest-UV
selected BM/BX galaxies also have a correlation between these values
but the correlation is not proportional
\citep{Reddy2006a}. Furthermore, the SFRs of 24\,$\mu$m-selected
galaxies do not depend on their stellar masses \citep{Caputi2006a}. We
need to note that SFRs estimated from UV and IR luminosities have
large uncertainties. UV light is affected by dust extinction and stars
with long lifetimes of $\sim1$\,Gyr. On the other hand, 24\,$\mu$m
radiation corresponds to the rest-frame 8\,$\mu$m wavelength for
galaxies at $z\sim2$. It is known that the far-infrared luminosity
estimated from rest-8\,$\mu$m luminosity has a large uncertainty due
to the complex SED at those wavelengths
\citep{Peeters2004,Chary2001,Daddi2007a,Papovich2007}.

H$\alpha$ luminosity is one of the most reliable indicators of
SFR. First, the emission line is reemitted from hydrogen gas ionized
by UV photons shorter than 912\,\AA, which are radiated by only the
most massive stars. These stars have shorter lifetimes than stars
emitting the UV continuum above 1500\,\AA. Second, the emission line
is less affected by dust extinction than UV emission. The amount of
dust extinction is larger at shorter wavelengths. Third, the H$\alpha$
luminosity can be measured without the uncertainty inherent in SED
estimates for the far-infrared luminosity. However, it is difficult to
measure H$\alpha$ luminosities of galaxies at high redshift, because
the H$\alpha$ emission line at $z\gtrsim0.4$ is out of the wavelength
range of optical spectrographs.

Recently, near-infrared spectroscopic studies of SFRs based on the
H$\alpha$ luminosity for large samples of star-forming galaxies at
$z\sim2$ have been conducted
\citep{Erb2006c,Erb2003,Hayashi2009,ForsterSchreiber2009}.
\citet{Erb2006c} demonstrate a weak correlation between SFR estimated
from H$\alpha$ luminosity and stellar mass. Their sample galaxies are
selected by $U_nG\cal{R}$ color and have redshifts confirmed using a
blue-sensitive optical spectrograph. The observed galaxies are biased
toward less attenuated galaxies. Since red galaxies account for the
significant fraction of stellar mass density at $z\sim2$
\citep{PerezGonzalez2008}, the galaxies at $z\sim2$ should be
re-examined with a technique more sensitive to red galaxies (e.g.,
$K$-selected galaxies).  The $K$-selected star-forming galaxies
observed by \citet{Hayashi2009} showed a flatter SFR-stellar mass
correlation than that of \citet{Daddi2007a}, but the number of
$K$-selected galaxies with H$\alpha$ detection is still rather limited
and observations for a larger sample are necessary.

In this work, we study the SFR and stellar mass of a large number of
$K$-selected star-forming galaxies at $z\sim2$ based on near-infrared
observations with the Multi-Object InfraRed Camera and Spectrograph
\citep[MOIRCS;][]{Suzuki2008}. MOIRCS has the capability for
multi-object near-infrared spectroscopy. We also utilize the deep
near-infrared imaging of the MOIRCS Deep Survey
\citep[MODS;][]{Kajisawa2009}, which covers $\sim 103.3\,{\rm
  arcmin}^2$ in the Great Observatories Origins Deep Survey North
(GOODS-N) region. The sample for the present work is selected from
$K$-selected galaxies in the MODS catalog and their properties are
studied using MODS and publicly available multi-wavelength data of
GOODS-N. Throughout the paper, we use the AB magnitude system and
adopt cosmological parameters of $h\equiv H_0/100 = 0.7$, $\Omega_M =
0.3$, $\Omega_\Lambda = 0.7$.

\section{Observation}
\label{sec:obs}
\subsection{Sample Selection}
\label{subsec:sample}
\figbzk

In order to select star-forming galaxies at $z\sim2$, we apply the BzK
color criterion for star-forming galaxies (sBzK; $BzK \equiv (z-K) -
(B-z) > -0.2$; \citealp{Daddi2004b}) to the $K$-selected galaxies with
$K_s < 23 $ from the MODS catalog, whose limiting magnitude is
$K_s=24.9$ ($5\sigma$, $\phi1.2\,{\rm arcsec}$). The BzK diagram of
the extracted galaxies is shown in Figure \ref{fig:bzk}. $B$ and $z$
band magnitudes are measured on the F435W and F850LP images of the
Advanced Camera for Surveys (ACS) on board the {\it Hubble Space
  Telescope} (HST) in the course of the GOODS Treasury Program
(Version 1.0; \citealp{Giavalisco2004}). In total, 508 galaxies are
identified as candidates.

We chose 37 sBzK galaxies from the candidates for the spectroscopic
observation. The coordinates and photometry of the target galaxies are
listed in Table \ref{tab:obscat}. Although two of the sample galaxies
(MODS12-0125 and MODS41-0194) are not detected in the $B$-band, we
include these galaxies because their $z-K$ colors suggest that they
are sBzK or passive BzK (pBzK; $BzK > -0.2 \bigcap z-K>2.5$), and
their photometric redshifts are consistent with the redshift range of
BzK galaxies. We did not select on the basis of spectroscopic
redshift, although spectroscopic redshifts were previously known for
some galaxies in the present sample (Table \ref{tab:obscat}).

To preferentially select actively star-forming galaxies, we put higher
priority on the galaxies identified in the publicly available {\it
  Spitzer} MIPS source catalog (the first delivery of the {\it
  Spitzer} data catalog from the GOODS Legacy Project; Chary et al. in
preparation). The flux limit of the catalog ($S_{24}>80\,\mu{\rm Jy}$)
corresponds to $L_{TIR} \sim 4.1\times 10^{11} - 3.6 \times
10^{12}{\rm L}_\odot$ at $z = 1.4 - 2.5$ using the infrared SED
templates by \citet{Chary2001}; thus the galaxies in the MIPS source
catalog are as luminous as (ultra-)luminous infrared
galaxies. MIPS sources in the public catalog are identified with 127
of the sample galaxies (we call them sBzK-MIPS galaxies,
hereafter). We selected 18 sBzK-MIPS galaxies. Furthermore, we added
19 sBzK galaxies not listed in the MIPS public catalog to the list of
targets (sBzK-non-MIPS galaxies, hereafter).

\fighistbzk

It is known that sBzK criterion picks out various types of
star-forming galaxies and passively evolving galaxies at $1.4 \leq z
\leq 2.5$ \citep{Grazian2007}. In order to check that our sample
represents the sBzK population, we compare the $B-K$ color
distribution of our target galaxies to that of sBzK-non-MIPS and
sBzK-MIPS galaxies with $K_s<23$ in the MODS field (Figure
\ref{fig:hist_bzk}). The $B-K$ color reflects both the age and the
amount of dust of a galaxy at this redshift range. Our sBzK-MIPS
sample is reasonable representative of those in the field, according
to both $\chi^2$-test (64.65\,\%) and Kolmorov-Smirnov (KS) test
(92.46\,\%). On the other hand, the sBzK-non-MIPS sample is possibly
biased toward blue galaxies ($\chi^2$-test : 35.35\,\%, KS-test:
17.74\,\%). Red sBzK-non-MIPS galaxies are relatively rare, so that
the scaled number in the redder bin is too small ($\lesssim1$) to
provide robust statistics given the present sample size.

We utilize publicly available multi-wavelength data of GOODS-N to
exclude possible AGN which may contaminate the optical star-formation
emission. The hard X-ray luminosity at the detection limit of the {\it
  Chandra} Deep Field-North 2\,Ms Point-Source Catalogs
\citep{Alexander2003} at $z\sim2$ ($1\times 10^{42}\,{\rm
  erg\,s}^{-1}$), suggests that the emission lines of the hard X-ray
detected sBzK galaxies are powered by an AGN rather than massive stars
\citep{Alexander2002}. X-ray sources are identified with 62 of the 508
sBzK galaxies. We exclude these sources from the present
sample. Rest-frame infrared colors are often used to identify AGN
candidates using {\it Spitzer} IRAC bands (3.6\,$\mu$m, 4.5\,$\mu$m,
5.8\,$\mu$m, and 8.0\,$\mu$m)
\citep[e.g.,][]{Caputi2006c,Hayashi2009}. A galaxy with emission
dominated by the stellar population has a strong bump in the SED at
1.6\,$\mu$m \citep{Sawicki2002}. We confirmed that all of the present
non-X-ray galaxies have strong 1.6\,$\mu$m bumps in the SED of the
IRAC bands.

\tabobscat

\subsection{Observations}
\label{subsec:obs}

\tabobslog

The spectroscopic observations for the selected galaxies were
performed with MOIRCS \citep{Suzuki2008} attached to the Subaru
Telescope \citep{Iye2004} during 2007 March 24--27 in the course of
the near-infrared spectroscopic observation of X-ray sources
(S07A-083, PI M. Akiyama). We used MOIRCS in MOS mode with the
low-resolution grism (HK500), which covers 1.3--2.3\,$\mu$m
wavelengths. The spectral resolution is $R\sim500$ ($\Delta
v\sim600~{\rm km}~{\rm s}^{-1}$) with 0\farcs8 slit width.  The 37
galaxies were observed with four slit masks as shown in Table
\ref{tab:obslog}. One slit mask was used for each night. The total
exposure time was 2.5--5\,hour for each mask.  The slit width was
fixed to 0\farcs8 for targets. Most of the slit lengths were 10\farcs0
or longer, depending on target positions. To monitor the flux
variation due to the atmospheric absorption and photon loss at the
slit mask, a bright point source was observed simultaneously on each
mask, except for mask CDFN3.

The FWHMs of seeing size in the $K_s$ band during the observations were
typically $\sim0\farcs6$.  According to the atmospheric attenuation
archived in the Canada France Hawaii Telescope Web
site\footnote{\url{http://nenue.cfht.hawaii.edu/Instruments/Elixir/skyprobe/home.html}},
the conditions were photometric for the first three nights.  For the
forth night, thin cirrus covered the sky.  The observation log is
summarized in Table \ref{tab:obslog}.

The observation followed the standard procedure for MOS observation with
MOIRCS described in \citet{Tokoku2007}. The target acquisition was done
using 6 to 7 alignment stars distributed over each slit mask. The
telescope was dithered along the slit (we call it an ``AB pattern'') by
3\farcs0, to achieve sufficient sky subtraction. The detector was read
eight times in non-destructive readout mode to reduce the read out
noise down to 16\,$e^-$.  The exposure time was set so that most of sky
emission lines, which were used later for the wavelength calibration,
were not saturated.  The typical exposure times were 300--900\,sec. The
relative position of the alignment star against the slit mask was
checked, and then the telescope pointing was adjusted every hour or two
without removing the slit masks from the focal plane. The systematic
guide error in one hour was measured to be $\lesssim$0\farcs4.  At the
beginning of each observing night, an A0V type star (HIP53735;
$K_s=8.856$; Two Micron All Sky Survey point source catalog;
\citealp{Skrutskie2006}) was observed for flux calibration with the same
instrument configuration as the targets.

\section{Data Reduction and Data Analysis}
\label{sec:ana}

\subsection{Data Reduction}

The data reduction was made, using a semi-automatic reduction script
(MOIRCS MOS Data Pipelines; MCSMDP\footnote{available at
  \url{http://www.naoj.org/Observing/DataReduction/}}), which
follows the standard procedure of IRAF\footnote{IRAF is distributed by
  the National Optical Astronomy Observatory, which is operated by the
  Association of Universities for Research in Astronomy (AURA) under
  cooperative agreement with the National Science
  Foundation.}:ONEDSPEC. At first, we make a map of bad pixels; bad
pixels are identified as those that show non-linearity. This includes
dead (always zero count) and hot (always high count) pixels. Cosmic
rays are detected by a moving block average filter using
IRAF:CRAVERAGE.  Each input frame for IRAF:CRAVERAGE is first divided
by its dithered pair so that cosmic rays stand out behind the sky
background. The cosmic-ray mask was combined with the bad pixel
mask. The values of bad pixels and cosmic-ray pixels are replaced with
values interpolated linearly from the pixels along the slit. Sky
emission lines were subtracted using the dithered pair frames (we call
the procedure ``A-B subtraction''). The dome-flat frames for flat
fielding were taken with the same slit mask and grism configuration as
used for the observation. For the first-order distortion correction,
the database for the imaging mode of MOIRCS was applied to the
spectral frames. The tilt of the spectrum against the detector rows
was corrected in the cases when it was measurable in the continuum of
bright objects.

Next, each spectrum was cut out based on a mask design file. For
wavelength calibration, OH sky emission lines were identified in each
frame using a night-sky spectral atlas \citep{Rousselot2000}. Using
the pixel coordinates of the sky emission lines, the sky-subtracted
frames were transformed such that the sky lines aligned along the
column of the frames and that the wavelength was a linear function of
pixel coordinates. Background emission still remained after A-B
subtraction due to time variation of OH emission during dithering
observations.  The residuals were subtracted by fitting with a
quadratic equation along the column. Finally, all frames were weighted
with the flux of the bright point source and stacked using the
weighted average to obtain the final two-dimensional spectra.

The response spectra, including atmospheric transmission and the
instrument efficiency, are calculated using the flux standard star to
convert the observed count to an absolute flux.  The data of the flux
standard star is reduced to a two-dimensional spectrum in the same way
for galaxy targets and then the continuum is traced with a polynomial
to extract a one-dimensional spectrum with IRAF:APALL. We calculate a
model spectrum of the A0V flux standard star using
SPECTRUM\footnote{\url{http://www.phys.appstate.edu/spectrum/spectrum.html}}
(SPECTRUM uses \citealp{Castelli2003} model). The model was scaled to
$K_s = 8.856$ and the spectral resolution of the model was degraded to
$R\sim500$ with a Gaussian kernel, so as to give the same hydrogen
line width as the observed spectra. Finally, the one-dimensional
spectrum of the standard star was divided by the model spectrum to
obtain the response spectrum.

Emission lines are detected by visual inspection on the
flux-calibrated two-dimensional spectra. The signal-to-noise ratio
evaluation described later shows that visual inspection corresponds to
a larger than $\sim3\sigma$ detection. The detected emission lines
were integrated along slit direction within an appropriate aperture
giving the largest S/N to obtain one dimensional spectra.

\subsection{Measurement of Emission Lines}
\label{subsec:emission} 

\fighaspec
\figothreespec
\tabsfrcat

We first consider the strongest emission line as H$\alpha$. Taking
account of the wavelength range where we observed and the expected
redshift range of sBzK galaxies, it is the most plausible candidate
for a strong emission line. The identification of H$\alpha$ is
confirmed in all cases when other emission lines, such as
\ntwo$\lambda\lambda$6548,6583, \othree$\lambda\lambda$4959,5007, and
H$\beta$ are also detected at the expected wavelength.  Otherwise, we
check the consistency of the redshift of the H$\alpha$ emission line
with that of the SED model fitting using broad-band data (see
\S\ref{subsec:sedfit}).

Although one of the conventional methods to estimate random error is
to obtain the rms of the scattering of background counts at both sides
of the object, the slit length for each object is not long enough. The
readout noise is suppressed down to 16\,$e^-$ by non-destructive
readout mode, so that the readout noise is about 1/5 of the background
noise. Because the readout noise and the photon noise of objects are
negligible, we estimated the background noise by assuming that the
random error is dominated by sky-background noise, which follows
Poisson statistics. In parallel to the reduction of the object
spectra, sky-background spectra were reduced without the background
subtraction, and the Poisson error of each pixel is estimated from
them. We found that the Poisson error is systematically lager than the
rms error by a factor of $\sim1.5$, though the scatter is large. The
rms errors can be underestimated, because counts in the
two-dimensional spectra correlate with the neighboring pixels due to
the sub-pixel shift in distortion correction and wavelength
calibration.

To measure the flux and central wavelength of the emission lines, we
used SPECFIT \citep{Kriss1994}, which fits multiple emission lines and
continuum simultaneously. Gaussian profiles were used for emission
lines. The continuum was assumed to be a linear function of wavelength.
The flux of \ntwo$\lambda6548$ was fixed to be 1/3 of \ntwo$\lambda6583$
flux. If H$\beta$ and/or \othree$\lambda\lambda$4959,5007 emission lines
were detected on the two-dimensional spectra, we measured the flux in
the same way. The random errors of the fluxes were estimated from the
best-fit parameters of the model fitting based on the sky-background
error described above. The emission lines detected with S/N$>$3 are
listed in Table \ref{tab:obscat}, and the spectra are shown in Figures
\ref{fig:haspec} and \ref{fig:o3spec}.  The measured flux and the
redshift derived from the central wavelength of H$\alpha$ emission line
are shown in Table \ref{tab:sfrcat}.

It is difficult to accurately evaluate systematic flux error since
part of the flux from the object will be lost by a slit mask; the
systematic error due to the flux loss is not negligible if we measure
the total flux of emission lines from slit spectroscopy. The flux loss
can be caused by a variety of things, for example, a larger size of
the object than the slit width, imperfect alignment of the slit mask
to the target, and misalignment of the dithering direction along the
slit.

First we estimated the flux loss for a point source. When a Gaussian
profile with FWHM = 0\farcs6, which is the typical seeing during the
observations, and a slit width = 0\farcs8 are assumed, fluxes passing
through the slit are 88.4\% of the integrated flux; the aperture
correction factor is $\sim1.13$. We applied this factor to the point
sources observed.  We found that, however, the fluxes of the point
sources were still underestimated by up to $\sim20$\,\%, even if we
also correct the effect of the seeing size variation and pointing
error, which are measured from alignment stars. This fact indicates
that imperfect alignment of dithering and/or short-time variation of
the seeing size during the observations can also add to the systematic
error.

On the other hand, it is expected that the flux loss for extended
sources varies with the size of the sources. We calculated aperture
correction factors from the ratio of total flux to that expected to
come in through slits, assuming that the regions emitting stellar
continuum and H$\alpha$ had the same spatial distribution. The flux
through a slit was measured on the $K_s$-band image with a rectangle
of the same aperture size as for extracting one-dimensional spectrum
and the same position angle of the slit, while the total flux was
obtained from MAG\_AUTO of SEXTRACTOR. The alignment error of the slit
was estimated from the alignment stars on the slit mask.  We found the
median of the aperture correction factors to be $\sim1.9$ with little
dependence on the size of the objects. The rms scatter of the factor
among the objects is $\pm0.5$. We apply the median value to all
objects to correct the flux loss. We checked the above aperture
correction method by comparing the continuum flux derived from the
corrected spectrum with total magnitude, which was measured from $K_s$
or $H$-band imaging data according to the redshift of each galaxy. We
used 7 galaxies in the sample whose stellar continuum around the
H$\alpha$ emission line was detected with sufficient S/N. We applied
the aperture correction to the continuum flux, then contrasted the
corrected flux with the total magnitude. The comparison shows that
both results are consistent and the uncertainty in the aperture
correction is within $\sim20\,\%$.

We need to note that the assumptions on the light distribution of
H$\alpha$ emission line would also cause systematic errors. The
H$\alpha$ and continuum emitting regions would not always be similar
for star-forming galaxies at $z\sim2$. Studies with the integral-field
spectrograph show that H$\alpha$ emission does not trace the stellar
continuum in some galaxies
\citep{Swinbank2005b,Swinbank2006b,Kriek2007}. However, this kind of
check on our slit-spectroscopy sample is difficult, because both the
continuum and emission lines are so faint that the light distribution
along the slit is not able to be measured. As a lower limit of the
aperture correction, when the H$\alpha$ emitting region is point like,
the flux becomes smaller by $\sim40\%$.

\subsection{Stacking Analysis}
\label{subsec:stack} 

\figothreestack

The flux ratios of rest-optical emission lines such as
\ntwo$\lambda$6583/H$\alpha$, \othree$\lambda$5007/H$\beta$, and
H$\alpha$/H$\beta$ provide us information on physical properties of
ionized regions such as the energy sources for ionization
(\ntwo/H$\alpha$ and \othree/H$\beta$;
\citealp{Baldwin1981,Veilleux1987,Kewley2001a,Kauffmann2003}), the
amount of dust extinction (Balmer decrement, e.g., H$\alpha$/H$\beta$;
\citealp{Osterbrock1989}), and gas-phase metallicity (\ntwo/H$\alpha$;
\citealp{StorchiBergmann1994,Raimann2000,Denicolo2002,Pettini2004}).
However, except for H$\alpha$, the emission lines of the present
sample are so faint that we can not evaluate the line ratios of an
individual galaxy. Therefore, we stacked the calibrated spectra in the
\ntwo-H$\alpha$, and \othree-H$\beta$ wavelength ranges to obtain high
S/N spectra, and discuss the averaged physical properties.

The stacked spectrum $\bar{F_{\lambda_e}}$ was calculated with
\begin{equation}
 \bar{F_{\lambda_e}} = \sum\limits_n \frac{F_{H\alpha, 0}}{F_{H\alpha, n}}  F_{\lambda_e,n}(\lambda_e),
 \label{eq:stack}
\end{equation}
where $F_{\lambda_e,n}(\lambda_e)$ and $F_{H\alpha, n}$ are the
spectrum and the total H$\alpha$ flux of $\rm n_{th}$ object,
respectively. The spectra were re-sampled in the rest-frame wavelength
scale ($\lambda_e$). The fluxes were scaled with total H$\alpha$ flux,
though the variety of fluxes among the samples are not large. Because
our main purpose of stacking is to examine the average physical
properties of galaxies from their line ratio, weighted mean (by either
inverse square of noise or luminosity) is not applied; the line ratios
of weighted mean flux could be biased toward strong emission line
galaxies.

We selected the objects whose spectra covered the wavelengths of
redshifted H$\alpha$, \ntwo, H$\beta$ and \othree. The emission lines
of the objects which had at least H$\beta$ or \othree\, with $S/N>3$
were stacked for the measurements of the line ratios of
\ntwo/H$\alpha$ and \othree/H$\beta$. As a result, 9 galaxies were
used for the stacking. The stacked H$\alpha$-\ntwo\, and
H$\beta$-\othree\, spectra are shown in Figure \ref{fig:o3stack}. The
emission lines of the stacked spectra were fitted with a combination
of the same Gaussian profiles and linear continuum measurements used
in \S\ref{subsec:emission}. The best-fit spectra are also shown in
Figure \ref{fig:o3stack}. We also made stacked spectra including the
objects with neither H$\beta$ nor \othree\, detection, but the
resultant line ratio was not changed.

To evaluate the effect of galaxies which would happen to have a
significantly different line ratio from the others, we used a bootstrap
method. We built 1000 bootstrap sets, whose galaxies were randomly
selected, allowing duplicate sampling from all galaxies used in the
stacking, and stacked the spectra of the galaxies in each set to derive
the line ratios. We regard the 16\,\% and 84\,\% quantile of the
distribution as 1$\sigma$ uncertainty in line ratio.

\subsection{SED fitting}
\label{subsec:sedfit} 

\figsed

To evaluate the properties of stellar population such as stellar mass,
amount of dust extinction ($A_V$), and age, we use multiband
photometric data from the MODS $JHK_s$ images and publicly available
images. The publicly available images we use are: $U$(KPNO/MOSAIC;
\citealp{Capak2004}), $BViz$ (ACS/HST; GOODS Treasury Program Version
1.0; \citealp{Giavalisco2004}), 3.6\,$\mu$m, and 4.5\,$\mu$m
(IRAC/SST; DR2 of the Spitzer Legacy Science program for GOODS;
Dickinson et al. in preparation). These images are deep enough to
detect our galaxies with sufficient S/N and to obtain the properties
of stellar population. The flux of the galaxies were obtained with
${\rm S/N}=8$ for $>80\%$ of the objects in all bands, except for $B$
(60\,\%) and $U$ (24\,\%) bands. The aperture size for photometry is
set to 1\farcs5, which is approximately $2\times{\rm FWHM}$ of the PSF
size in $K_s$ band.  Aperture corrections were applied to each band
properly \citep[see][]{Kajisawa2009}. The contribution of H$\alpha$
emission to the flux in $K_s$ or $H$ band, which depends on redshift,
was subtracted. The corrections are 0.01 - 0.34\,mag with a median of
$\sim0.08$\,mag.

We fitted the photometry data with stellar population SED models. The
SED libraries of GALAXEV \citep{Bruzual2003} are used for the
fitting. We adopted the libraries with metallicities of 1.0\,${\rm
  Z}_\odot$, 0.4\,${\rm Z}_\odot$, 0.2\,${\rm Z}_\odot$ and a
\citet{Chabrier2003} IMF. Exponentially declining star formation
histories ($\tau$-model) are assumed: SFR $\propto e^{-t/\tau}$ with
$\tau = $ 0.01, 0.02, 0.05, 0.10, 0.20, 0.50, 1.0, 2.0, and 5.0
Gyr. The starburst attenuation law of \citet{Calzetti2000} was applied
to the model with 0.1 steps from $A_V$ = 0.0 to 3.0.

Estimations of the stellar mass are affected by the adopted IMF. Many
authors employ a \citet{Salpeter1955} IMF, which is known to
overestimate stellar mass due to the steep faint-end slope of the
IMF. In addition to the \citet{Chabrier2003} IMF, we fitted the SEDs
to the model with the \citet{Salpeter1955} IMF and measured stellar
mass for comparison. For our sample, the stellar mass with the
\citet{Salpeter1955} IMF is larger by a factor of $\sim 1.7$ than that
with the \citet{Chabrier2003} IMF.

Recent evolutionary population synthesis models, for example
\citet{Maraston2005}, which take account of contributions of
thermal-pulsating asymptotic giant branch stars to SEDs, result in
changes in the inferred SED properties. We also fitted the SEDs with
\citet{Maraston2005} model and found that the model made the stellar
masses smaller by a factor of $\sim1.6$.

HYPERZ \citep{Bolzonella2000} was used to determine best-fit SED models.
The redshift for the fitting was fixed to that derived from the
H$\alpha$ emission line. We derived total stellar masses of galaxies,
scaling the stellar masses and SEDs with the total magnitude (MAG\_AUTO)
of the $K_s$ band. The parameter matrix, which gives a $\chi^2$ value
for each grid of SED models, was used to evaluate the uncertainty in the
model parameters. The uncertainty is obtained from the probability
distributions of the parameters. For example, the probability
distribution as a function of stellar mass ($P(M_\star)$) was calculated
from a minimum $\chi^2$ value for each stellar mass ($\chi^2(M_\star)$).
We determined a 68\% confidence level of the distribution as 1$\sigma$
error of the parameter. The stellar population parameters and the
1$\sigma$ errors derived from the SED fitting are listed in Table
\ref{tab:sedcat}. Some examples of the best fit SEDs and the probability
distributions of stellar mass, $A_V$, and age are shown in Figure
\ref{fig:sed}.

\tabsedcat

\section{Results}
\label{sec:res}

\subsection{Properties of H$\alpha$ Emission Lines}
\label{subsec:haemission}

\fighalum

H$\alpha$ emission lines are detected from 23 out of the 37 galaxies
(62\,\%). The detection rate does not depend on the mid-IR flux. The
rate for sBzK-MIPS and sBzK-non-MIPS galaxies are 11/18 (61\,\%) and
12/19 (63\,\%), respectively.

The spectroscopic redshifts are consistent with their photometric
redshifts in the MODS catalog \citep{Kajisawa2009}; the median value
of the relative errors is $\left<dz\right>\sim -0.026$ with a scatter
of 0.035, where $dz \equiv (z_{\rm phot} - z_{\rm H\alpha})/(1+z_{\rm
  H\alpha})$. The spectroscopic redshifts are newly determined for 10
galaxies. Out of the remaining 13 galaxies with spectroscopic
redshifts in the literature
\citep{Cohen2001,Cohen2000,Dawson2001,Wirth2004,Cowie2004,Treu2005,Chapman2005,Reddy2006b,Barger2008},
two galaxies have slightly different spectroscopic
redshift. MODS32-0116 has $z_{\rm H\alpha} = 1.488$ ($z_{\rm
  literature} = 1.315$) and MODS21-2321 has $z_{\rm H\alpha} = 1.734$
($z_{\rm literature} = 1.879$).

The left panel of Figure \ref{fig:halum} shows the apparent H$\alpha$
luminosity and the redshift distribution of the H$\alpha$-detected
sBzK galaxies. The galaxies lie at redshifts between 1.49 and 2.49
with a median of 2.12. The redshift distribution is consistent with
that expected for BzK galaxies ($1.4\lesssim z \lesssim 2.5$;
\citealp{Daddi2004b}). The lack of galaxies at $1.8<z<2.0$ is due to
an atmospheric absorption band between $H$ and $K$ bands.

The luminosity distances of these objects are calculated with their
redshifts measured using the H$\alpha$ emission lines. The fluxes of
H$\alpha$ emission lines ($F_{H\alpha}$) range between 1 and 17 $\times
10^{-17}\,{\rm ergs\,s}^{-1}\,{\rm cm}^{-2}$, which corresponds to an
H$\alpha$ luminosity ($L_{H\alpha}$) of 0.5 to 7.0 $\times 10^{42}\,{\rm
ergs\,s}^{-1}$. $F_{H\alpha}$ and $L_{H\alpha}$ are listed in Table
\ref{tab:sfrcat}.

Possible reasons for non-detections of emission lines from the
remaining 14 galaxies are; (1) their H$\alpha$ emission lines are too
faint for the present observations; (2) the redshift of those galaxies
are out of the $1.2\lesssim z \lesssim 2.5$ range, where the grism
(HK500) we used covers the H$\alpha$ wavelength; (3) the galaxies lie
at the redshift where H$\alpha$ emission line falls on strong sky
emission lines or atmospheric absorption bands. Because the
photometric redshifts and the errors of the galaxies with emission
lines undetected are at $1.2<z<2.5$, the second possibility is
unlikely except for MODS21-2047, whose $z_{\rm literature}$ is
0.999. In order to check the third possibility, we calculate the
detection limit of H$\alpha$ emission as a function of redshift using
the sky background noise and the response spectrum. A Gaussian profile
is assumed for H$\alpha$ with $\rm FWHM=540\,km\,s^{-1}$, which
corresponds to the wavelength resolution of the HK500 grism with a
0\farcs8 slit width. Provided that H$\alpha$ emitters have a uniform
redshift distribution at $1.4\leq z \leq 2.5$ and an apparent
H$\alpha$ luminosity of $1.8 \times10^{42}\,{\rm ergs}\,{\rm s}^{-1}$,
which corresponds to the 84\,\% quantiles of the distribution of
apparent H$\alpha$ luminosity, the expected detection rate of the
H$\alpha$ emission line with ${\rm S/N}>4$ is 87.7\,\%, which
corresponds to $\sim32$ objects: objects at $1.8\lesssim z \lesssim
2.0$ are rarely detected. Therefore, for 5 of the observed samples,
H$\alpha$ emission lines are undetected due to the strong sky emission
lines or atmospheric absorption bands, while the remaining 9 galaxies
have smaller SFR and/or larger attenuation than the galaxies with
emission line detection.

\subsection{\othree$\lambda$5007/H$\beta$ and \ntwo$\lambda$6583/H$\alpha$ Line Ratios}
\label{subsec:bpt}

Diagnostic diagrams based on optical emission-line ratios are commonly
used to determine the energy source of ionization. Possible sources
are massive stars, AGNs and energetic shocks
\citep{Baldwin1981,Veilleux1987}. In this paper, we use H$\alpha$
luminosity as an indicator of SFR, assuming that its energy source is
massive stars. We excluded AGN candidates from the sample using
multi-wavelength data (see \S\ref{subsec:sample}). In addition to
these diagnostics, we use optical emission-line ratios to determine
the energy sources of the present emission line objects.

In Figure \ref{fig:bpt}, we plotted \othree$\lambda$5007/H$\beta$ and
\ntwo$\lambda$6583/H$\alpha$ line ratios of the stacked spectra and
the galaxy with the four emission lines detected
(MODS31-0033). Results for $\sim347000$ galaxies from SDSS DR6
\citep{SDSSDR6} and local emission line galaxies
\citep{Moustakas2006a} are also shown. A classification line
empirically determined from the SDSS objects \citep{Kauffmann2003} and
a theoretically determined classification line \citep{Kewley2001a} are
also shown. Emission-line galaxies located below and to the left of
the lines are expected to have massive stars for their energy
sources. The distribution of local SDSS galaxies exhibits the two
expected sequences: the populations below (star-forming galaxies) and
above (AGNs) the solid line.

The emission line of the stacked sample place it on the theoretical
line, while MODS31-0033 is located to the lower left of the empirical
line. Our results are located above the sequence of local
star-forming galaxies, closer to the AGN sequence. In fact, a similar
offset is seen in star-forming galaxies at $z\sim2$
\citep{Erb2006a,Kriek2007}. One possible reason for the offset is a
higher electron density and a harder ionizing spectrum than for local
star-forming galaxies \citep{Erb2006a}.

We note that AGNs can be also identified based on \othree-H$\alpha$
line ratios for objects without an H$\beta$ detection. AGNs typically
have \othree/H$\beta$ $\gtrsim3$ on Figure \ref{fig:bpt}. On the other
hand since the H$\alpha$/H$\beta$ line ratio must be $\sim3$ (the
Balmer decrement) or larger (if there is a significant dust
extinction) the \othree-H$\alpha$ line ratio of an AGN candidate will
be larger than 1. Only one object (MODS41-0194) has \othree-H$\alpha$
$>1$ (1.67). When we exclude MODS41-0194 from the stacking, the
stacked line ratios slightly move below the empirical line (Figure
\ref{fig:bpt}). We regard this object as a possible candidate of AGN
in the later discussion.

\figbpt

\subsection{Extinction}
\label{subsec:ext} 

\figextinction

In order to evaluate SFR from intrinsic H$\alpha$ luminosity, the
extinction due to interstellar dust needs to be corrected. Moreover,
the samples are expected to have large amounts of dust, because about
half of the present galaxies are luminous infrared galaxies, in which
strong infrared emission is thought to originate from dust thermal
emission.

The extinction of emission lines in star-forming regions can be
estimated with the reddening of the Balmer decrement through an
extinction curve. The intrinsic line ratios of the Balmer emission lines
are nearly constant with respect to electron temperature of ionized
gas. The amount of reddening is estimated from the ratio of the
attenuated-to-intrinsic Balmer line ratio,
\begin{equation}
  E(B-V)_{\rm gas} = \frac{2.5}{k(\lambda_{H\alpha}) - k(\lambda_{H\beta})} \log
   \frac{F_i^{\lambda_{H\alpha}}/F_i^{\lambda_{H\beta}}}{F_o^{\lambda_{H\alpha}}/F_o^{\lambda_{H\beta}}},
  \label{eq:bd}
\end{equation}
where ``i'' and ``o'' denote intrinsic and observed flux, respectively,
and $k(\lambda)$ is an extinction curve. 
We assume \citet{Calzetti2000} law for $k(\lambda)$.

The Balmer decrement method can only be applied to the 14 galaxies at
$z\gtrsim1.9$ in the present sample, because the H$\beta$ wavelength
is not covered by the HK500 grism for galaxies at lower redshift. Both
H$\alpha$ and H$\beta$ are detected in four galaxies (MODS11-0390,
MODS12-0255, MODS22-1411, MODS31-0033), so that $E(B-V)_{\rm gas}$ is
estimated using the above equation.  For the rest of the sample, lower
limits on the $E(B-V)_{\rm gas}$ are evaluated using H$\beta$
emission-line upper limits. The resultant $E(B-V)_{\rm gas}$ values
are $\gtrsim0.5$. We also used the stacked spectrum to measure the
average amount of reddening and found it to be $E(B-V)_{\rm
  gas}=0.75^{+1.19}_{-0.02}$.

We estimate the amount of reddening for the ionized gas from the SED
fitting and apply these extinction corrections to the H$\alpha$
luminosities for all of the galaxies. Figure \ref{fig:extinction}
shows the comparison between the amounts of reddening derived from the
Balmer decrement and SED fitting. Our result shows that the gas is
more attenuated than the stars. From SED fitting, the amount of
reddening for stars in the samples used for stacking are found to be
$E(B-V)_{{\rm stellar}}=0.20^{+0.24}_{-0.08}$, thus $E(B-V)_{{\rm
    stellar}}/E(B-V)_{{\rm gas}}\sim 0.26^{+0.42}_{-0.10}$. It should
be noted that the galaxies which we used to determine the relation
suffer relatively less attenuation since we would not be able to
detect H$\beta$ emission in the more highly attenuated galaxies. The
hatched region in the figure also shows the lower limit of
$E(B-V)_{\rm gas}$ corresponding to the 3$\sigma$ detection limit of
H$\beta$. The height of the region calculated from the 68\%
distribution of the observed H$\alpha$ fluxes.

From studies of local starburst galaxies, it is known that ionized gas
is more attenuated than stars \citep{Calzetti2000,Calzetti2001}. The
difference is thought to be caused by changes in the geometry of the
ionized gas and stars in a typical galaxy. \citet{Calzetti2001} give
the relation between the color excess observed for gas and stars:
\begin{equation}
 E(B-V)_{{\rm stellar}} = 0.44 E(B-V)_{{\rm gas}}.
 \label{eq:ebv}
\end{equation}
This relation is shown with the dashed line in Figure
\ref{fig:extinction}. Our stacked result is consistent with the
equation within errors. On the other hand, H$\beta$ detected galaxies
are also consistent with the case that extinction for gas and stars is
equal (defined as ``equal extinction'', hereafter), though these
galaxies would lie at the lower end of reddening as noted above. We,
therefore, employ Equation \ref{eq:ebv} as an averaged nature of dust
extinction to correct for the effect of dust extinction on the
emission lines and compare these with that calculated assuming equal
extinction. We discuss the reliability of the application of these
extinction correction further in the \S \ref{subsec:extcor}.

The right panel of Figure \ref{fig:halum} shows the luminosity of the
H$\alpha$ emission lines of the galaxies after the extinction is
corrected. The extinction correction factor of our samples ranges from
1.2 to 87, resulting in a wider spread in extinction corrected
H$\alpha$ luminosities ($\sim0.6$\,dex for the \citet{Calzetti2001}
relation and $\sim0.4$\,dex for equal extinction) than in the apparent
H$\alpha$ luminosities ($\sim0.3$\,dex). This result indicates that
proper extinction correction is important to reliably calculate the
intrinsic H$\alpha$ luminosity.

\subsection{Star Formation Rate and Stellar Mass}
\label{subsec:sedms} 

\figsedms
\figagests

To convert the extinction-corrected H$\alpha$ luminosity to SFR, we
derive a relation between them using stellar population synthesis
models, following \citet{Kennicutt1998}. We employ GALAXEV models
\citep{Bruzual2003} with the \citet{Chabrier2003} IMF and the same
parameters as used in \S\ref{sec:ana} to calculate ionizing photons
($N_{{\rm Ly}}$). $N_{{\rm Ly}}$ is converted to $L(H\alpha)$ based on
a case-B recombination model \citep{Osterbrock1989}.  If a constant
SFR model and solar metallicity are assumed, the $L(H\alpha)$-SFR
relation with an age older than 20\,Myr is given by,
\begin{equation}
 {\rm SFR}(H\alpha) ({\rm M}_\odot~{\rm yr}^{-1})
 = 4.9 \times 10^{-42} L(H\alpha) ({\rm erg}~{\rm s}^{-1}).
\label{eq:hasfr}
\end{equation}
Using this equation, we derived the SFR with extinction-corrected
H$\alpha$ luminosity. The result is listed in Table
\ref{tab:sfrcat}. The SFR estimate is larger by a factor of $\sim1.5$,
if the \citet{Salpeter1955} IMF is adopted.

The stellar masses derived from the SED fitting are compared with the
SFRs calculated using both extinction correction approaches in Figure
\ref{fig:sed_ms}. For the extinction correction using the
\citet{Calzetti2001} relation, the SFRs range from 10 to several
hundreds ${\rm M}_\odot\,{\rm yr}^{-1}$. Some massive galaxies ($\sim
10^{11}\,{\rm M}_\odot$) and a significant fraction of low mass
galaxies show SFR larger than $100 \,{\rm M}_\odot\,{\rm yr}^{-1}$. As
a result, there is no clear correlation between SFR and stellar
mass. The present sample contains many sBzK-MIPS galaxies, whose SFRs
are expected to be high. Although sBzK-MIPS galaxies lie at the
massive and high SFR end of the distribution, there still exist
low-mass and high-SFR galaxies among sBzK-non-MIPS galaxies.

The specific SFR (SSFR), which is the SFR divided by stellar mass, is
one of the indicators showing the level of star formation; higher SSFR
indicates that a galaxy is accumulating large stellar mass compared to
its own stellar mass on a short timescale. In Figure \ref{fig:sed_ms},
SSFR is depicted by dashed lines. SSFR does not depend on MIPS
detection but on stellar mass. There exist actively star-forming
galaxies (${\rm SSFR}>10^{-8}\,{\rm yr}^{-1}$) among the less-massive
galaxies, while the SSFR of massive galaxies ($\gtrsim 10^{10}\,{\rm
  M}_\odot$) is nearly constant ($\sim 2\times10^{-9}\,{\rm
  yr}^{-1}$). It can imply that some low mass galaxies at $z\sim2$ are
actively forming stars.

On the other hand, for the SFRs corrected assuming the
equal-extinction case, the SFRs and SSFRs of the high-SSFR galaxies
become lower, but they still have higher SSFR when compared to other
galaxies. The SSFR of the original high-SSFR galaxies become ${\rm
  SSFR}>3.5\times10^{-9}\,{\rm yr}^{-1}$. Consequently, there is still
no correlation between them. The distribution of SFRs is nearly
constant ($\sim 30\,{\rm M}_\odot\,{\rm yr}^{-1}$).

The possible AGN candidate (MODS41-0194; denoted by a open square)
shows lower SFR than other galaxies with similar stellar masses. We
note that this object does not change our findings of no correlation
between SFR and stellar mass and low mass galaxies with active star
formation.

The inverse of SSFR (SSFR age) gives the estimated age for a galaxy to
build up all of its stellar mass with the current SFR. On the
contrary, mass-weighted age, which is calculated by integrating the
ages of stars weighted by SFR at that time, is an indicator of the
average age of the stars in a galaxy. The mass-weighted age is given
from integration of the star formation history of galaxy as
\begin{equation}
 \tilde{T}(t) = \frac{\int^t_0 \phi (t') \times (t-t')dt'}{\int^t_0 \phi (t') dt'},
\label{eq:mwage}
\end{equation} 
where $\phi(t)$ is the SFR at age t. Comparing the two ages, we can
roughly check consistency between the SED fitting result and the SFR
estimated from extinction corrected H$\alpha$ luminosity. The
comparison between the SSFR age for the both cases of extinction
correction and mass-weighted age of our sample is shown in Figure
\ref{fig:age_sts}. For extinction correction with the
\citet{Calzetti2001} relation, the SSFR ages agree with the
mass-weighted ages to within 1\,$\sigma$ error; a galaxy with a higher
SSFR has a younger stellar population. This fact suggests that the
majority of the stellar mass for the galaxies in our emission line
samples were formed in a recent starburst. On the other hand, for the
extinction correction assuming the equal extinction case, the ages of
stars are young compared to the SFR especially for high-SSFR
galaxies. In this case the galaxies must have experienced much more
active star formation in the recent past and terminated the star
formation in a very short timescale ($\lesssim 10\,$Myr).

\section{Discussion}
\label{sec:discuss}

\subsection{Extinction Correction and Star Formation Rate}
\label{subsec:extcor}
The median value of the reddening derived from the SED fitting is
$E(B-V)_{\rm stellar}=0.42,\,0.21$ ($A_V=1.7,\,0.85$) for sBzK-MIPS
and sBzK-non-MIPS galaxies, respectively. Both values are relatively
larger than those for local galaxies (e.g., $\left<A_V\right>_{\rm
  median}\sim0.43$ for $r$-band selected SDSS galaxies;
\citealp{Kauffmann2003}). The extinction corrections are key to the
discussion on the luminosities of H$\alpha$ emission line, and thus
the SFRs and SSFRs; larger extinction correction makes
extinction-corrected SFR and SSFR larger. In this section, we discuss
the reliability of our extinction correction through comparisons of
the H$\alpha$ SFRs with other SFR indicators, and the reason why the
higher SSFR galaxies are found by our observation.

\subsubsection{UV Luminosity}

\figsfrcompuv

The UV continuum is used as a SFR indicator, because the stellar
continuum at 1500-2800\,\AA\ is sensitive to emission from massive
stars. We compare the SFRs with those estimated from the UV
continuum. We calculate a relation between UV continuum at
$\lambda=1500$\,\AA, $L_{\nu,1500\AA}$, and SFR as for Equation
\ref{eq:hasfr}. $L_{\nu,1500\AA}$ is measured from the SED templates
of GALAXEV with the same parameters as in \S\ref{subsec:sedms} with
the $B$-band filter of ACS/HST redshifted to $z=2$. The relation is
given by,
\begin{equation}
 {\rm SFR(UV)} (M_\odot~{\rm yr}^{-1}) = 7.50 \times 10^{-29} L_{\nu,1500\AA} 
({\rm erg}~{\rm s}^{-1}~{\rm Hz}^{-1}).
\label{eq:uvsfr}
\end{equation}
For our samples, $L_{\nu,1500\AA}$ is measured with $\phi$1\farcs5 in
$B$-band with an aperture correction obtained from the $K_s$-band
total magnitude. The dust extinction is corrected with that derived
from the SED fit.

The comparison between SFR(H$\alpha$) and SFR(UV) is shown in Figure
\ref{fig:sfrcompuv}. For the equal-extinction case, the SFR(H$\alpha$) is
underestimated for the galaxies with higher SFR(UV). On the other
hand, using the \citet{Calzetti2001} relation the SFR(H$\alpha$) and
SFR(UV) are roughly consistent for a broad range of SFR, but are
systematically larger by 0.3\,dex for both sBzK-MIPS and
sBzK-non-MIPS galaxies.

We note that the luminosity ratio of the UV continuum to H$\alpha$
varies with the population of a galaxy. For example, because the UV
continuum is contaminated by less massive stars than stars emitting
ionizing photon, galaxies populated by younger massive stars are more
luminous in H$\alpha$. For our template, the ratio is larger by factor
$\sim 3$ for age\,$\lesssim10\,$Myr and becomes constant for
age\,$\gtrsim 100\,$Myr. In fact, high-SSFR galaxies have a
mass-weighted age\,$\lesssim 100\,$Myr. If a top-heavy IMF is assumed,
moreover, the ratio gets larger, so that more ionizing photons are
emitted than UV continuum.

\subsubsection{MIPS Flux}

\figsfrcomp

In order to examine the reliability of the extinction corrections
further, we compare the SFR(H$\alpha$) estimates with those derived
from the thermal infrared luminosities (SFR(IR)), because the infrared
luminosity is an SFR indicator almost unaffected by dust
extinction. We use the rest-8\,$\mu$m luminosity ($\nu L_{\nu,8\mu
  m}$) derived from the MIPS 24\,$\mu$m flux as representative of the
thermal infrared luminosity. Independently from the public catalog,
the 24\,$\mu$m fluxes of the present galaxies are measured using the
$Spitzer$ MIPS image (Dickinson et al. in preparation) with the
IRAF/DAOPHOT package \citep{Stetson1992} for both sBzK-MIPS and
sBzK-non-MIPS samples. Some galaxies have 24$\mu$m fluxes severely
contaminated by nearby bright 24$\mu$m sources because of the large
PSF of MIPS (${\rm FWHM}\sim5\farcs4$). In order to reduce the effect
of the contamination, the fluxes are measured by fitting crowded
objects simultaneously with the PSF measured from isolated point
sources. The objects are centered using the $K_s$-band light
distribution. The photometric error is estimated on a blank-sky region
on the image. In order to take into account the residuals of the
fitting, the image from which the fitted models are subtracted is
measured and the residual is added to the error. Then, $\nu
L_{\nu,8\mu m}$ is estimated from the spectroscopic redshift of
H$\alpha$ emission line and the observed 24\,$\mu$m flux densities
with $k$-corrections derived from the luminosity dependent SED models
of \citet{Chary2001}. The measured MIPS-24\,$\mu$m flux and estimated
rest-8\,$\mu$m luminosity are also listed in Table \ref{tab:sfrcat}.

The derived $\nu L_{\nu,8\mu m}$ and H$\alpha$ luminosities are
plotted in Figure \ref{fig:sfrcomp}. The H$\alpha$ luminosities are
corrected using both the \citet{Calzetti2001} relation ({\it left})
and the equal extinction case ({\it right}) for comparison. The
expected relation between $\nu L_{\nu,8\mu m}$ and H$\alpha$
luminosity is also shown using a solid line in the figure. The
relation is derived based on that expected between $\nu L_{\nu,8\mu
  m}$ and SFR. The $\nu L_{\nu,8\mu m}$ is converted to total infrared
luminosity ($L_{\rm IR}$) in the wavelength range between 8 and
1000\,$\mu$m with the luminosity dependent SED models
\citep{Chary2001}, and the relation between the total infrared
luminosity and SFR,
\begin{eqnarray}
 {\rm SFR}(IR)\,({\rm M}_\odot~{\rm yr}^{-1})
  &=& 3.0 \times 10^{-44} L_{\rm IR}~ ({\rm erg\,s}^{-1}) \label{eq:irsfr}\\
  &=& 1.2 \times 10^{-10} L_{\rm IR}~ (L_\odot) \nonumber
\end{eqnarray}
is used.  The relation is derived by dividing that in
\citet{Kennicutt1998} by 1.5 to convert the \citet{Salpeter1955} IMF
to the \citet{Chabrier2003} IMF.  As we see in Figure
\ref{fig:sfrcompuv}, it is suggested that the extinction correction
between the stellar continuum and ionized gas depends on the SFR of a
galaxy. For higher (lower) SFR, the H$\alpha$ luminosity corrected
with the \citet{Calzetti2001} relation (equal-extinction case) is
consistent with $\nu L_{\nu,8\mu m}$. We divide the present sample
into two groups, LIRGs and ULIRGs, at $\nu L_{\nu,8\mu
  m}=2.8\times10^{44}\,{\rm erg}\,{\rm s}^{-1}$, which corresponds to
$L_{\rm IR} = 10^{12}\,{\rm L}_\odot$. For ULIRGs, the systematic
offset between SFR(IR) and SFR(H$\alpha$) (SFR(IR)/SFR(H$\alpha$)) is
0.31 (0.72) dex, with a scatter of 0.47 (0.29) dex for the
\citet{Calzetti2001} relation (equal-extinction case); although the
\citet{Calzetti2001} relation results better consistency,
SFR(H$\alpha$) is still underestimated. For LIRGs, the systematic
offset is -0.42 (0.02) dex with a scatter of 0.29 (0.43) dex for the
\citet{Calzetti2001} relation (equal-extinction case); using the
equal-extinction case results in a good agreement between the two SFR
indicators.

We need to note that SFR estimated from the rest-8\,$\mu$m luminosity
has a large uncertainty. Rest-frame luminosity at $\sim8\,\mu$m
contains a very small fraction of thermal infrared emission, so that
the $L_{\rm IR}$ is estimated by extrapolating the SED to longer
wavelengths. The rest-8\,$\mu$m luminosity could also be affected by
Polycyclic Aromatic Hydrocarbons (PAHs) emission lines, whose energy
source is dominated by late short-lived B-type stars rather than
O-type stars \citep{Peeters2004}. These factors can produce large
random error in the estimation of SFR from the rest-8\,$\mu$m
luminosity. For local galaxies, \citet{Chary2001} demonstrated that
the 1$\sigma$ error for the $L_{\rm IR}$ estimated from
rest-6.7\,$\mu$m was a factor of two. The uncertainty is consistent
with the observed scatter around the solid line in Figure
\ref{fig:sfrcomp}.

\subsubsection{Radio and Far-Infrared}
A correlation between FIR luminosity and radio continuum is known for
local galaxies \citep{Condon1992,Yun2001}. Non-thermal radio continuum
is mainly emitted by synchrotron emission from the remnants of
supernovae, whose frequency is proportional to the number of massive
stars, and thus the SFR. The relation between these luminosities for
local galaxies is given by
\begin{equation}
 L_{\rm IR}\,(L_\odot) = 3.5\times 10^{-12}\,L_{\rm 1.4\,GHz}\,({\rm W\,Hz^{-1}})
\label{eq:ir_radio}
\end{equation}
with about 40\% of uncertainty \citep{Yun2001}.

We use a publicly available 1.4\,GHz radio catalog and images
\citep{Biggs2006}, whose 1$\sigma$ rms noise in the GOODS-N region is
5.8\,$\mu$Jy\,${\rm beam}^{-1}$ (the beam size $=1.52\times1.51\,{\rm
  arcsec}^2$). The catalog contains sources with a $>5\sigma$
detection. Although our H$\alpha$-emission-line galaxies are not
matched with the sources in the catalog, some galaxies are marginally
detected ($\lesssim3\sigma$), according to our eye-ball check on the
1.4\,GHz radio images. Therefore, their 1.4\,GHz fluxes are lower than
29\,$\mu$Jy (5\,$\sigma$), and some galaxies possibly have $\sim$ a
few 10\,$\mu$Jy.

If we assume the local radio-IR correlation, we can calculate an
expected $L_{\rm 1.4\,GHz}$ from SFR using Equation \ref{eq:irsfr} and
\ref{eq:ir_radio}. We assume a spectral index of ($\alpha$; $f_\nu
\propto \nu^{-\alpha}$) of the non-thermal radio continuum
$\alpha=0.8$ when we calculate the expected 1.4\,GHz flux density. The
expected 1.4\,GHz flux estimated from the MIPS-24\,$\mu$m flux exceeds
the detection limit only for one emission line galaxy (MODS12-0125),
even if the uncertainty in the estimation of $L_{\rm IR}$ from
rest-8\,$\mu$m luminosity is considered. This object is a sBzK-MIPS
galaxy and has a low SSFR. For this object the SFR(H$\alpha$) is also
smaller than SFR(IR), so that only the rest-8\,$\mu$m luminosity
exceeds that at other wavelengths. This infrared excess when compare
to the radio luminosity can be explained by a warm dust component,
which is thought to be caused by a compact nuclear starburst or
dust-enshrouded AGN \citep{Yun2001}.

On the other hand, when we calculate the radio flux from the
extinction-corrected H$\alpha$ luminosity using the
\citet{Calzetti2001} relation, the radio emission of 5 emission line
galaxies (MODS22-1411, MODS21-2612, MODS22-5133, MODS11-0390, and
MODS22-2658) exceed the detection limit. The first four galaxies are
high-SSFR galaxies. Although most of these galaxies have marginal
detection of 1.4\,GHz flux, the expected radio fluxes ( $L_{\rm
  1.4\,GHz}\sim50-150\,\mu{\rm Jy}$) are higher than the observed
radio flux, even if the uncertainty in the radio-IR correlation is
taken into account. On the other hand, for the H$\alpha$ luminosity
calculated assuming the equal-extinction case, none of the galaxies
exceed the detection limit. This would suggest that the amounts of
extinctions in the ionized gases are overestimated by using the
\citet{Calzetti2001} relation, though the uncertainties in their
extinctions are large. We also note that, equal extinction in turn
underestimates the extinction for ULIRGs (see Figure
\ref{fig:sfrcomp}). These H$\alpha$-excess galaxies consist of three
ULIRGs (MODS11-0390, MODS22-1411, and MODS22-2658).

Thermal emission from dust heated by massive stars ($\sim 30\,$K) is
observed at sub-millimeter wavelengths. The GOODS-N region is also
observed with SCUBA/JCMT \citep[850\,$\mu$m;][]{Holland1999},
AzTEC/JCMT \citep[1.1\,mm;][]{Wilson2008}, and MAMBO/IRAM
\citep[1.2\,mm;][]{Kreysa1998} down to $\sim2$\,mJy
\citep[e.g.,][]{Pope2005,Pope2006}, $\sim1$\,mJy
\citep{Perera2008,Chapin2009}, and $\sim0.7$\,mJy \citep{Greve2008},
respectively. One of our sample, MODS31-0033, is identified with an
AzTEC source, AzGN27, whose 1.4\,GHz and 24\,$\mu$m counterparts are
also identified by \citet{Chapin2009}, although this object is not
detected at both 850\,$\mu$m and 1.2\,mm wavelengths.

Interestingly, the 1.1\,mm flux (2.31$^{+1.16}_{-1.30}$\,mJy) of
MODS31-0033 is consistent with that expected from the 24\,$\mu$m flux
(1.6\,mJy), when the \citet{Chary2001} SED templates are assumed, as
though the SFR(H$\alpha$) is lower than the SFR(IR). None of our
samples including MODS31-0033 are detected at 850\,$\mu$m, in spite
that some of them have SFR(H$\alpha$) and/or SFR(IR) exceeding several
hundred ${\rm M}_\odot\,{\rm yr}^{-1}$. It is known that 850\,$\mu$m
selected galaxies seem to have cooler thermal emission than local
ULIRGs, which results from the fact that the 850\,$\mu$m flux is lower
than that expected from total IR luminosity
\citep{Pope2006}. Therefore, it is possible that 850\,$\mu$m emission
is not detected from all ultra-luminous starburst galaxies at $z\sim2$.

\subsubsection{X-ray}

The linear relation between X-ray luminosity and SFR is calibrated
with local galaxies \citep[e.g.,][]{Grimm2003,Ranalli2003}. X-ray
emission of star-forming galaxies is dominated by high mass X-ray
binaries (HMXB), whose progenitor masses are $\gtrsim 8\,{\rm
  M}_\odot$ and thus they are short-lived. The correlation of X-ray
luminosity with UV \citep{Daddi2004b} and H$\alpha$ \citep{Erb2006c}
luminosity have also been investigated for high-z galaxies.

Due to the sample selection, the present samples are not detected in
the {\it Chandra} 2\,Ms catalog, whose limiting flux is
$2.5\times10^{-17}\,{\rm erg\,cm}^{-2}\,{\rm s}^{-1}$ in soft X-ray
band (0.5--2.0\,keV), which detects the redshifted hard X-ray fluxes
of objects at $z\sim2$. According to the calibration by
\citet{Grimm2003}, the flux corresponds to SFR(X-ray)$=74-186\,{\rm
  M}_\odot\,{\rm yr}^{-1}$ for our galaxies (the IMF is converted to
the \citet{Chabrier2003} IMF). Even if the errors of the SFRs are
taken into account, five of our samples (MODS11-0390, MODS21-2612,
MODS22-1411, MODS42-0112, MODS22-2658) have the extinction corrected
SFR(H$\alpha$)s with the \citet{Calzetti2001} relation larger than the
upper limit of the SFR(X-ray). The first four galaxies are high-SSFR
galaxies, and the galaxies except MODS42-0112 are also H$\alpha$
excess galaxies against radio flux. This would also suggest that the
amounts of extinctions are overestimated by using the
\citet{Calzetti2001} relation. We note that \citet{Erb2006c} reported
the higher H$\alpha$ luminosities of their BM/BX galaxies than the
X-ray luminosities. The evolution of the first massive stars to HMXB
takes $\sim10-100$\,Myr, which is comparable to the age of our
high-SSFR galaxies.

\subsubsection{Extinction and Age}
Extinction and age could be degenerate in SED fitting, because both
parameters make the SED redder. For example, degeneracy in extinction
and age is found in the probability distribution for some galaxies;
see the bottom panels of Figure \ref{fig:sed}. However, if we limit
the age to be larger than 50\,Myr in the SED fit, for example, then
the estimated value of extinction is almost the same as that of the
original fit, showing that the degeneracy has only a marginal effect
on the estimate of extinction. The choice of star-formation history
applied to the SED model also affects the estimate. We use
$\tau$-models in the fitting, but the constant star-formation (CSF)
model tends to result in larger attenuation. In fact, if we apply the
CSF model to our galaxies, the $E(B-V)$s for about half of the
galaxies with ${\rm age}\gtrsim1$\,Gyr become larger by $\Delta
E(B-V)>0.1$ than that with the $\tau$-model, while the $E(B-V)$s are
unchanged for younger galaxies. In short, the CSF model gives similar
or larger best-fit extinction than the $\tau$-model in
general. Although a detailed analysis of the rest-optical continuum
would reduce further the possible range of solutions for best-fit
parameters \citep{Kriek2006a}, it is difficult to measure the
continuum from the present spectrum data.

\subsubsection{Comparison with Other Studies}

\figsedmscomp
\figsedmsbz

Our result shows no correlation between stellar mass and SFR. However,
\citet{Daddi2007a} demonstrated that BzK-selected star-forming
galaxies have a tight correlation within 0.2\,dex using SFR derived
from UV and mid-IR. We compare our result for SFR and stellar mass
with the \citet{Daddi2007a} relation (Figure
\ref{fig:sed_ms_comp}). For the \citet{Calzetti2001} relation,
significant fractions of sBzK galaxies (4/18 and 4/19 for sBzK-MIPS
and sBzK-non-MIPS, respectively) deviate from the correlation. Massive
galaxies ($\gtrsim 10^{10}\,{\rm M}_\odot$) in our sample are
distributed around the relation given by \citet{Daddi2007a}, while
some of the less massive galaxies ($\lesssim 10^{10}\,{\rm M}_\odot$)
have much higher SFR than that expected from the correlation. Even for
the equal-extinction case, which makes the SFRs of the higher SFR
galaxies lower, there is still no correlation between stellar mass and
SFR like that of \citet{Daddi2007a}.

One of the reasons for the discrepancy could be the different
extinction estimates used in \citet{Daddi2007a}, who assumed a linear
correlation between $B-z$ color and $E(B-V)$ derived from their SED
fitting result \citep{Daddi2004b} to estimate extinction-corrected UV
luminosities. In fact, if we apply the $E(B-V)$ values estimated from
the $B-z$ color, the extinction corrected SFR(H$\alpha$) becomes
smaller than the $E(B-V)$ values estimated from our stellar population
analysis (Figure \ref{fig:sed_ms_bz}). Especially when the
equal-extinction case is applied, the extinction corrected
SFR(H$\alpha$)s are similar to the \citet{Daddi2007a}
relation. Although our $E(B-V)$ values correlate with $B-z$ color, the
correlation has larger scatter than that of \citet{Daddi2004b} (Figure
\ref{fig:ebv_bz}). The high-SSFR galaxies have larger $E(B-V)$ values
than the galaxies with similar $B-z$ color.

The \citet{Daddi2007a} relation is also supported by
\citet{Pannella2009}, who shows the averaged properties of extinction
from the radio-to-UV luminosity ratio using a large numbers ($\sim
12000$) of sBzK galaxies. They also find a strong correlation between
the stellar mass and the amount of extinction, and thus the
correlation accentuates the \citet{Daddi2007a} relation. In Figure
\ref{fig:ms_ebv}, the amount of extinction of our galaxies is shown as
a function of the stellar mass. Our results show a larger scatter than
the errors of the correlation by \citet{Pannella2009}. Although the
extinction values of the low-SSFR galaxies correlate with the stellar
masses, consistent with that of \citet{Pannella2009}, the high-SSFR
galaxies have higher extinctions than the low-SSFR galaxies with
similar stellar masses. Because our emission line galaxies would be
biased towards H$\alpha$ bright systems, galaxies that deviate from
the average relation shown by \citet{Pannella2009} can be found, which
probably results in the larger scatter.

\citet{Daddi2007a} also shows the same correlation using a SFR
indicator unrelated to extinction. They estimated the SFR from the sum
of the extinction uncorrected UV and mid-IR luminosities. On the other
hand, \citet{Caputi2006a} reported that there exist significant
fraction of lower mass ($\sim 10^{10}\,M_\odot$) galaxies with high
SSFRs ($>10^{-8}\,{\rm yr}^{-1}$) at $z\sim2$ using SFR inferred from
the 24\,$\mu$m flux.  \citet{Daddi2007a} argues that the correlation
is not accurately recovered with SFRs inferred only from mid-IR, due
to a substantial presence of MIPS-excess galaxies, which are excluded
from their sample. A MIPS-excess galaxy could harbor a heavily
obscured AGN \citep{Daddi2007b}. However, our results on the SFR
inferred from H$\alpha$ do not recover the correlation even if
MIPS-excess galaxies are excluded from the sample. We calculate the
MIPS excesses of our sample following \citet{Daddi2007a}, and found
that only a fraction of the high-SSFR galaxies show MIPS-excess. In
Figure \ref{fig:sed_ms_comp}, MIPS-excess galaxies are denoted by
white dots, which found to be more typically low-SSFR galaxies rather
than high-SSFR galaxies.

Previous studies on the relation between SFR and stellar mass using
the SFR derived from the H$\alpha$ luminosity show a similar trend to
ours. The SFRs and stellar masses of BM/BX galaxies \citep{Erb2006c}
and sBzK galaxies \citep{Hayashi2009} are also plotted on Figure
\ref{fig:sed_ms_comp}. We convert the SFR and stellar mass of these
studies to those calibrated with a \citet{Chabrier2003} IMF. In the
literature, they applied different extinction corrections to estimate
the H$\alpha$ luminosities from that used to estimate the stellar
extinction. The equal-extinction case is applied to the BM/BX
galaxies, while the \citet{Calzetti2001} relation to the sBzK
galaxies. For comparison, we applied the same extinction correction to
the samples in the literature as that for the present sample in each
panel. There are no correlations between SFR and stellar mass of the
samples in the both panels. However, some high-SSFR galaxies with
SSFR$>10^{-8}\,{\rm yr}^{-1}$ appear when the \citet{Calzetti2001}
relation is applied.

In the studies for BM/BX and sBzK galaxies, validity of the extinction
correction is confirmed with the consistency between SFR(H$\alpha$)
and SFR(UV) \citep{Erb2006c,Hayashi2009}. We also confirmed the
consistency between them in the previous section. The extinction
corrected SFR(H$\alpha$) values assuming the equal-extinction case are
consistent only for lower SFR galaxies (SFR$\lesssim100\,{\rm
  M}_\odot\,{\rm yr}^{-1}$). SFRs calculated assuming the
\citet{Calzetti2001} relation are broadly consistent, but the
SFR(H$\alpha$)s are systematically larger than the SFR(UV)s by
0.3\,dex. These are almost consistent with the previous SFR(H$\alpha$)
works.  Although \citet{Erb2006c} applied only the equal-extinction
case, galaxies with SFR(UV) $>100\,{\rm M}_\odot\,{\rm yr}^{-1}$ are
rare in their sample. On the other hand, \citet{Hayashi2009}, who used
the \citet{Calzetti2001} relation, also showed that their
SFR(H$\alpha$) values are systematically larger than SFR(UV) by a
factor of 3.

Recently, the relation between Balmer decrement and stellar continuum
attenuation has been investigated for galaxies at low-to-medium
redshift. \citet{Cowie2008} studied spectroscopically confirmed
$K$-selected sample at $0.05<z<0.475$. Their extinctions from the
Balmer decrement and the continuum extinctions are equal on
average. On the other hand, \citet{Moustakas2006a} studied nearby
galaxies, including starburst galaxies such as infrared-luminous
galaxies. Comparison between their extinctions shows large scatter,
but the average stellar-to-nebular extinction ratio is consistent with
the \citet{Calzetti2001} relation. The dust properties of the samples
in these two studies are quite different. For example, the median
reddening of \citet{Moustakas2006a} sample is $E(B-V)\sim0.24$, while
about half of the galaxies of \citet{Cowie2008} have
$E(B-V)<0.01$. The values of reddening of the high-z samples discussed
above and our samples are similar to that of \citet{Moustakas2006a},
but vary with the sample selection. The median values of reddening are
$E(B-V)\sim0.15,\,0.38$ for BM/BX galaxies by \citet{Erb2006c} and
sBzK galaxies by \citet{Hayashi2009}, respectively. This can imply
that the \citet{Calzetti2001} relation gives better estimation of the
value of reddening for ionized gas in the active star-forming galaxies
at high redshift.

\figebvbz
\figmsebv

\subsection{Metallicity}
\label{subsec:mzr}

The gas-phase metallicity reflects the past star-formation history of
a galaxy since the heavy elements are produced by nuclear fusion in
stars and fed back into gas. Therefore, it is expected that the
gas-phase metallicity of galaxies in the high-z universe is lower than
that in the local Universe. The metallicities were found to be lower
in high-z galaxies than those in local galaxies
\citep[e.g.,][]{Shapley2004,Savaglio2005,Erb2006a,Maiolino2008,Hayashi2009,PerezMontero2008,Lamareille2008}.
We measured the gas-phase metallicity of our samples to investigate
their star-formation histories.

The gas-phase metallicity is estimated from the [N\,{\sc
ii}]$\lambda$6583-H$\alpha$ line ratio (N2 index), defined by
\begin{equation}
 N2 \equiv \log{\{[\mathrm{N_{II}}]\lambda 6583/\mathrm{H\alpha}\}},
\end{equation}
which correlates with oxygen abundance \citep{StorchiBergmann1994}. The
relation between the N2 index and gas-phase oxygen abundance is given by
\begin{equation}
 12+\log{(\mathrm{O/H})} = 8.90 + 0.57 \times N2
\end{equation}
\citep{Pettini2004}. [N\,{\sc ii}] and H$\alpha$ are so close to each
other in wavelength that the ratio is not affected by
reddening. Therefore, the N2 index has been used to estimate the
oxygen abundances of distant galaxies
\citep[e.g.,][]{Shapley2004,Erb2006a,Hayashi2009}.  We note that the
N2 index saturates at $\sim 1\,{\rm Z}_\odot$ \citep[$12+\log(\rm
  O/H)=8.66;$][]{Asplund2004}.

In order to gain a sufficiently high S/N to reliably measure the line
flux ratio, we stacked the [N\,{\sc ii}]-H$\alpha$ emission lines. We
divided the present sample into three groups based on stellar masses
at $9\times10^9\,{\rm M}_\odot$ and $2.5\times10^{10}\,{\rm
  M}_\odot$. In addition, we also divide the sample into two groups
based on a SSFR of $10^{-8}\,{\rm yr}^{-1}$. The stacking, measurement
of flux, and estimates of error are performed in the same manner as
described in \S\ref{subsec:stack}. The results are plotted in Figure
\ref{fig:mzr} as a function of the median stellar mass of each
group. For comparison, $\sim39500$ local objects from SDSS
DR4\footnote{\url{http://www.mpa-garching.mpg.de/SDSS/DR4/}}, local
Lyman Break Analog (LBA) galaxies \citep{Overzier2010}, BM/BX galaxies
\citep{Erb2006a}, and sBzK galaxies \citep{Hayashi2009} are also
plotted with the same calibration.

The mass-metallicity properties of our high-SSFR galaxies are possibly
different from that of rest-UV selected galaxies. The mass-metallicity
relation of the two largest mass bins is consistent with that of BM/BX
galaxies described by \citet{Erb2006a}, while the smallest mass bin
has higher metallicity than the similar mass bins of \citet{Erb2006a};
our sample is closer to the local galaxies. The discrepancy in the
smallest mass bin could be due to the galaxies with high SSFR. The
high-SSFR group has higher metallicity, close to that of the local
galaxies with similar stellar mass. As shown in Figure
\ref{fig:sed_ms}, high-SSFR galaxies dominate our sample in the
smallest mass bin, while most of the galaxies in \citet{Erb2006a} have
SSFR smaller than $10^{-8}\,{\rm yr}^{-1}$.

The $K$-band selected samples would provide many galaxies with high
SSFR and metallicity. The existence of higher metallicity galaxies in
$K$-selected sample is also supported by the observations of sBzK
galaxies by \citet{Hayashi2009}, as shown in Figure \ref{fig:mzr}. On
the other hand, the local LBAs, which are local analogs of high-z
UV-selected galaxies, tend to have a lower mass-metallicity
relation. This would suggest that the sample by \citet{Erb2006a} is
biased toward lower metallicities due to the rest-UV selection. The
results indicate that in order to reveal the unbiased cosmic evolution
of mass-metallicity relation, we need to evaluate metallicities of a
larger sample of $K$-band selected (i.e., mass-selected) galaxies.

It is surprising that the high-SSFR galaxies have already acquired
metallicities as high as local galaxies at $z\sim2$.  The age of the
stellar population of the high-SSFR galaxies is young, so that the
metal enrichment must have occurred in a short timescale. The N2 index
is known to be sensitive to a hard ionizing radiation field from an
AGN or strong shock excitation and ionization parameter
\citep{Kewley2002}. In fact, our stacked \othree$\lambda$5007/H$\beta$
and \ntwo$\lambda$6583/H$\alpha$ line ratios show evidence for a
harder ionizing spectrum. If the metallicity is really high, the
result is in conflict with the observations of local galaxies. These
galaxies can not keep their stellar masses and metallicities after the
observed epoch, because the studies for local galaxies demonstrate
that galaxies in a similar mass range formed a significant fraction
of their stars in the low redshift universe \citep{Heavens2004},
suggesting that these local galaxies must have increased their
metallicity recently. It is possible that such high-SSFR galaxies
evolved into larger stellar mass galaxies with low star-forming
efficiency in the local universe, depleting the gas and/or merging
with higher mass galaxies.

\figmzr

\subsection{Contribution to the Cosmic SFR Density}
\label{subsec:csfrd}

We estimate the contribution of the observed galaxies to the cosmic
SFR density at $z\sim2$ with the SFR(H$\alpha$)s, dividing the sample
into two groups with SSFR $>10^{-8}\,{\rm yr}^{-1}$ and
$<10^{-8}\,{\rm yr}^{-1}$ for the extinction corrected SFRs with the
\citet{Calzetti2001} relation. Since we observed only a limited number
of sBzK galaxies, we need to assume that the SFR and redshift
distributions of the other galaxies are the same as the observed
galaxies; this is not unreasonable since we randomly selected the
present sample from the sBzK-MIPS and sBzK-non-MIPS population. First,
we further divide the sample based on the MIPS detection in the public
catalog. Second, we calculated the total SFR in a certain group; we
multiply the total SFR in the group by the ratio of the number of the
entire galaxies in the group to that of the number of galaxies
observed. In the calculation of total SFR, we assumed that the SFR of
the galaxies without H$\alpha$ detection is negligible; we set those
SFRs to zero. Then the total SFRs of galaxies with and without MIPS
detection are summed. Third, we divide the total SFR by the expected
detection rate of H$\alpha$ emission line (87.7\%; see
\S\ref{subsec:haemission}) to recover the H$\alpha$ emission lines not
detected due to strong sky emission lines or atmospheric absorption
bands.  Finally, we divide the total SFRs by comoving volume of the
survey region for $1.4<z<2.5$.

The resultant total cosmic SFR densities of the sBzK galaxies are
$0.136^{+0.020}_{-0.028}$, $0.029^{+0.002}_{-0.002}$\, ${\rm
  M}_\odot\,{\rm yr}^{-1}\,{\rm Mpc}^{-3}$ for extinction correction
calculated using the \citet{Calzetti2001} relation and the
equal-extinction case, respectively. As we see in the previous
discussions, the relation between extinction estimated from the
stellar continuum and the emission lines probably depend on the
intrinsic luminosities, and thus the SFRs. To examine this effect,
when we apply extinction correction with the \citet{Calzetti2001}
relation for the sBzK-MIPS galaxies and equal extinction for the
sBzK-non-MIPS galaxies (differential correction), the result is
$0.089^{+0.008}_{-0.021}$. In any case of the extinction correction,
high-SSFR galaxies have a large contribution to the SFR
density. Although these galaxies account for only $\sim 20\,\%$ of all
of the galaxies, they contribute to $\sim 75\,\%$ of the SFR density
for the extinction correction with the \citet{Calzetti2001} relation
and the differential extinction. Even for the equal-extinction case,
they contribute to no less than $\sim 50\,\%$ of the SFR density.

We compare our SFR density at $z\sim2$ with previous works (e.g.,
\citealp{Hopkins2004,Hopkins2006} and the references therein;
\citealp{Wang2006,Caputi2007}). If we use the same calibration of IMF
as that of these studies, the SFR densities are 0.203, 0.044,
0.134\,${\rm M}_\odot\,{\rm yr}^{-1}\,{\rm Mpc}^{-3}$ for the
\citet{Calzetti2001} relation, the equal-extinction case, and the
differential correction, respectively. The values compiled by
\citet{Hopkins2004} and \citet{Hopkins2006} are 0.05-0.38\,${\rm
  M}_\odot\,{\rm yr}^{-1}\,{\rm Mpc}^{-3}$, which are consistent with
both of those with the \citet{Calzetti2001} relation and differential
extinction. Recent studies on the SFR density at $z\sim2$ using IR
luminosities (850\,$\mu$m; \citealp{Wang2006}; 24\,$\mu$m;
\citealp{Caputi2007}) show slightly lower SFR densities
(0.08-0.09\,${\rm M}_\odot\,{\rm yr}^{-1}\,{\rm Mpc}^{-3}$), which are
still consistent with the differential extinction correction. This
would also suggest that the relation between the dust properties of
the stellar continuum and the emission lines are different depending
on the intrinsic SFR. The SFR density using the equal-extinction case
is lower than that of any study. The lower SFR density implies that
applying the equal-extinction case to all galaxies underestimates
SFR(H$\alpha$).

\section{Summary and Conclusions}
\label{sec:sum}
We performed near-infrared spectroscopic observations for 37 $BzK$
color selected star-forming galaxies with MOIRCS on the Subaru
Telescope. The sample galaxies are picked from the deep $K_s$-selected
catalog of MODS. We select 18 of our sample from the publicly
available 24\,$\mu$m source catalog of MIPS on board the $Spitzer$
Space Telescope. In order to avoid contamination by AGNs, we exclude
the BzKs detected in the {\it Chandra} Deep Field-North 2\,Ms
Point-Source Catalogs. All of the sample galaxies have strong
1.6\,$\mu$m bumps in their SEDs, so that their optical emission is
thought to be dominated by the stellar continuum. The H$\alpha$
emission lines are detected from 23 galaxies, of which the median
redshift is $z=2.12$. In the diagnostic diagram with
\othree$\lambda$5007/H$\beta$ and \ntwo$\lambda$6583/H$\alpha$, the
line ratios of the stacked spectra are above that found for local
star-forming galaxies but closer to the expectations from
star-formation activity rather than AGN activity.

We derived the SFRs from the extinction-corrected H$\alpha$
luminosities. The extinction correction for the H$\alpha$ luminosity
is estimated from the SED fitting of multi-band photometric data with
stellar population models of exponentially declining SFR and the
\citet{Calzetti2000} extinction curve. To estimate the extinction
correction for the ionized gas, we assumed the \citet{Calzetti2001}
relation ($E(B-V)_{\rm stellar} = 0.44\times E(B-V)_{\rm gas}$) as
well as the equal-extinction case $E(B-V)_{\rm stellar} = E(B-V)_{\rm
  gas}$, and compared the extinction corrected SFRs. Although the
Balmer decrement of the stacked emission lines is consistent with the
\citet{Calzetti2001} relation, a fraction of the H$\beta$ detected
galaxies are consistent with the equal-extinction case. We further
investigated these galaxies by comparing the extinction corrected
H$\alpha$ luminosities with other SFR indicators. The comparison with
the UV and MIPS 24\,$\mu$m flux shows that the SFRs of lower SFR
galaxies ($\lesssim100\,{\rm M}_\odot\,{\rm yr}^{-1}$) are in good
agreement for the equal-extinction case, while those of the higher SFR
galaxies agree for the \citet{Calzetti2001} relation. This implies
that the relation between the dust properties of the stellar continuum
and emission lines are different depending on the intrinsic SFR
(differential extinction correction). If the Calzetti relation is
assumed, a fraction of the high-SFR galaxies should be detected in
radio, (sub-)mm, and X-ray catalogs, but none of them except one AzTEC
source are detected. We could not explain the reason for this
discrepancy consistently. However, most of the discrepancy could be
explained by the short timescale of star-formation activity, because
the younger ages of the stellar population are derived from our SED
analysis. We note that other mechanisms (e.g., top-heavy IMF, lower
dust temperature, etc.) could also explain a part of the discrepancy.

We compared the extinction-corrected H$\alpha$ luminosities with the
stellar masses obtained from the SED fitting. When we apply the
\citet{Calzetti2001} relation, the comparison shows no correlation
between SFR and stellar mass. A fraction of the galaxies with stellar
mass smaller than $\sim10^{10}\,{\rm M}_\odot$ show SFRs higher than
$\sim100\,{\rm M}_\odot\,{\rm yr}^{-1}$. The SSFRs of these galaxies
are higher than $\sim10^{-8}\,{\rm yr}^{-1}$. Even if equal extinction
is applied, their SSFRs are still high (${\rm
  SSFR}>3.5\times10^{-9}\,{\rm yr}^{-1}$). This is inconsistent with
the correlation demonstrated by \citet{Daddi2007a}. Our value of
extinction correction shows large scatter against $B-z$ color and
stellar mass, with which \citet{Daddi2004b} and \citet{Pannella2009}
show strong correlations. From the best-fit parameters of the SED
fitting for these high-SSFR galaxies, we found that the average age of
the stellar population is younger than 100\,Myr, which are consistent
with SFR age (${\rm SSFR}^{-1}$) with the \citet{Calzetti2001}
relation. We also calculate the contribution of the present H$\alpha$
emission to the cosmic SFR density at $z\sim2$. Our results are
consistent with previous works, if we apply the \citet{Calzetti2001}
relation or differential extinction correction. The high-SSFR galaxies
significantly contribute ($50-75\%$) to the cosmic SFR density.

The metallicity of the high-SSFR galaxies, which is estimated from the
N2 index, is larger than that expected from the mass-metallicity
relation of UV-selected galaxies at $z\sim2$ by \citet{Erb2006a}.
Although metallicities estimated from the N2 index is known to have
significant uncertainties due to the harder ionizing radiation field
and ionization parameter, the high metallicity comparable to galaxies
in the same mass range is inconsistent with the known star-formation
history of such galaxies in the local universe, which formed a
significant fraction of their stars in the low redshift universe. The
high-SSFR galaxies, therefore, possibly evolve into larger stellar
mass galaxies with low star-forming efficiency, depleting the gas.

\acknowledgements
We thank the anonymous referee for valuable comments.
We thank the staff of the Subaru Telescope for their assistance with
the development and observation of MOIRCS.
MA is supported by the Ministry of Education, Culture, Science
and Technology (MEXT) Grant-in-Aid for Young Scientist (B), 18740118,
2006-2008.
DMA is supported by the Royal Society and Leverhulme Trust for
financial support.
Some of the data presented in this paper were obtained from the
Multimission Archive at the Space Telescope Science Institute
(MAST). STScI is operated by the Association of Universities for
Research in Astronomy, Inc., under NASA contract NAS5-26555.  Support
for MAST for non-HST data is provided by the NASA Office of Space
Science via grant NAG5-7584 and by other grants and contracts.
This work is based in part on archival data obtained with the Spitzer
Space Telescope, which is operated by the Jet Propulsion Laboratory,
California Institute of Technology under a contract with NASA.
This work makes use of data products from the Two Micron All Sky Survey,
which is a joint project of the University of Massachusetts and the
Infrared Processing and Analysis Center/California Institute of
Technology, funded by the National Aeronautics and Space Administration
and the National Science Foundation.
This work makes use of archival data provided by the Sloan Digital Sky
Survey (SDSS). The SDSS Web Site is \url{http://www.sdss.org/}.

{\it Facilities:} \facility{Subaru (MOIRCS, Suprime-Cam)}, \facility{HST (ACS)},
\facility{Spitzer (MIPS, IRAC)}, \facility{2MASS ()},
\facility{Sloan ()}


\clearpage

\clearpage


\begin{thebibliography}{999}
\bibitem[Adelman-McCarthy et al.(2008)]{SDSSDR6} Adelman-McCarthy,
  J.~K., et al.\ 2008, \apjs, 175, 297


\bibitem[Alexander et al.(2002)]{Alexander2002} Alexander, D.~M.,
  Aussel, H., Bauer, F.~E., Brandt, W.~N., Hornschemeier, A.~E.,
  Vignali, C., Garmire, G.~P., \& Schneider, D.~P.\ 2002, \apjl, 568,
  L85


\bibitem[Alexander et al.(2003)]{Alexander2003} Alexander, D.~M., et
  al.\ 2003, \aj, 126, 539


\bibitem[Asplund et al.(2004)]{Asplund2004} Asplund, M., Grevesse, N.,
  Sauval, A.~J., Allende Prieto, C., \& Kiselman, D.\ 2004, \aap, 417,
  751


\bibitem[Baldwin et al.(1981)]{Baldwin1981} Baldwin, J.~A., Phillips,
  M.~M., \& Terlevich, R.\ 1981, \pasp, 93, 5


\bibitem[Barger et al.(2008)]{Barger2008} Barger, A.~J., Cowie, L.~L.,
  \& Wang, W.-H.\ 2008, \apj, 689, 687


\bibitem[Biggs \& Ivison(2006)]{Biggs2006} Biggs, A.~D., \& Ivison,
  R.~J.\ 2006, \mnras, 371, 963


\bibitem[Bolzonella et al.(2000)]{Bolzonella2000} Bolzonella, M.,
  Miralles, J.-M., \& Pell{\'o}, R.\ 2000, \aap, 363, 476


\bibitem[Bruzual \& Charlot(2003)]{Bruzual2003} Bruzual, G., \&
  Charlot, S.\ 2003, \mnras, 344, 1000


\bibitem[Calzetti(2001)]{Calzetti2001} Calzetti, D.\ 2001, \pasp, 113,
  1449


\bibitem[Calzetti et al.(2000)]{Calzetti2000} Calzetti, D., Armus, L.,
  Bohlin, R.~C., Kinney, A.~L., Koornneef, J., \& Storchi-Bergmann,
  T.\ 2000, \apj, 533, 682


\bibitem[Capak et al.(2004)]{Capak2004} Capak, P., et al.\ 2004, \aj,
  127, 180

\bibitem[Caputi et al.(2006a)]{Caputi2006a} Caputi, K.~I., et
  al.\ 2006a, \apj, 637, 727

\bibitem[Caputi et al.(2006b)]{Caputi2006c} Caputi, K.~I., Dole, H.,
  Lagache, G., McLure, R.~J., Dunlop, J.~S., Puget, J.-L., Le Floc'h,
  E., \& P{\'e}rez-Gonz{\'a}lez, P.~G.\ 2006b, \aap, 454, 143

\bibitem[Caputi et al.(2007)]{Caputi2007} Caputi, K.~I., et al.\ 2007,
  \apj, 660, 97


\bibitem[Castelli \& Kurucz(2004)]{Castelli2003} Castelli, F., \&
  Kurucz, R.~L.\ 2004, arXiv:astro-ph/0405087


\bibitem[Chabrier(2003)]{Chabrier2003} Chabrier, G.\ 2003, \pasp, 115,
  763


\bibitem[Chapin et al.(2009)]{Chapin2009} Chapin, E.~L., et al.\ 2009,
  \mnras, 398, 1793


\bibitem[Chapman et al.(2005)]{Chapman2005} Chapman, S.~C., Blain,
  A.~W., Smail, I., \& Ivison, R.~J.\ 2005, \apj, 622, 772


\bibitem[Chary \& Elbaz(2001)]{Chary2001} Chary, R., \& Elbaz,
  D.\ 2001, \apj, 556, 562


\bibitem[Cohen(2001)]{Cohen2001} Cohen, J.~G.\ 2001, \aj, 121, 2895


\bibitem[Cohen et al.(2000)]{Cohen2000} Cohen, J.~G., Hogg, D.~W.,
  Blandford, R., Cowie, L.~L., Hu, E., Songaila, A., Shopbell, P., \&
  Richberg, K.\ 2000, \apj, 538, 29


\bibitem[Condon(1992)]{Condon1992} Condon, J.~J.\ 1992, \araa, 30, 575


\bibitem[Cowie \& Barger(2008)]{Cowie2008} Cowie, L.~L., \& Barger,
  A.~J.\ 2008, \apj, 686, 72

\bibitem[Cowie et al.(2004)]{Cowie2004} Cowie, L.~L., Barger, A.~J.,
  Hu, E.~M., Capak, P., \& Songaila, A.\ 2004, \aj, 127, 3137


\bibitem[Daddi et al.(2004)]{Daddi2004b} Daddi, E., Cimatti, A.,
  Renzini, A., Fontana, A., Mignoli, M., Pozzetti, L., Tozzi, P., \&
  Zamorani, G.\ 2004, \apj, 617, 746


\bibitem[Daddi et al.(2007a)]{Daddi2007a} Daddi, E., et al.\ 2007a,
  \apj, 670, 156


\bibitem[Daddi et al.(2007b)]{Daddi2007b} Daddi, E., et al.\ 2007b,
  \apj, 670, 173


\bibitem[Dahlen et al.(2007)]{Dahlen2007} Dahlen, T., Mobasher, B.,
  Dickinson, M., Ferguson, H.~C., Giavalisco, M., Kretchmer, C., \&
  Ravindranath, S.\ 2007, \apj, 654, 172


\bibitem[Dav{\'e}(2008)]{Dave2008} Dav{\'e}, R.\ 2008, \mnras, 385,
  147


\bibitem[Dawson et al.(2001)]{Dawson2001} Dawson, S., Stern, D.,
  Bunker, A.~J., Spinrad, H., \& Dey, A.\ 2001, \aj, 122, 598


\bibitem[Denicol{\'o} et al.(2002)]{Denicolo2002} Denicol{\'o}, G.,
  Terlevich, R., \& Terlevich, E.\ 2002, \mnras, 330, 69


\bibitem[Erb et al.(2006a)]{Erb2006a} Erb, D.~K., Shapley, A.~E.,
  Pettini, M., Steidel, C.~C., Reddy, N.~A., \& Adelberger,
  K.~L.\ 2006a, \apj, 644, 813


\bibitem[Erb et al.(2003)]{Erb2003} Erb, D.~K., Shapley, A.~E.,
  Steidel, C.~C., Pettini, M., Adelberger, K.~L., Hunt, M.~P.,
  Moorwood, A.~F.~M., \& Cuby, J.-G.\ 2003, \apj, 591, 101


\bibitem[Erb et al.(2006b)]{Erb2006c} Erb, D.~K., Steidel, C.~C.,
  Shapley, A.~E., Pettini, M., Reddy, N.~A., \& Adelberger,
  K.~L.\ 2006b, \apj, 647, 128


\bibitem[F{\"o}rster Schreiber et al.(2009)]{ForsterSchreiber2009}
  F{\"o}rster Schreiber, N.~M., et al.\ 2009, \apj, 706, 1364


\bibitem[Giavalisco et al.(2004)]{Giavalisco2004} Giavalisco, M., et
  al.\ 2004, \apjl, 600, L93


\bibitem[Grazian et al.(2007)]{Grazian2007} Grazian, A., et al.\ 2007,
  \aap, 465, 393


\bibitem[Greve et al.(2008)]{Greve2008} Greve, T.~R., Pope, A., Scott,
  D., Ivison, R.~J., Borys, C., Conselice, C.~J., \& Bertoldi,
  F.\ 2008, \mnras, 389, 1489


\bibitem[Grimm et al.(2003)]{Grimm2003} Grimm, H.-J., Gilfanov, M., \&
  Sunyaev, R.\ 2003, \mnras, 339, 793


\bibitem[Hayashi et al.(2009)]{Hayashi2009} Hayashi, M., et al.\ 2009,
  \apj, 691, 140


\bibitem[Heavens et al.(2004)]{Heavens2004} Heavens, A., Panter, B.,
  Jimenez, R., \& Dunlop, J.\ 2004, \nat, 428, 625


\bibitem[Holland et al.(1999)]{Holland1999} Holland, W.~S., et
  al.\ 1999, \mnras, 303, 659


\bibitem[Hopkins(2004)]{Hopkins2004} Hopkins, A.~M.\ 2004, \apj, 615,
  209


\bibitem[Hopkins \& Beacom(2006)]{Hopkins2006} Hopkins, A.~M., \&
  Beacom, J.~F.\ 2006, \apj, 651, 142


\bibitem[Iye et al.(2004)]{Iye2004} Iye, M., et al.\ 2004, \pasj, 56,
  381


\bibitem[Kajisawa et al.(2009)]{Kajisawa2009} Kajisawa, M., et
  al.\ 2009, \apj, 702, 1393


\bibitem[Kauffmann et al.(2003)]{Kauffmann2003} Kauffmann, G., et
  al.\ 2003, \mnras, 346, 1055


\bibitem[Kennicutt(1998)]{Kennicutt1998} Kennicutt, R.~C., Jr.\ 1998,
  \araa, 36, 189


\bibitem[Kewley et al.(2001)]{Kewley2001a} Kewley, L.~J., Dopita,
  M.~A., Sutherland, R.~S., Heisler, C.~A., \& Trevena, J.\ 2001,
  \apj, 556, 121


\bibitem[Kewley \& Dopita(2002)]{Kewley2002} Kewley, L.~J., \& Dopita,
  M.~A.\ 2002, \apjs, 142, 35


\bibitem[Kreysa et al.(1998)]{Kreysa1998} Kreysa, E., et al.\ 1998,
  \procspie, 3357, 319


\bibitem[Kriek et al.(2006)]{Kriek2006a} Kriek, M., et al.\ 2006,
  \apj, 645, 44


\bibitem[Kriek et al.(2007)]{Kriek2007} Kriek, M., et al.\ 2007, \apj,
  669, 776


\bibitem[Kriss(1994)]{Kriss1994} Kriss, G.\ 1994, Astronomical Data
  Analysis Software and Systems, 3, 437


\bibitem[Lamareille et al.(2009)]{Lamareille2008} Lamareille, F., et
  al.\ 2009, \aap, 495, 53


\bibitem[Maiolino et al.(2008)]{Maiolino2008} Maiolino, R., et
  al.\ 2008, \aap, 488, 463


\bibitem[Maraston(2005)]{Maraston2005} Maraston, C.\ 2005, \mnras,
  362, 799


\bibitem[Moustakas \& Kennicutt(2006)]{Moustakas2006a} Moustakas, J.,
  \& Kennicutt, R.~C., Jr.\ 2006, \apjs, 164, 81


\bibitem[Osterbrock(1989)]{Osterbrock1989} Osterbrock, D.~E.\ 1989,
  Research supported by the University of California, John Simon
  Guggenheim Memorial Foundation, University of Minnesota, et al.~Mill
  Valley, CA, University Science Books, 1989, 422 p.,


\bibitem[Overzier et al.(2010)]{Overzier2010} Overzier, R.~A.,
  Heckman, T.~M., Schiminovich, D., Basu-Zych, A., Gon{\c c}alves, T.,
  Martin, D.~C., \& Rich, R.~M.\ 2010, \apj, 710, 979


\bibitem[Pannella et al.(2009)]{Pannella2009} Pannella, M., et
  al.\ 2009, \apjl, 698, L116

\bibitem[Papovich et al.(2007)]{Papovich2007} Papovich, C., et
  al.\ 2007, \apj, 668, 45


\bibitem[P{\'e}rez-Gonz{\'a}lez et al.(2005)]{PerezGonzalez2005}
  P{\'e}rez-Gonz{\'a}lez, P.~G., et al.\ 2005, \apj, 630, 82


\bibitem[P{\'e}rez-Gonz{\'a}lez et al.(2008)]{PerezGonzalez2008}
  P{\'e}rez-Gonz{\'a}lez, P.~G., et al.\ 2008, \apj, 675, 234


\bibitem[P{\'e}rez-Montero et al.(2009)]{PerezMontero2008}
  P{\'e}rez-Montero, E., et al.\ 2009, \aap, 495, 73


\bibitem[Peeters et al.(2004)]{Peeters2004} Peeters, E., Spoon,
  H.~W.~W., \& Tielens, A.~G.~G.~M.\ 2004, \apj, 613, 986


\bibitem[Perera et al.(2008)]{Perera2008} Perera, T.~A., et al.\ 2008,
  \mnras, 391, 1227


\bibitem[Pettini \& Pagel(2004)]{Pettini2004} Pettini, M., \& Pagel,
  B.~E.~J.\ 2004, \mnras, 348, L59


\bibitem[Pope et al.(2005)]{Pope2005} Pope, A., Borys, C., Scott, D.,
  Conselice, C., Dickinson, M., \& Mobasher, B.\ 2005, \mnras, 358,
  149


\bibitem[Pope et al.(2006)]{Pope2006} Pope, A., et al.\ 2006, \mnras,
  370, 1185


\bibitem[Raimann et al.(2000)]{Raimann2000} Raimann, D.,
  Storchi-Bergmann, T., Bica, E., Melnick, J., \& Schmitt, H.\ 2000,
  \mnras, 316, 559


\bibitem[Ranalli et al.(2003)]{Ranalli2003} Ranalli, P., Comastri, A.,
  \& Setti, G.\ 2003, \aap, 399, 39


\bibitem[Reddy et al.(2006a)]{Reddy2006a} Reddy, N.~A., Steidel,
  C.~C., Fadda, D., Yan, L., Pettini, M., Shapley, A.~E., Erb, D.~K.,
  \& Adelberger, K.~L.\ 2006a, \apj, 644, 792


\bibitem[Reddy et al.(2006b)]{Reddy2006b} Reddy, N.~A., Steidel,
  C.~C., Erb, D.~K., Shapley, A.~E., \& Pettini, M.\ 2006b, \apj, 653,
  1004


\bibitem[Rousselot et al.(2000)]{Rousselot2000} Rousselot, P., Lidman,
  C., Cuby, J.-G., Moreels, G., \& Monnet, G.\ 2000, \aap, 354, 1134


\bibitem[Salpeter(1955)]{Salpeter1955} Salpeter, E.~E.\ 1955, \apj,
  121, 161


\bibitem[Savaglio et al.(2005)]{Savaglio2005} Savaglio, S., et
  al.\ 2005, \apj, 635, 260


\bibitem[Sawicki(2002)]{Sawicki2002} Sawicki, M.\ 2002, \aj, 124, 3050


\bibitem[Shapley et al.(2004)]{Shapley2004} Shapley, A.~E., Erb,
  D.~K., Pettini, M., Steidel, C.~C., \& Adelberger, K.~L.\ 2004,
  \apj, 612, 108


\bibitem[Skrutskie et al.(2006)]{Skrutskie2006} Skrutskie, M.~F., et
  al.\ 2006, \aj, 131, 1163


\bibitem[Smartt et al.(2009)]{Smartt2009} Smartt, S.~J., Eldridge,
  J.~J., Crockett, R.~M., \& Maund, J.~R.\ 2009, \mnras, 395, 1409


\bibitem[Stetson(1992)]{Stetson1992} Stetson, P.~B.\ 1992,
  Astronomical Data Analysis Software and Systems I, 25, 297


\bibitem[Storchi-Bergmann et al.(1994)]{StorchiBergmann1994}
  Storchi-Bergmann, T., Calzetti, D., \& Kinney, A.~L.\ 1994, \apj,
  429, 572


\bibitem[Suzuki et al.(2008)]{Suzuki2008} Suzuki, R., et al.\ 2008,
  \pasj, 60, 1347


\bibitem[Swinbank et al.(2006)]{Swinbank2006b} Swinbank, A.~M.,
  Chapman, S.~C., Smail, I., Lindner, C., Borys, C., Blain, A.~W.,
  Ivison, R.~J., \& Lewis, G.~F.\ 2006, \mnras, 371, 465


\bibitem[Swinbank et al.(2005)]{Swinbank2005b} Swinbank, A.~M., et
  al.\ 2005, \mnras, 359, 401


\bibitem[{Tokoku(2007)}]{Tokoku2007} Tokoku~C. 2007, PhD thesis,
  Tohoku University


\bibitem[Treu et al.(2005)]{Treu2005} Treu, T., Ellis, R.~S., Liao,
  T.~X., \& van Dokkum, P.~G.\ 2005, \apjl, 622, L5


\bibitem[Veilleux \& Osterbrock(1987)]{Veilleux1987} Veilleux, S., \&
  Osterbrock, D.~E.\ 1987, \apjs, 63, 295


\bibitem[Wang et al.(2006)]{Wang2006} Wang, W.-H., Cowie, L.~L., \&
  Barger, A.~J.\ 2006, \apj, 647, 74


\bibitem[Wilson et al.(2008)]{Wilson2008} Wilson, G.~W., et al.\ 2008,
  \mnras, 386, 807


\bibitem[Wirth et al.(2004)]{Wirth2004} Wirth, G.~D., et al.\ 2004,
  \aj, 127, 3121


\bibitem[Yun et al.(2001)]{Yun2001} Yun, M.~S., Reddy, N.~A., \&
  Condon, J.~J.\ 2001, \apj, 554, 803

\end{thebibliography}
\end{document}